\documentclass[letterpaper,twocolumn,showpacs,amsmath,amssymb]{rt4}

\makeatletter
\def\ignorecitefornumbering#1{%
     \begingroup
         \@fileswfalse
         #1
    \endgroup
}
\makeatother

\usepackage{graphicx,color}
\usepackage[colorlinks,plainpages=false,linkcolor=black,urlcolor=blue,citecolor=blue,pdfpagemode=UseNone,pdfstartview=FitH]{hyperref}

\begin{document}

\title{Pseudogap from ARPES experiment: three gaps in cuprates \\and topological superconductivity}

\author{A.~A.~Kordyuk}
\affiliation{Institute of Metal Physics of National Academy of Sciences of Ukraine, 03142 Kyiv, Ukraine}

\begin{abstract}
A term first coined by Mott back in 1968 a `pseudogap' is the depletion of the electronic density of states at the Fermi level, and pseudogaps have been observed in many systems. However, since the discovery of the high temperature superconductors (HTSC) in 1986, the central role attributed to the pseudogap in these systems has meant that by many researchers now associate the term pseudogap exclusively with the HTSC phenomenon. Recently, the problem has got a lot of new attention with the rediscovery of two distinct energy scales ('two-gap scenario') and charge density waves patterns in the cuprates. Despite many excellent reviews on the pseudogap phenomenon in HTSC, published from its very discovery up to now, the mechanism of the pseudogap and its relation to superconductivity are still open questions. The present review represents a contribution dealing with the pseudogap, focusing on results from angle resolved photoemission spectroscopy (ARPES) and ends up with the conclusion that the pseudogap in cuprates is a complex phenomenon which includes at least three different `intertwined' orders: spin and charge density waves and preformed pairs, which appears in different parts of the phase diagram. The density waves in cuprates are competing to superconductivity for the electronic states but, on the other hand, should drive the electronic structure to vicinity of Lifshitz transition, that could be a key similarity between the superconducting cuprates and iron based superconductors. One may also note that since the pseudogap in cuprates has multiple origins there is no need to recoin the term suggested by Mott.
\end{abstract}


\maketitle

\tableofcontents

\section{Introduction}

The term pseudogap was suggested by Nevill Mott in 1968 \cite{MottRMP1968} to name a minimum in the electronic density of states (DOS) of liquid mercury at the Fermi level. Later he had shown that when this pseudogap is deep enough the one-electron states become localized \cite{MottPM1969}.

\begin{figure}
\begin{center}
\includegraphics[width=0.36\textwidth]{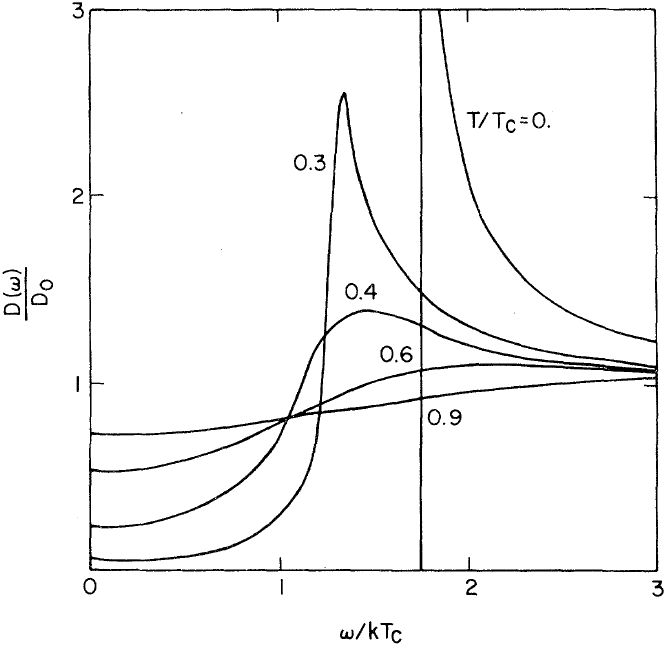}
\caption{The electronic density of states normalized to the metallic density of state, plotted versus $\omega/k T_c$, for various temperatures. The $T/T_c = 0$ curve is the mean-field result. After \protect\ignorecitefornumbering{\cite{1973_PRL_Lee}}.
\label{FluctPG}}
\end{center}
\end{figure}

Next, the term pseudogap was narrowed to `\emph{fluctuating band gap}', the gap formed by fluctuating charge density wave (CDW) at a Peierls transition \cite{1930_AdP_Peierls} in quasi-one-dimensional (1D) metals \cite{1973_SSC_Rice, 1973_PRL_Lee, 1974_SP_Sadovskii, 1978_PR_Toombs}, as shown in Fig.\;\ref{FluctPG}.

\begin{figure*}
\begin{center}
\includegraphics[width=1\textwidth]{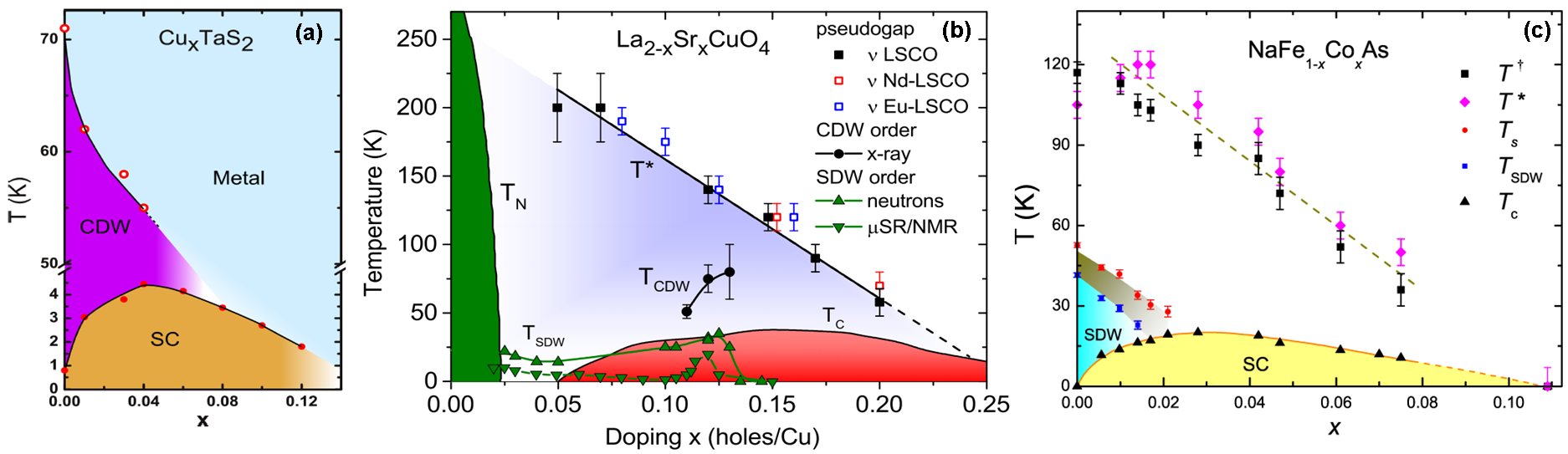}
\caption{Examples of the phase diagrams of quasi 2D metals in which the charge or spin ordering compete or coexist with superconductivity and a pseudogap phase: a transition metal dichalcogenide \protect\ignorecitefornumbering{\cite{2008_PRB_Wagner}} (a), a high-$T_c$ cuprate \protect\ignorecitefornumbering{\cite{2014_PRB_Croft}} (b), and an iron based superconductor \protect\ignorecitefornumbering{\cite{2013_NJP_Wang}} (c).
\label{PhDs}}
\end{center}
\end{figure*}

In fact, the systems with fluctuating CDW can be described similarly to disordered systems without long-range order \cite{Sadovskii1974}, so, the pseudogap should not be necessarily related with low dimensionality. Indeed, in quasi-two-dimensional (2D) metals with CDW ordering, such as transition metal dichalcogenides (TMD) \cite{1975_AiP_Wilson} (see also recent review \cite{2011_JPCM_Rossnagel}), the fluctuation effects are considered negligible but a partial gap, which can be called `pseudogap' according to the Mott's definition, appears in a number of CDW phases. Two such kinds of pseudogaps have been discussed: traditional Peierls gap but smeared out due to \emph{incommensurability} \cite{BorisenkoPRL2008, 2009_PRL_Borisenko} (or, may be, short-range-order CDW fluctuations \cite{2012_JETP_Kuchinskii} as in `nearly commensurate' \cite{1977_JPSJ_Nakanishi, 2014_PRB_Arguello} or `quasicommensurate' \cite{1999_PRL_Pillo} CDW state); and a `\emph{correlation gap}' of Mott-Hubbard insulating phase in a commensurate CDW state \cite{1992_PRB_Dardel, 1999_PRL_Pillo, 2003_PRL_Perfetti}.

Curiously, except the study of fluctuating effects in 1D CDW compounds, the pseudogap phenomena in 2D CDW systems, despite a variety of the aforementioned possibilities, had not earned so much attention \cite{1975_AiP_Wilson, Gruner} as it had done later in the field of high temperature superconductors (HTSC) \cite{TimuskRPP1999, LoktevPR2001, 2001_P_Sadovskii, GabovichSST2001, GabovichPR2002, 2004_xxx_Sadovskii, NormanAP2005, Huefner2008, plakida2010high} for which it is often considered unique \cite{Huefner2008}. On one hand, the discovery of the superconducting cuprates (Cu-SC) slowed down noticeably the study of the CDW-materials. On the other hand, the role of the pseudogap in HTSC might be greatly exaggerated---partly due to real complexity of the phenomenon but partly because a lot of people struggling to find the mechanism of high temperature superconductivity needed a `guilty' why that has appeared to be so hard. In this sense, the well-turned definition of the pseudogap in cuprates as ``\emph{a not-understood suppression of excited states}" was given by Robert Laughlin in early years of HTSC era \cite{1996_PT_Levi}.

Nevertheless, the pseudogap phenomenon in cuprates has stimulated appearance of many fascinating theories (some of which will be briefly overviewed in Sec.\ref{theor}), and has been extended to a number of other materials, for example, A15 superconductors \cite{2014_LTP_Ekino}, manganites \cite{1998_PRL_Dessau, 2005_N_Mannella}, Kondo insulators \cite{1996_PRL_Susaki, 2014_xxx_Okawa}, thin films of conventional superconductors \cite{2010_NC_Sacepe} and nanoislands \cite{2009_PRL_Wang, 2013_SST_Liu}, Co-Fe-based half metals \cite{2012_PRX_Mann}, ultracold Fermi gases \cite{2011_PRL_Magierski}.

In many of those systems the pseudogap phenomenon is discussed as a pseudogap phase on the \emph{phase diagrams} of temperature \emph{vs} charge carrier concentration (also called `doping') or \emph{vs} pressure, where the pseudogap phase neighbors both the density wave and superconducting phases. Fig.\;\ref{PhDs} shows three recent examples of such phase diagrams for a transition metal dichalcogenide \cite{2008_PRB_Wagner} (a), a high-$T_c$ cuprate \cite{2014_PRB_Croft} (b), and an iron based superconductor \cite{2013_NJP_Wang} (c).

In present review we mostly discuss these three families of quasi 2D superconductors from an empirical point of view, focusing on results from the angle resolved photoemission spectroscopy (ARPES), which is the most direct tool to access the electronic density of states at the Fermi level \cite{lynch1999, DamascelliRMP2003, 2014_LTP_Kordyuk}. We end up with the conclusion that the pseudogap in cuprates is a complex phenomenon which includes at least three different `intertwined' orders: spin and charge density waves (similar to the 2D CDW compounds) and preformed pairs, which appears in different parts of the phase diagram. The density waves in cuprates are competing to superconductivity for the electronic states but, on the other hand, should drive the electronic structure to vicinity of Lifshitz topological transition, the proximity to which is shown to correlate to $T_c$ maximum in all the iron based superconductors (Fe-SC) \cite{2012_LTP_Kordyuk}.

\begin{figure*}
\begin{center}
\includegraphics[width=0.8\textwidth]{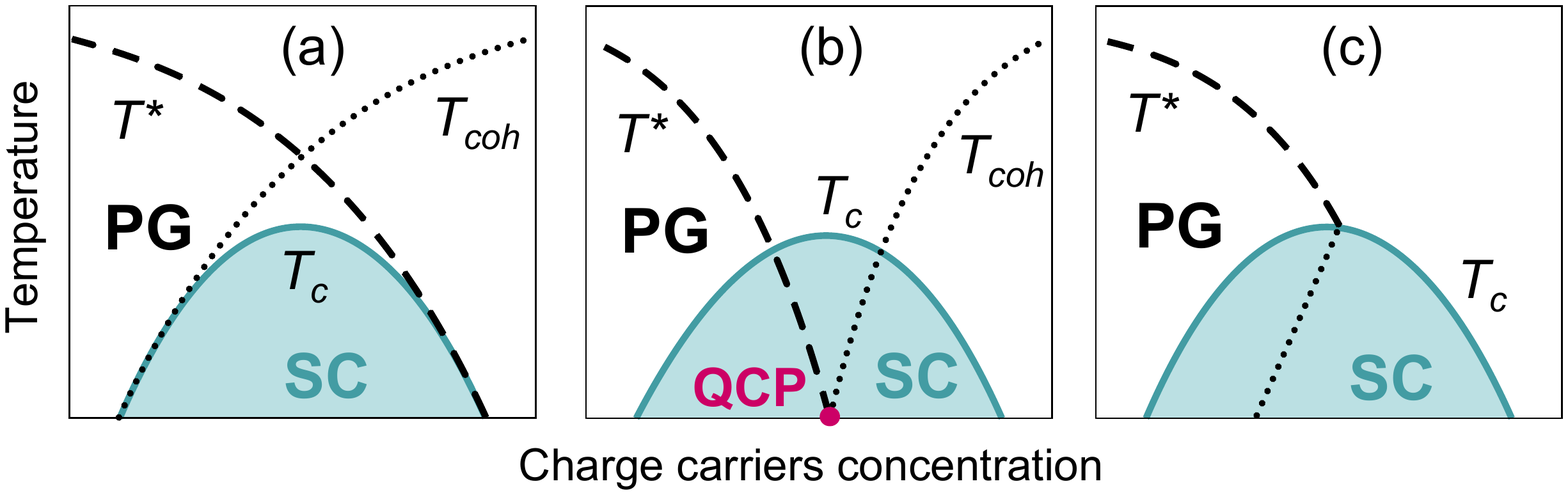}
\caption{Three theoretical idealizations for the interplay of pseudogap (PG) and superconductivity (SC) in the temperature-doping phase diagram of the HTSCs. $T_c$, $T$*, and $T_{coh}$ temperatures represent the phase transition to the SC state and crossovers to the PG and a coherent states, respectively.
\label{ThreePhDs}}
\end{center}
\end{figure*}

The paper is organized as following. Sec.\;\ref{theor} gives a short overview of selected theories of the pseudogap in cuprates. The manifestations of the pseudogap in different experiments are briefly discussed in Sec.\;\ref{exp}. Then, in the rest of the paper, the focus is made on ARPES results, starting from a short introduction to ARPES data analysis and gap extraction methods (Sec.\;\ref{ARPES}), the pseudogap phenomenon is considered in HTSC cuprates and CDW bearing TDM in Sec.\;\ref{HTSC}. The growing evidence for the pseudogap in Fe-SC are reviewed in Sec.\;\ref{Fe-SC}. Possible relation of the pseudogap to superconductivity is discussed in Sec.\;\ref{PG-SC}.

\section{Theories of pseudogap}
\label{theor}

The theories of the pseudogap in cuprates are reviewed in a number of papers \cite{LoktevPR2001, 2001_P_Sadovskii, GabovichPR2002, NormanAP2005, 2006_RMP_Lee, 2006_LTP_Tremblay, 2012_RPP_Rice, 2014_PRB_Wang} and textbooks \cite{plakida2010high, manske2004theory}. Most of these theories can be classified by their predictions about a crossover line, $T$*, which borders the pseudogap phase from a normal metal (or a `strange metal') on $T-x$ phase diagram (see Fig.\;\ref{ThreePhDs}). Here I briefly recall some of the most discussed models.

Diagram (a) is for the models which consider the pseudogap phase as a precursor to the superconducting state, the \textbf{preformed pairs scenarios} \cite{LoktevPR2001, 2005_Larkin}.

The fluctuations in bulk clean superconductors are extremely small. It is evident from very sharp transitions of thermal and electrical properties and has been shown theoretically by Levanyuk and Ginzburg back in 1960 \cite{2005_Larkin}. The corresponding Ginzburg number $Gi = \delta T / T_c \sim (T_c / E_F)^4 \sim 10^{-12} - 10^{-14}$, where $\delta T $ is the range of temperatures in which the fluctuation corrections are relevant and $E_F$ is the Fermi energy. In thin dirty superconducting films the fluctuations should be increased drastically \cite{1968_PLA_Aslamasov}: $Gi = T_c / E_F$ for clean 2D superconductor and $Gi \sim \tau^{-1}/E_F$ for dirty 2D superconductor \cite{2005_Larkin}, where $\tau^{-1}$ is the quasiparticle scattering rate at $E_F$. Thus, the width of the superconducting transition became experimentally measurable, but still $Gi \ll 1$.

The said behavior was deduced for the conventional superconductors to which the mean-field BCS theory or the Ginzburg-Landau model of the second-order phase transition is applicable. In superconductors with very small correlation length $\xi$ ($\xi k_F \ll 1$) the Bose-Einstein condensation (BEC) of local pairs takes place at $T_c$ while the formation of singlet electron pairs (that could be bipolarons \cite{1994_RPP_Alexandrov}) is assumed at some higher temperature \cite{plakida2010high}. Therefore, soon after discovery of HTSC, when it became clear that these materials are quasi-2D and dirty, with extremely small $\xi$, the superconductive fluctuations was the first scenario for the pseudogap \cite{FriedelPC1988}. Moreover, in strictly 2D systems, the phase fluctuations of the order parameter destroy the long-range order at finite temperature and only the Berezinskii-Kosterlitz-Thouless (BKT)
superconducting instability may occur \cite{LoktevPR2001, LoktevLTP2000}.

The `phase fluctuation' scenario \cite{1994_PC_Chakraverty, 1995_N_Emery} stems from the empirical `Uemura relation', that $T_c$ is proportional to the zero-temperature superfluid density $n_s(0)$ (or `phase stiffness') \cite{1989_PRL_Uemura, 2006_PRL_Rufenacht}. It was suggested that HTSC with low superconducting carrier density are characterized by a relatively small phase stiffness for the superconducting order parameter and by poor screening, both of which imply a significantly larger role for phase fluctuations. So, the pseudogap state is a region where the phase coherence is destroyed, but the amplitude of the order parameter remains finite. Two crossover lines in Fig.\;\ref{ThreePhDs} (a), $T$* and $T_{coh}$, border the regions where pairs are formed and become coherent, respectively, while superconductivity appears only under both lines \cite{1995_N_Emery}.

\begin{figure*}
\begin{center}
\includegraphics[width=0.66\textwidth]{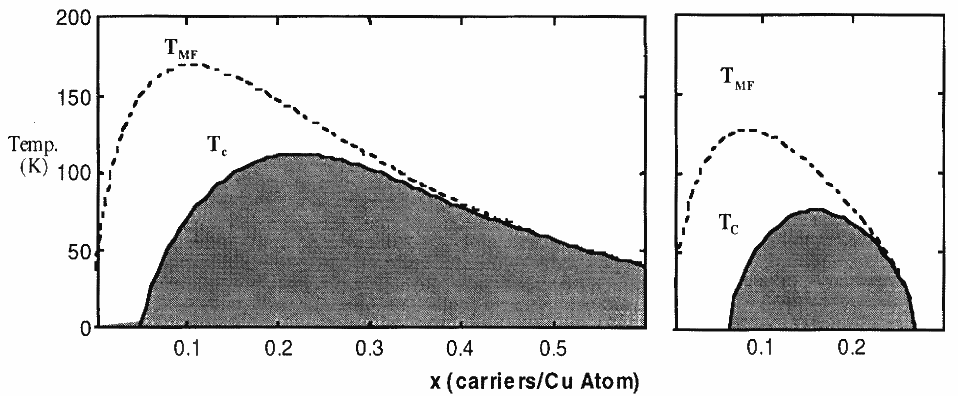}
\caption{Effect of thermal and quantum phase fluctuations (left), and of dimensional crossover (right) on the critical temperature for phase coherence $T_c$. Adapted from \protect\ignorecitefornumbering{\cite{1998_JPCS_Ariosa}}.
\label{Ariosa}}
\end{center}
\end{figure*}

One should note that calculated $T$*$(x)$ for either phase fluctuation \cite{manske2004theory} or BKT model \cite{1998_JPCS_Ariosa, LoktevPR2001} show decrease with lowering the charge carrier density, as in Fig.\;\ref{Ariosa}. Also, while the experimental $T$*$(x)$ dependence looks universal for all the hole doped cuprates, the fluctuation effects should be very sensitive to dimensionality and therefore different for different families. For example, the striking difference in the shape of the specific heat anomaly at $T_c$ is observed for quasi 2D Bi$_2$Sr$_2$CaCu$_2$O$_{8+x}$ (BSCCO or Bi-2212), where it follows the BEC phase transition, and for more 3D YBa$_2$Cu$_3$O$_{7-\delta}$ (YBCO), with classical BCS jump \cite{1999_PC_Junod}. Another problem of BEC models is that the Bose quasiparticles have no Fermi surface, while it is clearly observed by ARPES \cite{LoktevPR2001, plakida2010high}.

Nevertheless, recently, the `checkerboard' pattern observed in experiments \cite{2002_S_Hoffman, 2004_S_Vershinin, 2005_PRL_McElroy} has been explained by the model in which CDW is induced by superconducting fluctuations \cite{2014_PRB_Meier}.

The \textbf{spin singlet scenario} \cite{1992_PRB_Nagaosa, 2006_RMP_Lee} leads to the same phase diagram: the spin singlets play the role of preformed pairs, i.e.\;the pseudogap state is a liquid of spins without long range order (the original RVB idea of Anderson \cite{1987_S_Anderson}) and superconductivity occurs below two crossover lines due to spin-charge recombination. Similar considerations occur also for the SO(5) model \cite{2004_RMP_Demler} which attempts to unify antiferromagnetism and superconductivity. An important aspect of these scenarios is the general doping dependence of $T$*. Since the energy gain associated with spin singlet formation is the superexchange energy, $J$, the $T$* line is proportional to $J - tx$, where $t$ is the hopping energy of the doped hole \cite{NormanAP2005, 1988_SST_Zhang}.

Diagram (b) in Fig.\;\ref{ThreePhDs} is for scenarios in which another order with a \textbf{quantum critical point} (QCP) interplays with superconductivity. In QCP theories \cite{1976_PRB_Hertz, 1993_PRL_Sokol, 1995_PRL_Varma}, the transition between the ordered and disordered quantum phases transforms in a region of critical fluctuations which can mediate singular interactions between the quasiparticles, providing at the same time a strong pairing mechanism \cite{1996_ZPBCM_Castellani}. As for the nature of QCP, various proposals have been discussed.

In Ref.\;\onlinecite{1995_PRL_Castellani}, in which CDW and QCP were put together for the first time, it had been proposed that in the presence of the long-range Coulomb forces a uniform Fermi liquid can be made unstable by a moderate electron-phonon coupling (Hubbard-Holstein model) giving rise to incommensurate CDW in the form of ``frustrated phase separation", and the related QCP around optimal doping. Within this scenario, the static CDW compete (and kill) superconductivity like in some 1/8 doping systems, but, as long as CDW fluctuations stay dynamic, they can mediate superconductivity, and even the $d$-wave pairing can arise from CDW fluctuations without any spin interaction \cite{1996_PRB_Perali, 2001_PRL_Andergassen}. It was noted that CDW may also evolve into a spin-charge separation deeper in the charge-ordered phase as a consequence of modulation of charge density, anharmonic effects \cite{2008_PRL_Ortix}, closer proximity to the anti-ferromagnetic (AFM) phase, pinning, and so on.

The QCP determined by magnetic interaction \cite{Sachdev2000} leads to the spin-fluctuation scenarios. In the \textbf{spin-fermion model}, the pseudogap phase reflects the onset of strong AFM spin correlations, a spin-liquid without long-range order \cite{1995_PRB_Barzykin, 1996_ZPBCM_Pines, 1999_PRB_Schmalian, 1999_JETP_Kuchinskii}. The full analysis of the normal state properties of the spin-fermion model near the anti-ferromagnetic instability in two dimensions was given in \cite{2003_AiP_Abanov}. Recently, it has been shown \cite{2013_NP_Efetov, 2014_PRB_Wang} that within this model, a magnetically mediated interaction, which is known to give rise to $d$-wave superconductivity and charge order with momentum along zone diagonal \cite{2010_PRB_Metlitski}, also gives rise to the charge density wave with a `$d$-symmetry form factor' consistent with recent experiments \cite{2014_PNAS_Fujita}.

The anti-ferromagnetic scenario within the Hubbard model was also considered in the two-particle self-consistent approach \cite{2004_PRL_Senechal, 2006_LTP_Tremblay} and studied within a generalized dynamical mean-field theory \cite{2005_PRB_Sadovskii, 2008_JL_Kuchinskii}.

Several exotic scenarios of symmetry breaking, in which $T$* would be a true phase line, had been also suggested. For example, the orbital current state proposed by Varma \cite{1999_PRL_Varma} and the `flux-density wave' \cite{1990_PRB_Wang} or `$d$-density wave' current state \cite{2001_PRB_Chakravarty}.

Diagram (c) in Fig.\;\ref{ThreePhDs} is a result of similar competition between superconductivity and another ordering which does not require the QCP for its understanding. These could be either a spin-charge separation, predicted  \cite{1989_PRB_Zaanen, 1989_PC_Machida, 1990_JPSJ_Kato} and found  \cite{1995_N_Tranquada} long ago in some families of cuprates and known as `stripes', or `ordinary' (Peierls type) CDW or spin density wave (SDW) \cite{1997_JS_Eremin, 1997_PRB_Eremin, 1997_JPCM_Gabovich}, like in the transition metal dichalcogenides \cite{1975_AiP_Wilson, 2011_JPCM_Rossnagel}. The former can be responsible for the pseudogap in one-electron spectrum either due to density wave fluctuations \cite{2003_RMP_Kivelson} or by causing an electronic nematic order (quantum liquid-crystal) \cite{1998_N_Kivelson}. Broken rotational symmetry in the pseudogap phase of cuprates is really observed \cite{2010_N_Daou}. And nematic order becomes very fashionable today \cite{2003_RMP_Kivelson, 2009_AP_Vojta}.

A driving force for the Peierls type ordering is peculiarity of the electronic band structure: either the Fermi surface nesting \cite{BorisenkoPRL2008, 2012_JETP_Kuchinskii} or nesting of Van Hove singularities (VHs) \cite{1975_PRL_Rice, 1997_PRB_Markiewicz, 1997_JPCS_Markiewicz}. Nowadays, the FS nesting is considered responsible for CDW and pseudogap not only in cuprates and transition metal dichalcogenides but also in a number of other low dimensional metals such as manganites \cite{2001_S_Chuang, 2010_PRL_Evtushinsky}, binary and ternary molybdenum oxides \cite{1989_ACR_Whangbo},
Bi-dichalcogenide layered superconductors \cite{2012_PRB_Usui, 2013_JL_Shein}, etc. The competition between density wave and superconductivity is usually considered in frame of the Bilbro-McMillan relation \cite{1976_PRB_Bilbro}, according to which $\Delta_{SC}$ and $\sqrt{\Delta_{SC}^2+\Delta_{DW}^2}$ increase essentially identically with falling temperature, so, the density wave is suppressed by superconductivity and can be suppressed completely, as shown in Fig.\;\ref{ThreePhDs} (c). Recently it has been shown that for QCP models this relation will lead to $(T_{DW}(x)/T_{DW}^{max})^2 + (T_{c}(x)/T_{c}^{max})^2 = 1$ \cite{2015_JMMM_Regueiro}.

Many other possible reasons for pseudogap formation have been suggested, such as, for example, an intrinsic inhomogeneity \cite{2000_PC_Kresin}, $d$-wave-type Fermi surface deformations (Pomeranchuk instability) \cite{2007_PRB_Yamase}, or interaction with diatomic negative $U$ centers \cite{2008_JETP_Mitsen}, but it is hardly possible even to mention all of them here.

To conclude, there are many theories for the pseudogap phenomenon in HTSC and, may be consequently, there is no consensus on its origin. On the other hand, it seems that the main problem of the acceptance of these theories, until recently, was a general expectation that they should describe the whole pseudogap region on the phase diagram and all its experimental manifestations, briefly considered in the following section. Nowadays, there is growing evidence that the cuprates do indeed provide a complicated background for theorists revealing simultaneously a bunch of different phenomena: stripes, CDW, SDW, electronic fluctuations and localization. Thus, it seems that at least several of those models are related to reality of HTSC.

\begin{figure*}
\begin{center}
\includegraphics[width=0.96\textwidth]{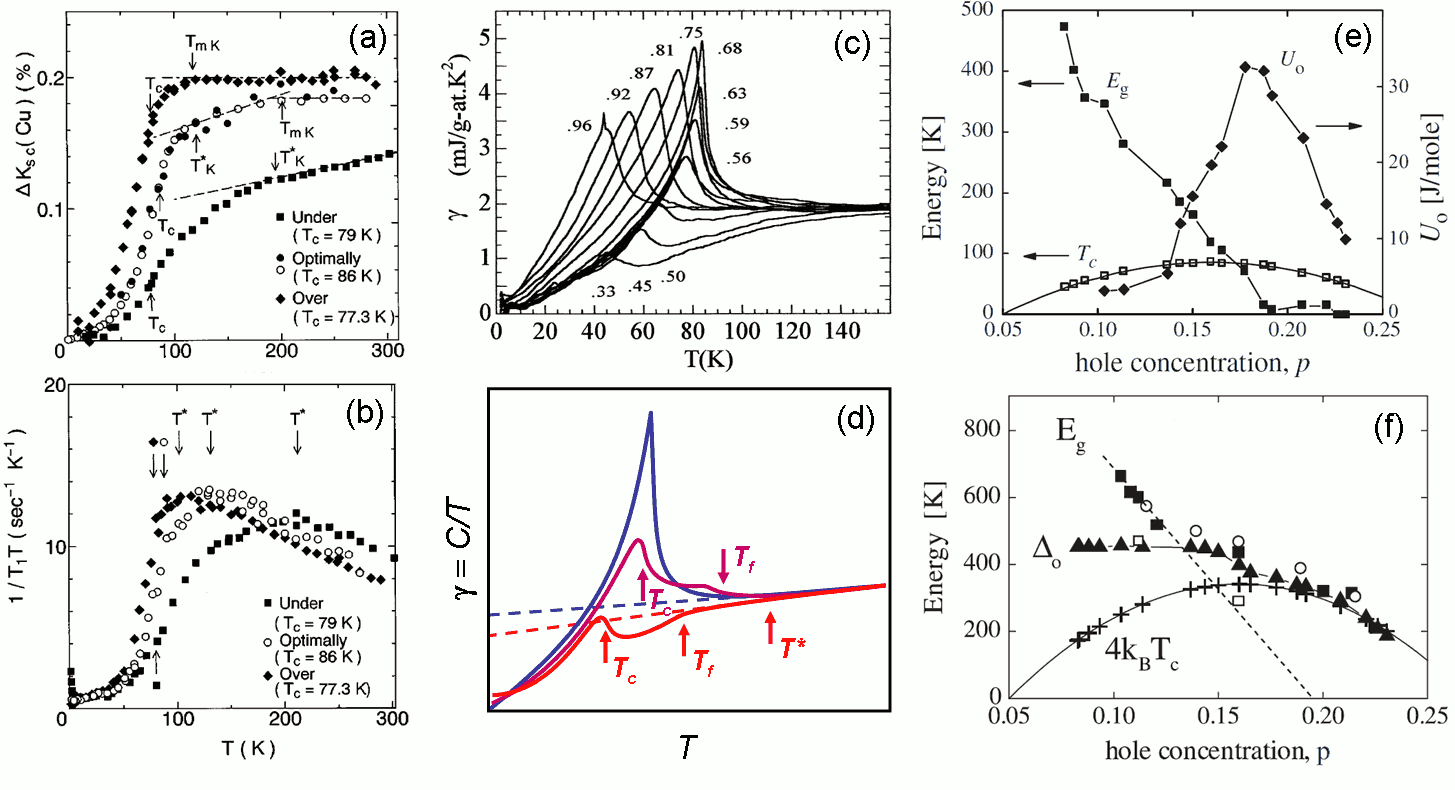}
\caption{The $T$-dependent Knight shift (a) and the spin-lattice relaxation rate (b) are shown for underdoped, optimally doped, and overdoped BSCCO \protect\ignorecitefornumbering{\cite{1998_PRB_Ishida}}. (c) Temperature dependence of the Sommerfeld constant for Y$_{0.8}$Ca$_{0.2}$Ba$_2$Cu$_3$O$_{6+x}$, labels show $x$ \protect\ignorecitefornumbering{\cite{1994_PC_Loram}}. (d) A sketch to indicate the transition temperatures $T_c$, $T$* and a crossover to superconducting fluctuations, $T_f$, for a optimally doped and two underdoped samples. (e) The doping dependence of the gap energy $E_g$, of the condensation energy $U_0$, and of $T_c$, the SC gap determined from heat capacity is shown on (f) \protect\ignorecitefornumbering{\cite{1999_pssb_Tallon}}.
\label{NMR+C}}
\end{center}
\end{figure*}

\section{Pseudogap in experiments}
\label{exp}

Opening of a gap or just a depletion of the electronic density of states at the Fermi level can hardly be missing by a number of experimental probes. Indeed, any transition to one of possible CDW states in, for example, transition metal dichalcogenides, left signatures in temperature dependences of different experimental parameters: heat capacity, resistivity, magnetic susceptibility, etc. Those signatures were usually accompanied by change of the diffraction patterns, so, the character of the symmetry change was more or less clear \cite{1975_AiP_Wilson}.

In cuprates, the pseudogap was observed in many experiments as something that starts to happen above $T_c$ \cite{TimuskRPP1999}, while any indication of new order could not be found by diffraction techniques. Then a depletion of the spectral weight was observed directly by ARPES \cite{1996_S_Loeser, 1996_N_Ding} and tunneling spectroscopy \cite{1997_PCS_Tao, 1998_PRL_Renner}, and some kind of CDW/SDW, a spin-charge separation in form of `stripes', was found in some HTSC compounds \cite{1995_N_Tranquada}. Nowadays, there are many experimental evidences for CDW in almost all families of cuprates, but the nature of the pseudogap remains puzzling.

In this section, before turning to the ARPES results, we briefly consider experimental manifestations of the pseudogap in cuprates by other experimental probes: spectroscopic methods such as nuclear magnetic resonance (NMR), infrared optical conductivity (IR), Raman scattering (RS), and tunneling spectroscopies (except STM/STS these are intrinsic tunneling, superconductor/insulator/normal-metal (SIN) and superconductor/insulator/superconductor (SIS) tunneling, and Andreev reflection tunneling (AR)), and inelastic neutron scattering (INS), as well as traditional thermodynamic/transport probes such as heat conductivity and resistivity measurements (or `dc conductivity'). ARPES and tunneling measure directly the density of single electronic states while other spectroscopies as well as thermodynamic/transport probe the two-particle spectrum.

\textbf{NMR.} The pseudogap in cuprates was first detected by NMR \cite{1989_PRL_Warren, 1990_PRB_Walstedt}, which measures the Knight shift, $K_s$, and spin-lattice relaxation rate, $1/T_1$. The Knight shift is a measure of the polarization of electrons by the applied magnetic field and is proportional to the real part of the paramagnetic (Pauli) susceptibility, $\chi'(\mathbf{q}=0,\omega)$, that, in the Fermi liquid model is proportional to the density of states at the Fermi level and should be independent on $T$. The spin-lattice relaxation rate is related to the imaginary part of susceptibility, such as $1/T_1 T \sim \sum_{\mathbf{q}} |F(\mathbf{q})|^2 \chi''(\mathbf{q},\omega)/\omega$, where $F(\mathbf{q})$ is the form factor for the particular nuclear site---by probing various nuclei in the unit cell one can probe different parts of momentum space \cite{TimuskRPP1999, plakida2010high}. In Fig.\;\ref{NMR+C} the $T$-dependent Knight shift (a) and the spin-lattice relaxation rate (b) are shown for underdoped, optimally doped, and overdoped BSCCO \cite{1998_PRB_Ishida}. The suppression of both quantities starts below $T$* but no additional anomaly is seen at $T_c$, that has been considered in support of the preformed pairs scenario \cite{NormanAP2005}.

\textbf{Specific heat.} If a gap, which lowers the kinetic energy of electrons, opens (or starts to develop) at $T$*, one should see a peculiarity in any thermodynamic/transport quantity at $T$* rather than at $T_c$ when the energy of the electrons does not change. Indeed, the specific heat jump at $T_c$ fades out with underdoping, see Fig.\;\ref{NMR+C} (c,d), but usually there is no jump at $T$* (though some measurements reveal a weak bump \cite{2004_JPSJ_Matsuzaki}). In general, the specific heat data have frequently been cited in support of diagram (b) of Fig.\;\ref{ThreePhDs} \cite{NormanAP2005} since the determined $T$* line cuts through the $T_c$ dome \cite{1994_PC_Loram, 1999_pssb_Tallon}. It is also consistent with the sharp decrease of the specific heat jump or the superconducting condensation energy $U_0$, defined as the entropy difference integrated from $T = 0$ to $T_c$, which is a constant $U_0^{BCS} = 0.24 \gamma_n T_c^2$ for a BCS superconductor with $d$-wave pairing \cite{plakida2010high} (see Fig.\;\ref{NMR+C} (e)). Based on those NMR and heat capacity data, it has been concluded \cite{1999_pssb_Tallon} that the pseudogap and superconductivity are `two gaps', independent and competing. So, smooth evolution of tunneling spectra from the pseudogap into superconductivity does not necessarily imply the pseudogap is a short-range pairing state with the same mean-field gap energy as superconductivity \cite{1999_PRL_Tallon, 2002_PRL_Markiewicz}. One should note that the interpretation of specific heat measurements is tricky because at transition temperatures the phonon contribution in cuprates is typically a hundred times stronger than the electronic one \cite{TimuskRPP1999} and differential techniques should be used.

\begin{figure*}
\begin{center}
\includegraphics[width=1\textwidth]{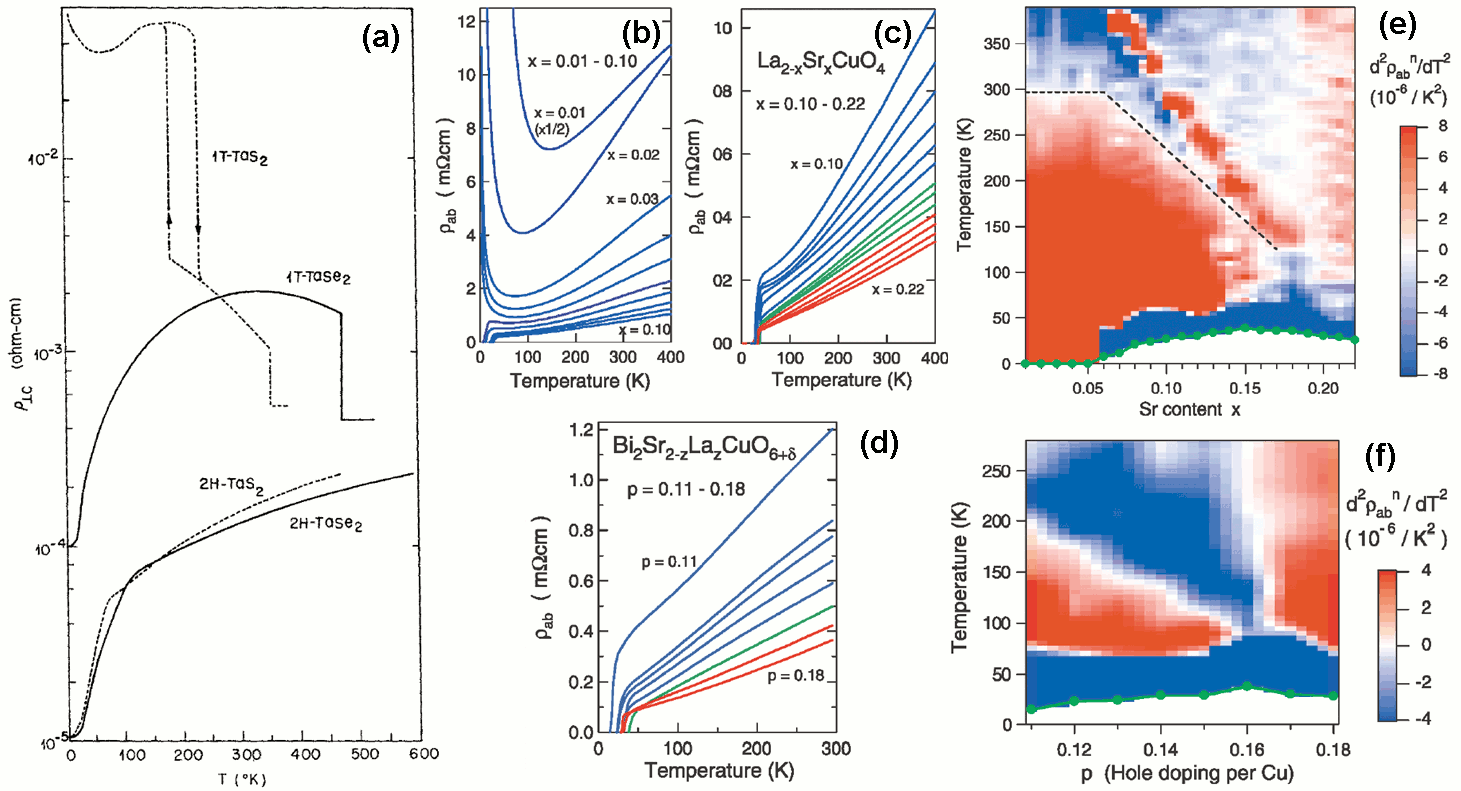}
\caption{Resistivity over phase transitions: (a) for selected transition metal dichalcogenides \protect\ignorecitefornumbering{\cite{1975_AiP_Wilson}}; (b-d) for high-$T_c$ cuprates. Resistivity curvature maps for LSCO (e) and BSLCO (f) \protect\ignorecitefornumbering{\cite{2004_PRL_Ando}}.
\label{Resistivity}}
\end{center}
\end{figure*}

\textbf{Transport properties.}
After discovery of a new superconductor, its transport properties, i.e., dc conductivity, Hall effect, thermal conductivity and thermopower, are the first quantities to study. Any phase transition which affects the electronic density of states at the Fermi level should be seen as a peculiarity on temperature dependences of transport properties. Though, it is often hard to say which peculiarity to expect. For example, the dc conductivity depends on both charge carrier concentration $n$ (or density of states at $E_F$, $N(0)$) and scattering time $\tau$ (in simplest Drude model, $\sigma = n \tau e^2/m$). If, due to CDW, a full gap opens, it is a Peierls type of metal-insulator transition and resistivity changes from metallic to insulating ($d\rho/dT < 0$). If a partial gap opens, than $n$ decreases but $\tau$ increases due to less space for electron to scatter. So, depending on Fermi surface geometry, the resistivity (see Fig.\;\ref{Resistivity} (a) \cite{1975_AiP_Wilson}) can show steps as for 1T-TaS$_2$, which has several subsequent transitions to an incommensurate at 550 K, quasicommensurate (or `nearly commensurate' \cite{1977_JPSJ_Nakanishi, 1989_S_Wu}) at 350 K and commensurate CDW at 180 K \cite{2000_PRB_Pillo}, or kinks as for 2H-TaSe$_2$ with transitions to an incommensurate at 122 K and commensurate CDW at 90 K.

In cuprates, the transition to the pseudogap state is less pronounced in resistivity (see Fig.\;\ref{Resistivity} (b-d) \cite{2004_PRL_Ando}) but still detectable and heavily discussed.
Soon after discovery of HTSC, a peculiar feature of cuprates, a quasi-linear dependence of resistivity over a wide temperature range has been found \cite{1987_PRL_Gurvitch}. It means that the experimental magnitude of the resistivity in cuprates at high temperatures is much larger than the Ioffe-Regel limit considered within the conventional semiclassical transport theory based on the Boltzmann equation \cite{plakida2010high}. This linear region on the phase diagram has inspired appearance of many new HTSC theories modeling this ``strange metal" behavior, such as fluctuating staggered currents \cite {1996_PRL_Wen} or the ``marginal" Fermi-liquid (MFL) model \cite{1999_PRL_Varma}. On the other hand, it has been shown \cite{2003_RMP_Gunnarsson} that within the $t-J$ model the saturation resistivity should be much larger than the Ioffe-Regel limit, so, the absence of saturation of resistivity at high temperatures is expected for strongly correlated systems.

The linear resistivity is observed only in a narrow region of temperatures near the optimal doping, as has been shown \cite{2004_PRL_Ando} by mapping of the in-plane resistivity curvature $(d^2\rho_{ab}/dT^2)$ of the La$_{2-x}$Sr$_{x}$CuO$_4$ (LSCO), YBCO, and Bi$_2$Sr$_{2-z}$La$_{z}$CuO$_{6+\delta}$ (BSLCO) crystals (see Fig.\;\ref{Resistivity}). The pseudogap temperature, determined on these maps as the inflection point of the resistivity $(d^2 \rho_{ab}/dT^2 = 0)$, decreases linearly with doping and terminates near the optimal value $p = 0.16$. Below $T$* the curvature is positive until the superconducting fluctuations make it negative again.

The idea of two pseudogaps has been confirmed by measurements of the $c$-axis resistivity and magnetoresistance \cite{2002_EPL_Lavrov}: while $T$* increases with decreasing hole doping and is field-insensitive, a field-sensitive gap is found at lower temperature, which scales with $T_c$, and may be considered therefore as a precursor to superconductivity. By applying magnetic field to Y$_{1-x}$Ca$_x$Ba$_2$(Cu$_{1-y}$Zn$_y$)$_3$O$_{7-\delta}$ thin films and changing the Zn concentration to suppress both the superconductivity and superconducting fluctuations, it has been shown that the pseudogap region persists below $T_c$ on the overdoped side and $T$* extrapolates to zero at about 0.19 holes concentration \cite{2005_PRB_Naqib}.

\begin{figure*}
\begin{center}
\includegraphics[width=0.32\textwidth]{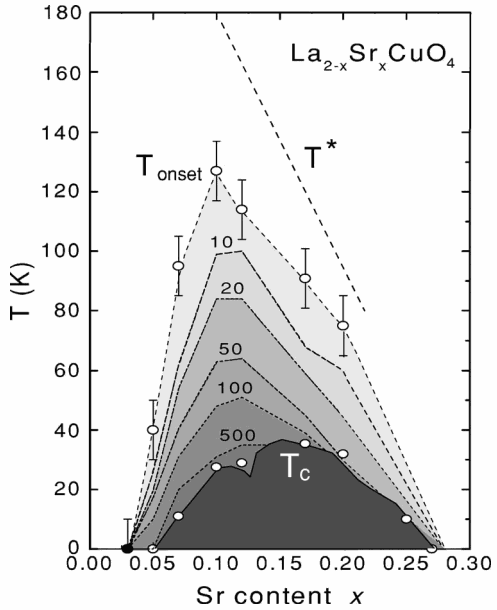}
\includegraphics[width=0.32\textwidth]{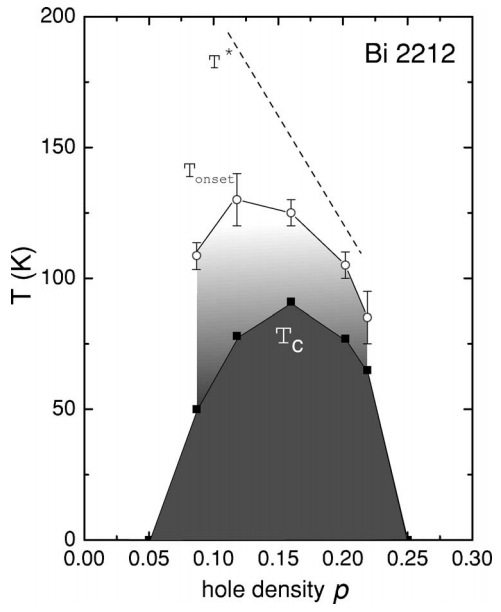}
\caption{The phase diagrams of LSCO (left) and BSCCO (right) showing the Nernst region between $T_c$ and $T_{onset}$ (numbers on the contour curves indicate the value of the Nernst coefficient). The $T_{onset}$-curves peak near $x = 0.10$. The dashed lines are $T$* estimated from heat-capacity measurements. After \protect\ignorecitefornumbering{\cite{2006_PRB_Wang}}.
\label{Nernst}}
\end{center}
\end{figure*}

\textbf{Nernst effect.}
The Nernst effect is considered as one of the most convincing evidences for the existence of the preformed pairs \cite{TimuskRPP1999, plakida2010high}. The Nernst effect in solids is the detection of an electric field $E$ perpendicular to orthogonally applied temperature gradient $\nabla T$ and magnetic field $H$ \cite{2006_PRB_Wang}. The Nernst signal, defined as $e_N (H,T) = E/ \nabla T$, is generally much larger in ferromagnets and superconductors than in nonmagnetic normal metals. In the superconducting state, the Nernst signal is the sum of the vortex and quasiparticle terms, $e_N = e_N^v + e_N^{qp}$, which can be distinguished with proper analysis, measuring the thermopower, Hall angle, and resistivity in addition to the Nernst effect \cite{2001_PRB_Wang}. In Fig.\;\ref{Nernst} the onset of $e_N^v$ is defined by temperature $T_{onset}$ on the phase diagrams of LSCO and Bi-2212 (numbers on the contour curves indicate the value of the vortex Nernst coefficient $\nu = e_N^v/\mu_0 H$ in nV/KT). The observation of a large vortex Nernst signal in an extended region above $T_c$ in hole-doped cuprates provides evidence that vortex excitations survive there \cite{2001_PRB_Wang, 2006_PRB_Wang}. The results support the preformed pairs scenario and suggest that superfluidity vanishes because long-range phase coherence is destroyed by thermally created vortices (in zero field). Interestingly, in electron doped cuprates (e.g. NCCO) where the PG is believed absent the vortex Nernst signal is also absent. So, the comparison of Nernst effect in hole and electron-doped cuprates shows that the `thermally created vortices' are not generic to any highly anisotropic layered superconductor but may be related to the physics of the pseudogap state in hole-doped cuprates \cite{2006_PRB_Wang}. The vortex Nernst signal above $T_c$ is analogous to an excess current observed in the same temperature range in the Andreev contacts \cite{2013_LTP_Dyachenko} that also indicates the presence of Cooper pairs.

\textbf{Optics.}
Like transport measurements, optical studies of electronic spectra \cite{1996_JPCM_Puchkov, 2005_RMP_Basov} provide information on the spectrum of collective electron-hole pair excitations, where a transition takes place from an initial state to a different final state. The difference is in final states. While in transport techniques the initial and final states have the same energy, in optics they hold the same momentum. Both the first- and second-order processes of light scattering are used. In the former, the light excites bosonic degrees of freedom: phonons, electron-hole pairs, spin waves or other electronic density fluctuations. These are studied by \textbf{infrared} and \textbf{optic absorption}. The second-order processes when a photon absorbed and reemitted are used in the Raman scattering.

The absorption spectroscopy methods measure reflectance on single crystals or transmission in thin-films, that allows one to study the complex dielectric function $\epsilon(\omega)=\epsilon_1(\omega)+i\epsilon_2(\omega)$ in the long-wave limit ($\mathbf{q} = 0$), from which the dynamical complex conductivity $\sigma(\omega) = \sigma_1(\omega) + i\sigma_2(\omega)$ can be derived: $4\pi\sigma_1 = \omega\epsilon_2$, $4\pi\sigma_2=\omega(1-\epsilon_1)$ \cite{plakida2010high, 2005_RMP_Basov}. The real part of conductivity, $\sigma_1(\omega)$, is proportional to the joint density of states (Kubo-Greenwood formula) and determines absorption of radiation at the frequency $\omega$. The real part of the inverse conductivity $\sigma^{-1}(\omega)$ is proportional to the quasiparticle scattering rate $\tau^{-1}$ while its imaginary part is proportional to mass renormalization $m^*/m = 1 + \lambda(\omega)$. The Kramers-Kronig (KK) relations allow one to calculate both the real and the imaginary parts of $\varepsilon(\omega)$ or $\sigma(\omega)$ from the raw experimental data. In the ellipsometric technique \cite{2004_S_Boris}, the real and imaginary parts of $\varepsilon(\omega)$ can be measured independently.

\begin{figure*}
\begin{center}
\includegraphics[width=0.9\textwidth]{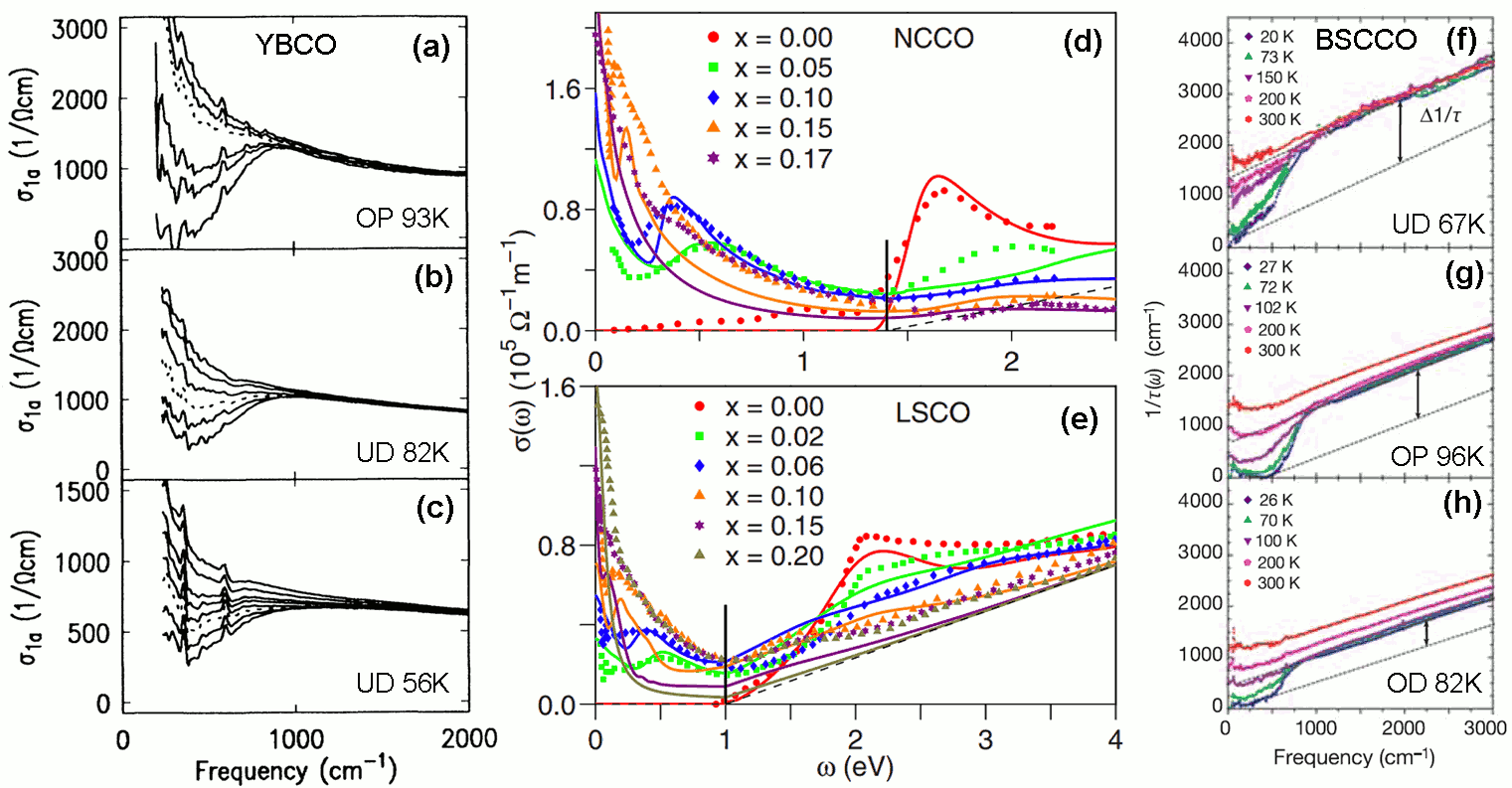}
\caption{Optical spectroscopy data which show pseudogap in different HTSC. (a-c) Inplane conductivity of YBCO for various temperatures in both the normal and superconducting states \protect\ignorecitefornumbering{\cite{1991_PRL_Rotter}}: (a) optimally doped with $T_c =$ 93 K at $T=$ 120, 100, 90 (dashed), 70, 20 K (from top to bottom); (b) underdoped 82 K at $T=$ 150, 120, 90, 80 (dashed), 70, 20 K; (c) underdoped 56 K $T=$ 200, 150, 120, 100, 80, 60 (dashed), 50, 20 K. (d,e) Conductivity of electron doped NCCO \protect\ignorecitefornumbering{\cite{2004_PRB_Onose}} and hole doped LSCO \protect\ignorecitefornumbering{\cite{1991_PRB_Uchida}} (symbols) in wider frequency range (1 eV $\approx$ 8066 cm$^{-1}$) compared to model calculations (solid lines) \protect\ignorecitefornumbering{\cite{2010_PRB_Das}}. (f-h) Doping and temperature dependence of the scattering rate of BSCCO \protect\ignorecitefornumbering{\cite{2004_N_Hwang}}.
\label{Optics}}
\end{center}
\end{figure*}

Simple Drude model predicts that reflectance decreases monotonically with frequency. In HTSC, a structure in the form of a `kink' was found. In underdoped materials, this kink starts to develop already in the normal state at temperatures similar to $T$* derived from other experiments and, therefore, was interpreted as a manifestation of the pseudogap. The corresponding changes in the optical conductivity appears as a depletion of the spectral weight in the range 300-700 cm$^{-1}$ (about 40-90 meV) \cite{1990_PRB_Orenstein, 1991_PRL_Rotter}, as one can see in Fig.\;\ref{Optics} for YBCO (a-c). Since $\sigma(\omega)$ just above this range looks not changing with temperature, it has been concluded that the gapped spectral weight is shifted to lower frequencies, resulting in a narrowing of the Drude peak \cite{1996_JPCM_Puchkov}. The measurements over much wider frequency range, as one can see in Fig.\;\ref{Optics} (d,e) \cite{1991_PRB_Uchida, 2004_PRB_Onose, 2010_PRB_Das}, shows that much higher energies could be involved. Similar depletion by the pseudogap is observed for the derived from conductivity scattering rate, as shown in Fig.\;\ref{Optics} (f-h) for BSCCO \cite{2004_N_Hwang}.

\begin{figure}
\begin{center}
\includegraphics[width=0.45\textwidth]{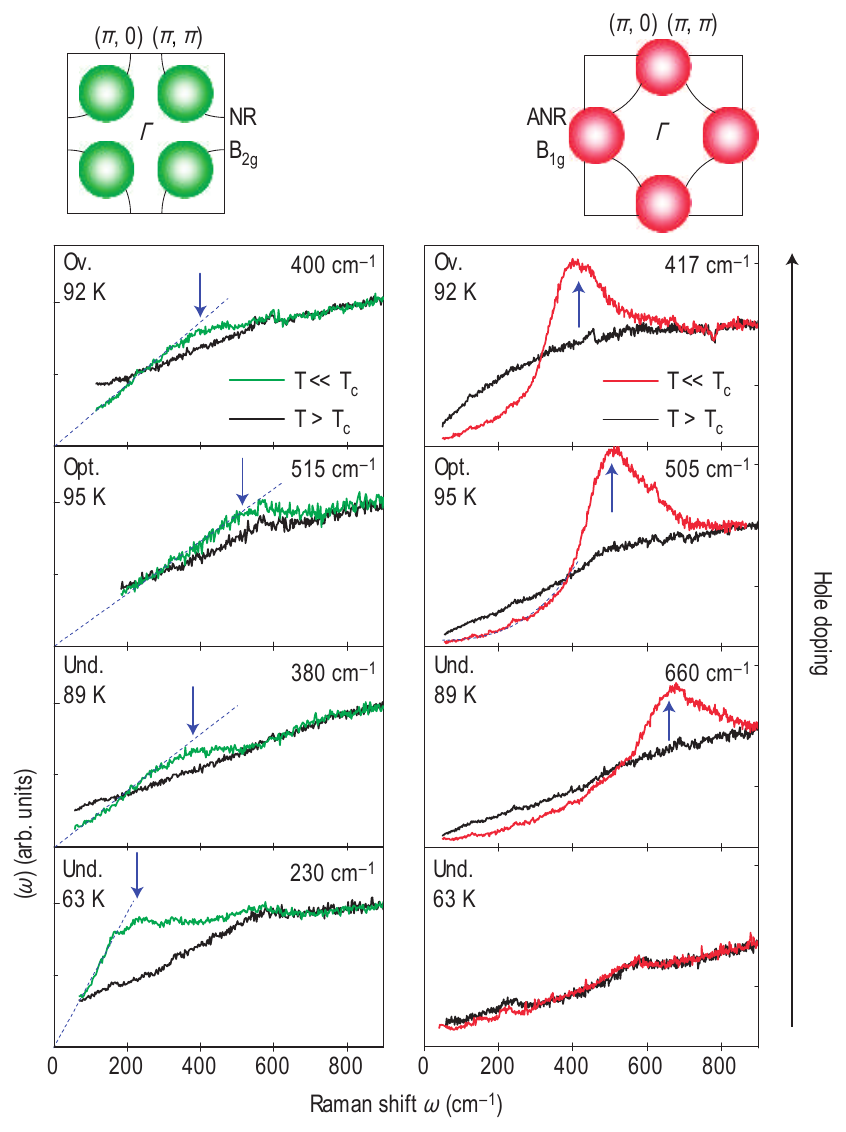}
\caption{Raman spectra for HgBa$_2$CuO$_{4+\delta}$ (Hg-1201) for B$_{2g}$ (left) and B$_{1g}$ (right) symmetries. The arrows indicate the position of the superconducting peak maxima. Ov.: overdoped; Opt.: optimally doped; Und.: underdoped. After \protect\ignorecitefornumbering{\cite{2006_NP_LeTacon}}.
\label{Raman}}
\end{center}
\end{figure}

Naturally, the origin of the pseudogap has been addressed in many optical studies. Most of that ideas can be found in the topical reviews \cite{1996_JPCM_Puchkov, TimuskRPP1999, 2005_RMP_Basov}, which, nevertheless, ended with the conclusions that there is no unified view on the nature of the pseudogap state. That was also noted on controversy between optical experiments and ARPES about coherence state \cite{2005_RMP_Basov}: from ARPES point of view, it is set only below $T_c$, but infrared methods provide evidence for coherence below the spin-gap temperature $T_s > T_c$. Also, an important role of magnetic correlations in the pseudogap state has been found by optical study of (Sm,Nd)Ba$_2$\{Cu$_{1-y}$(Ni,Zn)$_y$\}$_3$O$_{7-\delta}$ with magnetic (Ni) and nonmagnetic (Zn) impurities \cite{2005_PRL_Pimenov}. The broadband infrared ellipsometry measurements of the $c$-axis conductivity of underdoped RBa$_2$Cu$_3$O$_{7-\delta}$ (R = Y, Nd, and La) have separated energy scales due to the pseudogap and the superconducting gap and provided evidence that these gaps do not share the same electronic states \cite{2008_PRL_Yu}.

\begin{figure}
\begin{center}
\includegraphics[width=0.4\textwidth]{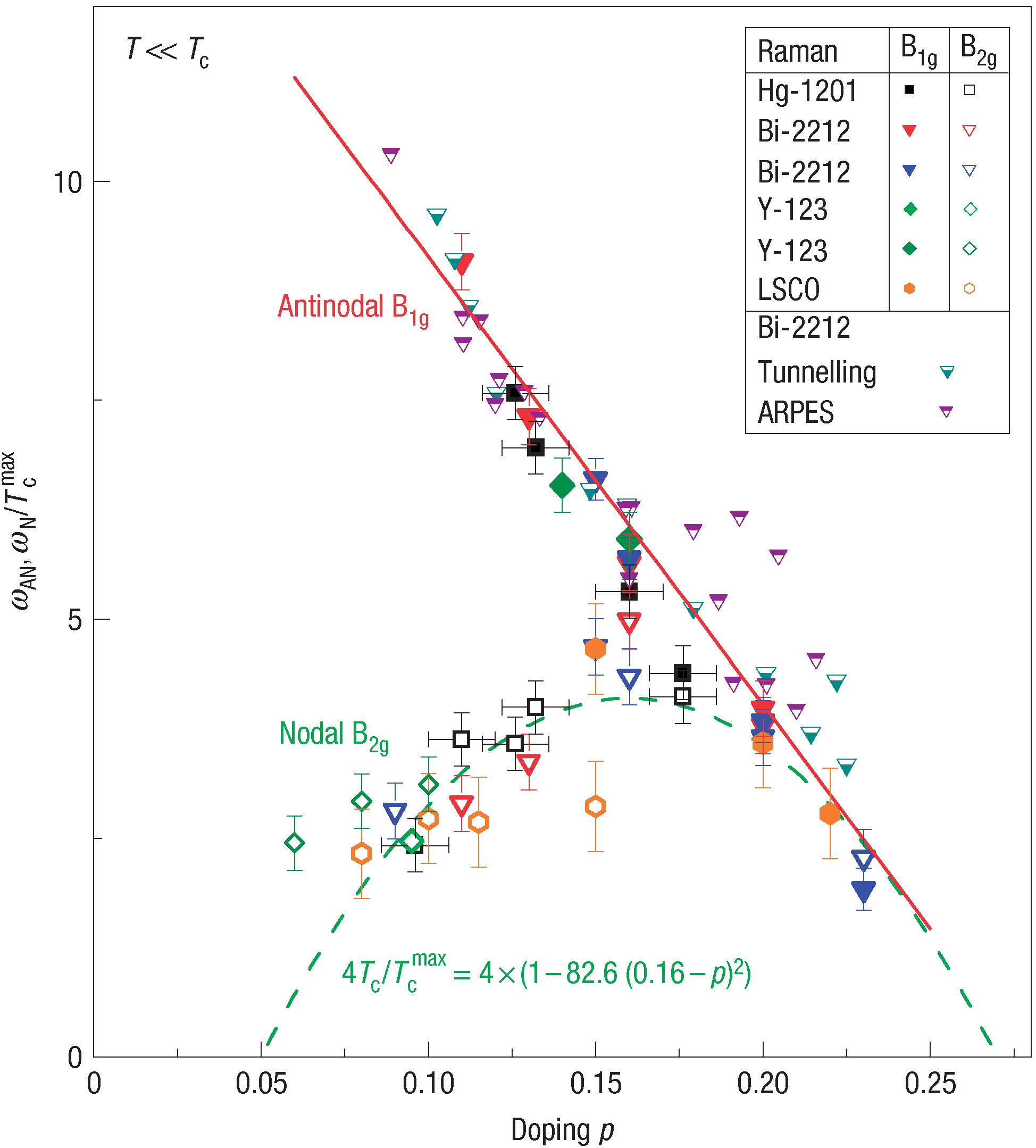}
\caption{Antinodal and nodal peak energies normalized to $T_c^{max}$ for Hg-1201 \cite{2006_NP_LeTacon}, Bi-2212 \cite{2002_JPCS_Venturini, 2003_PRB_Sugai}, Y-123 \cite{2003_PRB_Sugai} and LSCO  \cite{2003_PRB_Sugai}). The ratios $2\Delta/T_c^{max}$ determined by ARPES \cite{2002_PRB_Borisenko, 1998_N_Norman,
1999_PRL_Campuzano} and tunnelling spectroscopy \cite{1998_PRL_Miyakawa,
1998_PRL_DeWilde} are shown for comparison.
After \protect\ignorecitefornumbering{\cite{2006_NP_LeTacon}}.
\label{RamanPhD}}
\end{center}
\end{figure}

\begin{figure*}
\begin{center}
\includegraphics[width=0.96\textwidth]{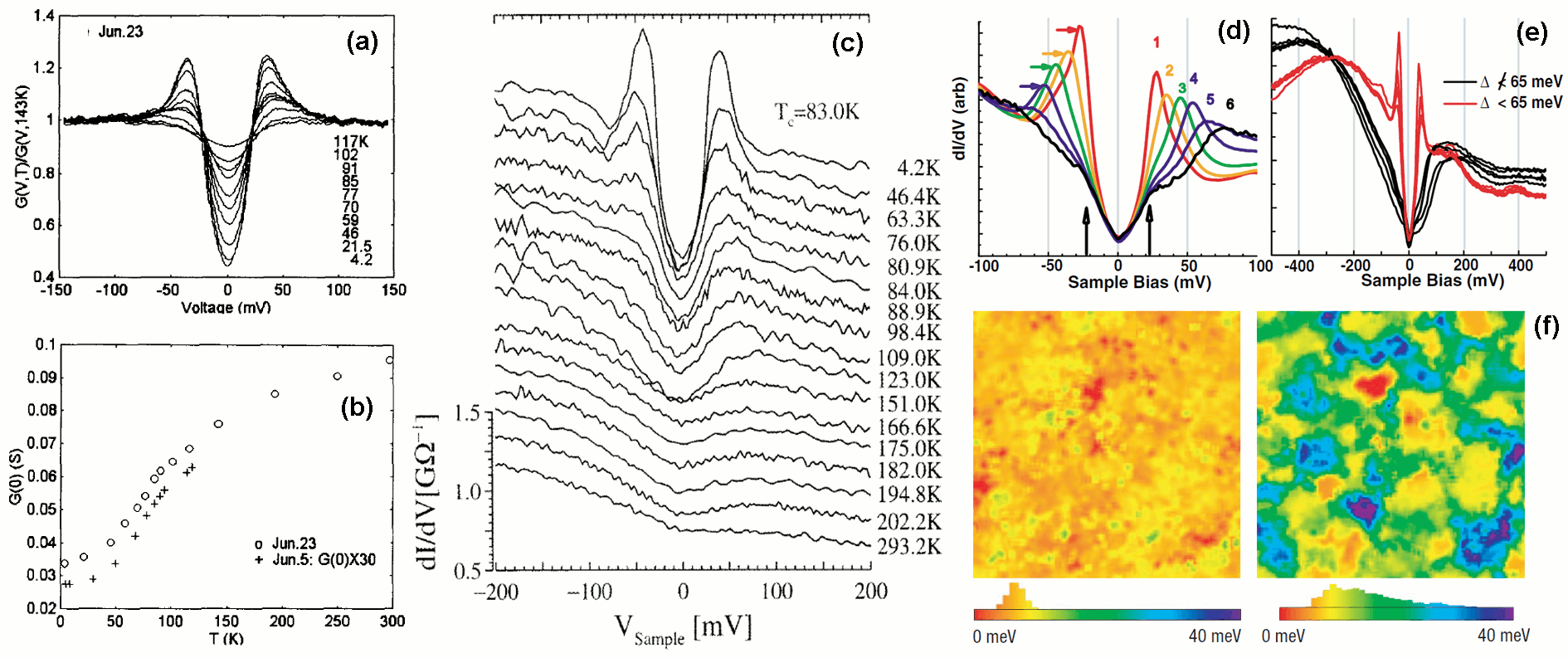}
\caption{Pseudogap in tunneling spectroscopy on BSCCO. (a) SIN tunneling spectra of optimally doped sample ($T_c =$ 85-90 K) \protect\ignorecitefornumbering{\cite{1997_PCS_Tao}}, note that zero bias tunneling conductance $G(0)$ does not saturates at $T$*$\approx$ 150 K (b). (c) STM spectra for underdoped sample (83 K) \protect\ignorecitefornumbering{\cite{1998_PRL_Renner}}, the depletion of the density of states at Fermi level is seen to persist in the normal state, the size of the pseudogap looks independent on temperature. (d-f) Inhomogeneity of the pseudogap: (d) each curve is STM spectrum integrated over many tip positions with the same gap value, (e) characteristic spectra from the two regions $\Delta <$ 65 and $\Delta \nless$ 65 meV \protect\ignorecitefornumbering{\cite{2005_PRL_McElroy}}. (f) 180 {\AA} square maps of gaps (defined as
half the distance between the edges of the gap) and corresponding histograms of the superconducting gap (left) and pseudogap (right) for underdoped sample (15 K) \protect\ignorecitefornumbering{\cite{2007_NP_Boyer}}.
\label{STM}}
\end{center}
\end{figure*}

\begin{figure*}
\begin{center}
\includegraphics[width=0.8\textwidth]{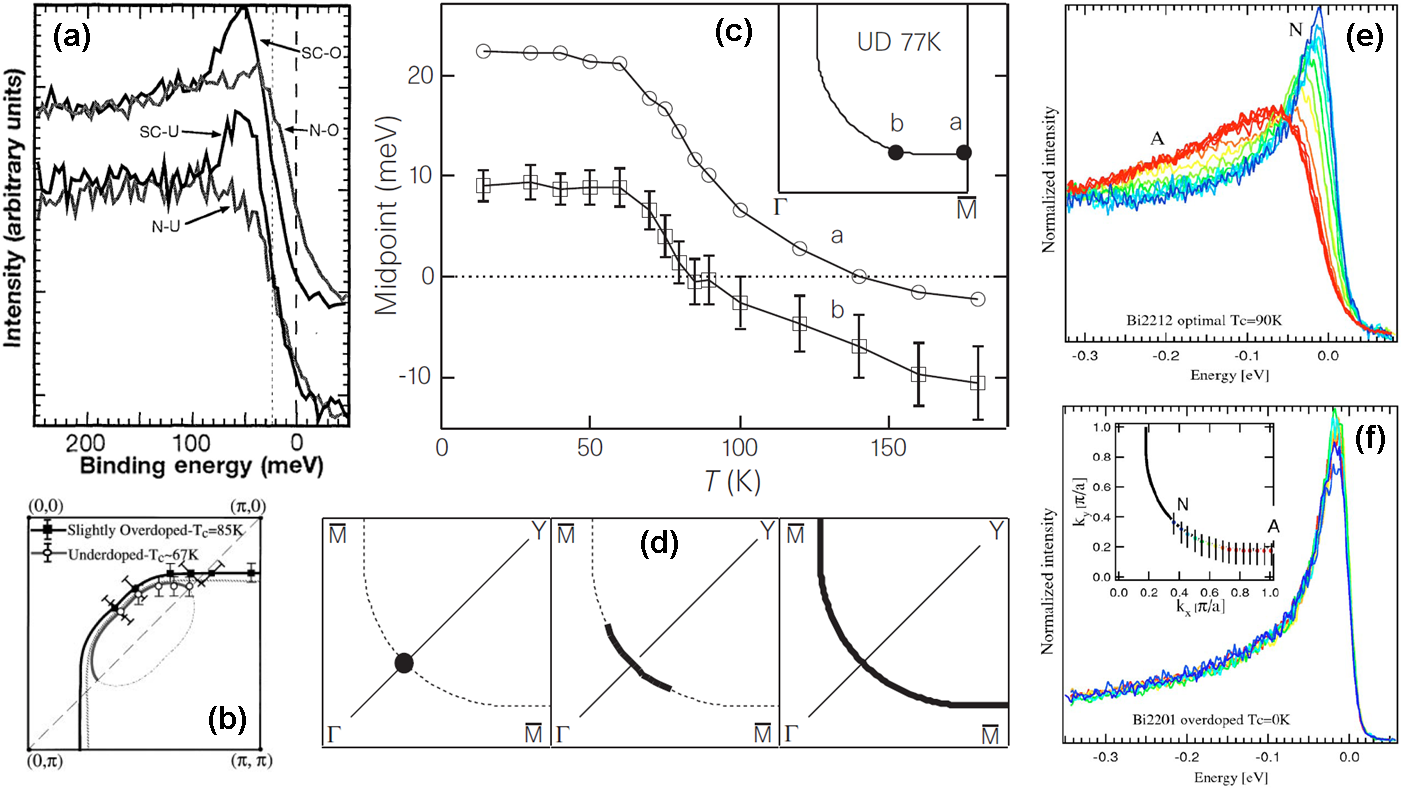}
\caption{Pseudogap anisotropy by ARPES. (a) Energy distribution curves (EDCs) from the antinodal region of underdoped (U) and overdoped (O) BSCCO in the normal (N) and superconducting states (SC) \protect\ignorecitefornumbering{\cite{1996_S_Loeser}}. (b) Hole pockets around $(\pi/2,\pi/2)$ as one of explanations of the gapped sections of Fermi surface \protect\ignorecitefornumbering{\cite{1996_PRL_Marshall}}. (c) Midpoints of the leading edge, the `leading edge gap' (LEG), of the EDCs of underdoped BSCCO vs. temperature, which inspired the `Fermi arc' idea, sketched in panel (d): $d$-wave node below $T_c$ becomes a gapless arc above $T_c$ which expands with increasing temperature to form the full Fermi surface at $T$*  \protect\ignorecitefornumbering{\cite{1998_N_Norman}}. More accurate set of EDCs along the Fermi surface from node (N) to antinode (A) for optimally doped two-layer BSCCO ($T_c$ = 90 K) at $T$ = 140 K(e) contrasted to heavily overdoped
Bi2201 ($T_c$ = 0) at $T=40$ K (f) \protect\ignorecitefornumbering{\cite{1998_N_Norman}}.
\label{FermiArcs}}
\end{center}
\end{figure*}

\textbf{Raman scattering}, like optical absorption, measures a two-particle excitation spectrum providing direct insight into the total energy needed to break up a two-particle bound state. In metals, the Raman effect is difficult to observe because of a small penetration depth and limited energy range \cite{TimuskRPP1999}. The signal is often riding on a high background, which might result in a considerable data scattering, and the nodal results need a numerical analysis \cite{Huefner2008}. But its big advantage, compared to the infrared spectroscopy, is that the symmetry selection rules enable to measure some momentum dependence of the spectrum \cite{2007_RMP_Devereaux}. For cuprates there are two useful momentum averages: B$_{1g}$ symmetry, that is peaked at $(\pi,0)$, and B$_{2g}$ symmetry, peaked at $(\pi/2,\pi/2)$. Fig.\;\ref{Raman} shows typical Raman spectra for HgBa$_2$CuO$_{4+\delta}$ (Hg-1201) for these two symmetries \cite{2006_NP_LeTacon}. One can see that the peaks in these two symmetries depend on doping in opposite directions.

These two energy scales are plotted on the phase diagram in Fig.\;\ref{RamanPhD} taken from Ref.\;\onlinecite{2006_NP_LeTacon}, which has reanimated the interest to the `two gaps' scenario discussed earlier \cite{1999_PRL_Tallon, 2002_PRL_Markiewicz}. Very similar diagrams have been suggested in Refs.\;\onlinecite{2007_RMP_Devereaux} and \onlinecite{Huefner2008}. It has been noted that the B$_{1g}$ peak coincides with the pseudogap values, $2\Delta_{PG}$ derived from other experiments, while the B$_{2g}$ peak follows the superconducting gap $2\Delta = 8 k_B T_c$.

\textbf{Resonant inelastic x-ray scattering} (RIXS) \cite{2001_RMP_Kotani, 2011_RMP_Ament} is similar to Raman spectroscopy but has the additional advantage of full-momentum-space resolution. Despite remarkable progress of this new spectroscopic technique in the past decade \cite{2013_JESRP_Schmitt}, the results of this experiment are not fully understood \cite{2011_RMP_Ament}. Nevertheless, many exciting RIXS measurements have already been reported, and it is generally believed that RIXS can be an extremely powerful tool to probe the interplay between charge, spin, orbital, and lattice degrees of freedom. In particular, it has been shown that RIXS is a suitable probe across all energy scales, including pseudogap, charge-transfer gap, and Mott gap in cuprates \cite{2012_PRB_Basak}. Recent RIXS experiments \cite{2005_NP_Abbamonte, 2011_PRB_Fink, 2012_S_Ghiringhelli, 2014_NP_LeTacon} together with X-ray diffraction \cite{ChangNP2012} has revealed CDW ordering in cuprates.

\textbf{Inelastic neutron scattering} (INS) method works similar to RIXS but with neutrons instead of photons. Due to large penetration depth it is the most `bulk' among the spectroscopies considered here but requires very large single crystals and has mainly been done on YBCO, LSCO, and HgBa$_2$CuO$_{4+\delta}$ (Hg-1201). Like optical methods, INS measures except phonons the two-particle (electron-hole) excitations but with spin flip and with momentum resolution---in joint momentum-energy space. The most prominent feature seeing by INS in cuprates is a `spin resonance' \cite{2006_AP_Eschrig} that is peaked at the antiferromagnetic wavevector and at energy about 40 meV. The resonance is a part of a `hourglass shape' spin excitation spectrum \cite{2004_N_Tranquada, 2004_PRL_Pailhes} which became incommensurate above and below the resonance energy but never extends to zero energy being limited at low energies by the so-called spin-gap \cite{2001_PRB_Dai}.
In the normal state both YBCO and LSCO show a much weaker spectrum, which is centered around $\mathbf{Q}=(\pi,\pi)$ and is broader in momentum than in the superconducting state. In the pseudogap state, some intermediate picture is observed, with a gradually sharpening response at the antiferromagnetic wavevector, which has been considered as a precursor of the magnetic resonance mode that starts to develop below $T$* \cite{1999_S_Dai, 2001_PRB_Dai}. Other authors believe that there is no justification for a separation of the normal state spin excitations spectrum into resonant and non-resonant parts \cite{2000_PRB_Fong}.

For the scope of this review, it is important to mention the role of INS in discovery \cite{1995_N_Tranquada} and study \cite{1997_PRL_Tranquada, 2003_RMP_Kivelson} of incommensurate SDW and CDW, called `stripes', in the hole doped cuprates. As mentioned, the pseudogap can be a consequence of fluctuating stripes \cite{2003_RMP_Kivelson} or an electronic nematic order \cite{1998_N_Kivelson}.

Commensurate AFM ordering has been observed in the superconducting YBCO by elastic neutron scattering \cite{2001_PRL_Sidis}. More recently, the polarized neutron diffraction experiments on YBCO \cite{2006_PRL_Fauque} and Hg-1201 \cite{2008_N_Li} have shown an existence of a magnetic order below $T$* consistent with the circulating orbital currents and QCP scenario. This has been further supported by INS observation of a 52–56 meV collective magnetic mode appearing below the same temperature \cite{2010_N_Li}. The idea of the intra-unit-cell magnetic order has been also supported by recent polarized elastic neutron scattering experiments on BSCCO \cite{2014_PRB_Mangin-Thro} which raise important questions concerning the range of the magnetic correlations and the role of disorder around optimal doping.

\textbf{Tunneling spectroscopies}, like ARPES, measure the single-particle density of states. So, it is the most direct probe to see the pseudogap in Mott's definition \cite{MottRMP1968}. There are a number of different tunneling probes: intrinsic tunneling spectroscopy \cite{2000_PRL_Krasnov, 2009_PRB_Krasnov}, Andreev reflection tunneling (ART) \cite{2000_RPP_Kashiwaya, 2005_RMP_Deutscher}, superconductor/insulator/superconductor (SIS) \cite{2001_PRL_Zasadzinski} tunneling (in fact, both ART and SIS probe the two-particle DOS) and superconductor/insulator/normal metal (SIN) \cite{1997_PCS_Tao, 1999_PRL_Miyakawa}, as well as scanning-tunneling microscopy/spectroscopy (STM/STS) \cite{1998_PRL_Renner, 2007_RMP_Fischer}. The latter provides sub-atomic the spatial resolution and, with the Fourier transformation \cite{2002_S_Hoffman_b, 2003_N_McElroy}, an access to the momentum space \cite{2003_PRB_Wang, 2004_PRB_Markiewicz, 2007_JESRP_Kordyuk}.

The most convincing tunneling results showing that the superconducting and pseudogaps represent
different coexisting phenomena were obtained by intrinsic tunneling from one- and two-layers BSCCO \cite{2000_PRL_Krasnov, 2002_PRB_Krasnov, 2003_PRL_Yurgens, 2009_PRB_Krasnov}. The data for $T$* presented in Fig.\;\ref{STM} have been obtained by SIS tunneling on break junctions \cite{1998_PRL_Miyakawa} and SIN point contact tunneling  \cite{1998_PRL_DeWilde} also support the two-gaps scenario. Andreev reflection is expected to be similar to SIN and STM, but appears to be sensitive to the superconducting energy scale only, that may be because the tunneling mechanisms are actually different \cite{Huefner2008}.

STM, despite more complicated theoretical justification \cite{Tersoff}, has appeared to be extremely useful for study the pseudogap phenomenon in cuprates allowing one to explore spatial inhomogeneity \cite{2002_N_Lang, 2005_PRL_McElroy, 2007_NP_Boyer} and detect new orderings. In superconducting state, STM/STS reveals intense and sharp peaks at the superconducting gap edges which smoothly transform to broad maxima at the pseudogap energy above $T_c$, as one can see in Fig.\;\ref{STM} from the result of early SIN tunneling (a,b) \cite{1997_PCS_Tao} and STM (c) \cite{1998_PRL_Renner} experiments. The depletion of the density of states at Fermi level is seen to persist in the normal state, even above $T$*, at which it evolves more rapidly. The visual smoothness of the gap transition over $T_c$ may suggest a common origin of the gaps \cite{1998_PRL_Renner}. On the other hand, the size of the pseudogap looks independent on temperature, that makes it markedly different from superconducting gap (see discussion in Ref.\onlinecite{1999_PRL_Tallon}). Also, the studies of the normalized differential conductance \cite{2007_NP_Boyer} have shown a coexistence of a sharp homogeneous superconducting gap superimposed on a large but inhomogeneous pseudogap, see Fig.\;\ref{STM} (f).

The much weaker inhomogeneity observed at low energies in the Fourier transform maps of the STM spectra shows two type of modulations. The first one is due to the quasiparticle interference \cite{2003_PRB_Wang, 2004_PRB_Markiewicz, 2007_JESRP_Kordyuk} on the $d$-wave gapped electronic structure \cite{2002_S_Hoffman_b}. It allows to recover the momentum dependence of the superconducting gap \cite{2003_N_McElroy, 2005_PRL_McElroy}. The second one is a nondispersive modulation at higher energies, which can be related to the incoherent pseudogap states at the antinodes \cite{2004_S_Vershinin}. They could be related to a short-range local charge ordering with periods close to four lattice spacing in the form of the square `checkerboard' \cite{2002_S_Hoffman, 2003_PRB_Howald, 2004_N_Hanaguri} or unidirectional domains \cite{2007_S_Kohsaka}. These two modulations coexist in the superconducting state but compete with each other for the electronic states.

\section{Pseudogap in C\lowercase{u}-SC and TMD}
\label{HTSC}

ARPES is the most direct tool to measure the one particle spectrum with momentum resolution \cite{lynch1999, DamascelliRMP2003, 2014_LTP_Kordyuk}. Naturally, it has been successfully used to show that both superconducting gap and pseudogap are anisotropic: absent along the nodal direction and maximal at the antinodal region, and doping dependent: vanishing with overdoping, but the pseudogap is vanishing earlier \cite{1996_N_Ding, 1996_S_Loeser}. Moreover, while the superconducting gap follows a $d$-wave like dependence being zero only at the nodes, the pseudogap behaves more unusually, leaving non-gapped sections of the Fermi surface around the nodes \cite{1996_PRL_Marshall} later called `Fermi arcs' \cite{1998_N_Norman}. It was also suggested \cite{1998_N_Norman} that `Fermi arc' gradually changes its length from zero at $T_c$ to the full Fermi surface at $T$*, as shown in Fig.\;\ref{FermiArcs}. Panel (e) shows that the pseudogap increases gradually from the node and stays constant in the whole antinodal region for optimally doped two-layer BSCCO that is in contrast to heavily overdoped ($T_c$ = 0) one-layer Bi2201 (f) \cite{1998_N_Norman}.

\subsection{Measuring gaps in ARPES}
\label{ARPES}

\begin{figure}
\begin{center}
\includegraphics[width=0.44\textwidth]{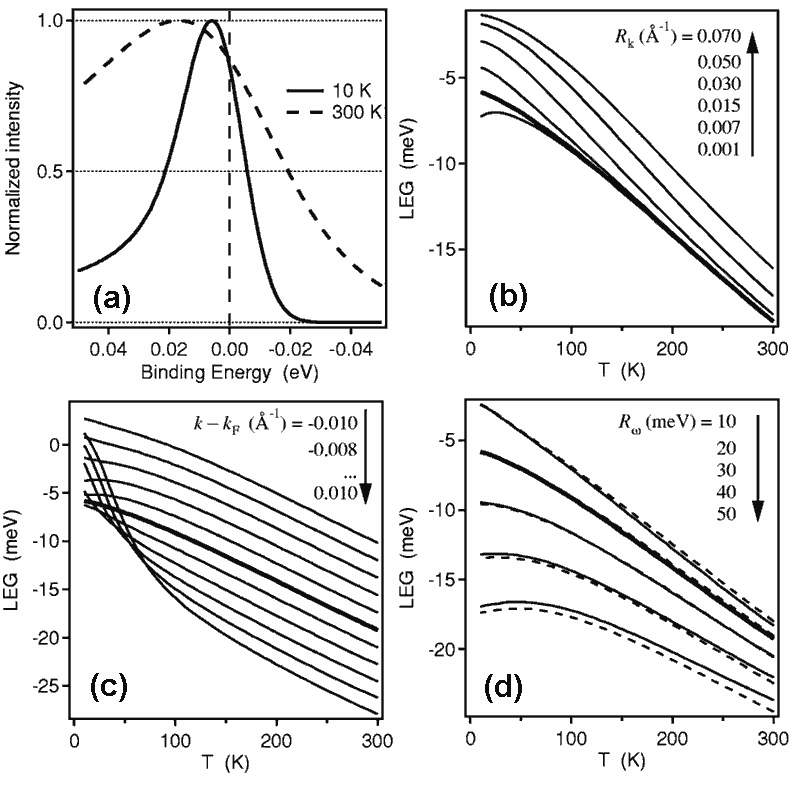}
\caption{'Leading edge gap' (LEG) in non-gapped ARPES spectra. (a) Leading edge midpoint of $k_F$-EDC depends on temperature, momentum (b) and energy (d) resolutions. False fast `opening' of the gap can be seen for EDCs slightly away from $k_F$ (c).
After \protect\ignorecitefornumbering{\cite{2003_PRB_Kordyuk}}.
\label{ModelLEG}}
\end{center}
\end{figure}

\begin{figure*}
\begin{center}
\includegraphics[width=0.9\textwidth]{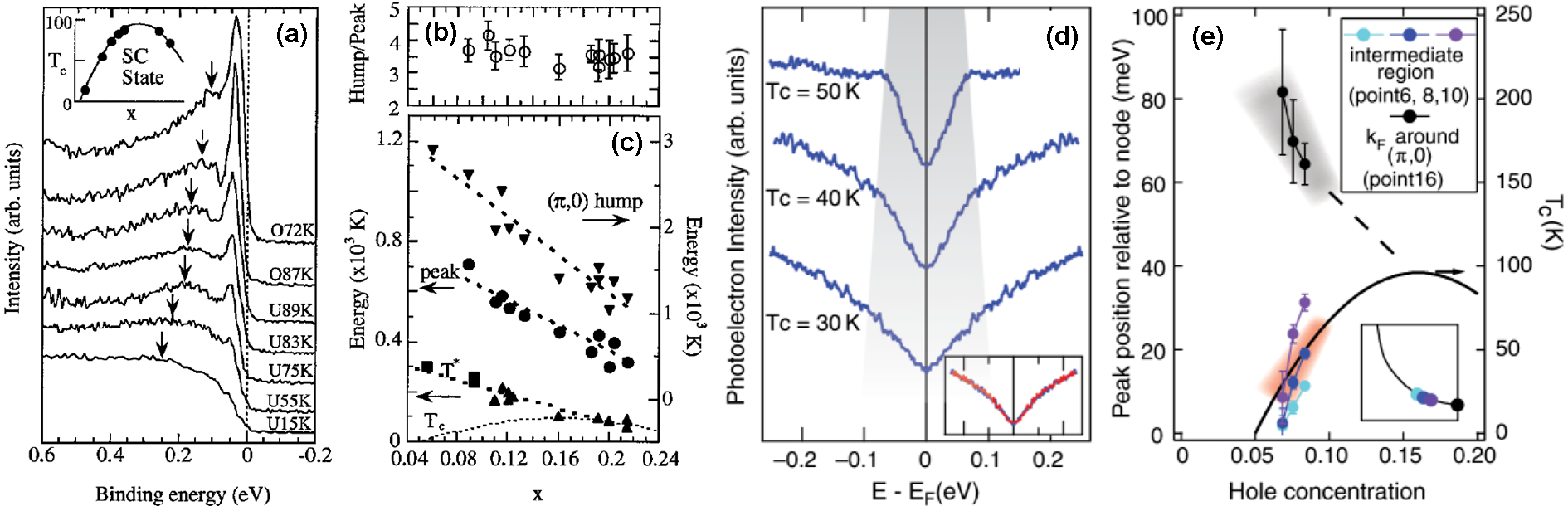}
\caption{Doping dependence of the pseudogap from ARPES. (a-c) EDCs from the antinodal region of BSCCO samples of different doping levels and two energy scales derived from them: positions of `peak' and `hump' \protect\ignorecitefornumbering{\cite{1999_PRL_Campuzano}}. (d) The symmetrized spectra of three underdoped BSCCO samples on which the `superconducting peak' is evolving into a kink, which can be defined as the second derivative maximum of the spectra; inset shows the absence of temperature dependence of these spectra for the
UD 30 K sample taken at 10 K (blue) and 50 K (red). (e) Doping dependence of the peak position for these three spectra (black symbols) and for three other momenta at the Fermi surface closer to the node, as marked in the inset \protect\ignorecitefornumbering{\cite{2006_S_Tanaka}}.
\label{PGfromARPES}}
\end{center}
\end{figure*}

Despite the clear evidences for the pseudogap anisotropy, the determination of the momentum resolved gap value in ARPES is far from being straightforward \cite{2003_PRB_Kordyuk}. First, one should distinguish a gap from a number of possible artifacts. Second challenge is to derive the gap value $\Delta$ that can be compared to other experiments and theoretical models.
Among possible artifacts in cuprates: charging by photocurrent, superstructure \cite{2004_N_Borisenko}, misalignment \cite{2003_PRB_Kordyuk}, bilayer splitting \cite{2002_PRB_Borisenko}, matrix elements \cite{2001_PRB_Borisenko}, photoemission background \cite{2004_PC_Borisenko}, and Van Hove singularity \cite{2002_PRL_Kordyuk}.
Most of them, if known, can be taken into account due to improved accuracy of the state-of-the-art ARPES technique \cite{2014_LTP_Kordyuk}.

If the gap model in known, as in the case of BCS-like superconducting gap or CDW gap, the best way to derive the gap value from experimental spectrum is to fit it to the model. And it seems that the most accurate method to extract the value of the BCS-like gap from ARPES spectra is fitting of a partial DOS (the momentum integrated EDCs along a cut perpendicular to the Fermi surface) to the formula derived by Evtushinsky \cite{EvtushinskyPRB2009}:
\begin{equation}
\text{IEDC}(\omega)=\Biggl[f(\omega, T)\cdot \Bigl|\text{Re}\frac{\omega+i\Sigma^{\prime\prime}}{E}\Bigr|\Biggr]\otimes R_{\omega},
\end{equation}
which coincides with the Dynes function \cite{1978_PRL_Dynes} multiplied by the Fermi function and convolved
with the energy resolution function $R_{\omega}$. Here $E=\sqrt{(\omega+i\Sigma^{\prime\prime})^2-\Delta_{\mathbf{k}}^2}$, $\Sigma^{\prime\prime}$ is the imaginary part of the self-energy, and $\Delta_{\mathbf{k}}$ is the momentum-dependent superconducting gap. This formula is obtained in approximation of linear bare electron dispersion, but there is also useful analytical solution for a shallow parabolic band \cite{EvtushinskyPRB2009}. A similar method of gap extraction is widely used in angle-integrated photoemission spectroscopy \cite{2001_PRL_Tsuda}. For our case it could be useful if the pseudogap in cuprates is due to either preformed pairs or Peierls like density waves.

If the model behind the gap is not known, other empirical methods could be used. The most straightforward one is to measure the \emph{peak position} of the gapped EDC and assume that $\Delta$ is the distance to $E_F \equiv 0$. It works well for momentum integrated spectra with BCS-like gap if such a `coherence peak' is well defined, that is usually not the case for the pseudogap in cuprates. Moreover, looking for a gap in momentum resolved ARPES spectrum, one deals with the $k_F$-EDC (EDC taken at Fermi momentum), which never peaks at $E_F$. In a normal non-gapped state this EDC is a symmetrical spectral function $A(\omega) = A(-\omega)$, which width is twice of the scattering rate $\Sigma''(0, T) = h \tau^{-1}$, multiplied by the fermi function: $I(\omega, T) = A(\omega, T) f(\omega, T)$. So, its peak position is temperature dependent, as one can see in Fig.\;\ref{ModelLEG} (a) \cite{2003_PRB_Kordyuk}.

Two procedures have been suggested to work around this problem, the \emph{symmetrization} \cite{1998_N_Norman} and \emph{division by Fermi function} \cite{2007_N_Lee}. If at $k_F$ the gaped spectral function obeys a particle-hole symmetry $A(\omega) = A(-\omega)$, both procedures should lead to the same result: $I(\omega) + I(-\omega) = I(\omega)/f(\omega) = A(\omega)$. This, however, does not help much to determine small gaps, when $\Delta < \Sigma''(\Delta)$: In this case two peaks below and above Fermi level are just not resolved and the symmetrized EDC is peaked at $E_F$. Thus, after symmetrization procedure, a smooth evolution of the gap with either temperature or momentum will look like a sharp gap opening when $\Delta(k, T) =$ $\Sigma''(\omega = \Delta, k, T)$.

The position (binding energy) of the midpoint of the leading edge of EDC is called the `leading edge shift' or `leading edge gap' (LEG) \cite{1996_N_Ding, 1996_ZPBCM_Pines}. Naturally, it is sensitive to the gap size but also depends on a number of parameters \cite{2003_PRB_Kordyuk}, as one can see in Fig.\;\ref{ModelLEG}: quasiparticle scattering rate and temperature, momentum and energy resolutions, displacement from $k_F$, etc. At 150 K, for example, for standard experimental resolutions (thicker middle curve on panels (b-d)) LEG is 10 meV above $E_F$, so, one can roughly say that LEG would be at $E_F$ if the pseudogap is about 10 meV.

The `Fermi arcs' story is illustrative in this respect. Initially, the `gapless arc' was defined as a set of Fermi momenta for which the leading edge midpoint is above $E_F$ (LEG $< 0$ in binding energy) \cite{1998_N_Norman}. Negative LEG is equivalent to a peak in the spectral function at $E_F$, so, the symmetrization procedure has been used instead of LEG in a number of detailed study of Fermi arcs evolution with temperature, see \cite{2006_NP_Kanigel}, for example. The observed dependence of the length of the arcs with temperature is consistent with temperature dependence of the $k_F$-EDC width, as explained above, or, in more theoretical language, as a consequence of inelastic scattering in a phase-disordered $d$-wave superconductor \cite{2007_PRB_Chubukov}. Thus, comparing a number of proposed models for the Fermi arcs, authors of Ref.\;\onlinecite{2007_PRB_Norman} have concluded that the best one to model the ARPES data is a $d$-wave energy gap with a lifetime broadening whose temperature dependence is suggestive of fluctuating pairs.

\begin{figure*}
\begin{center}
\includegraphics[width=0.9\textwidth]{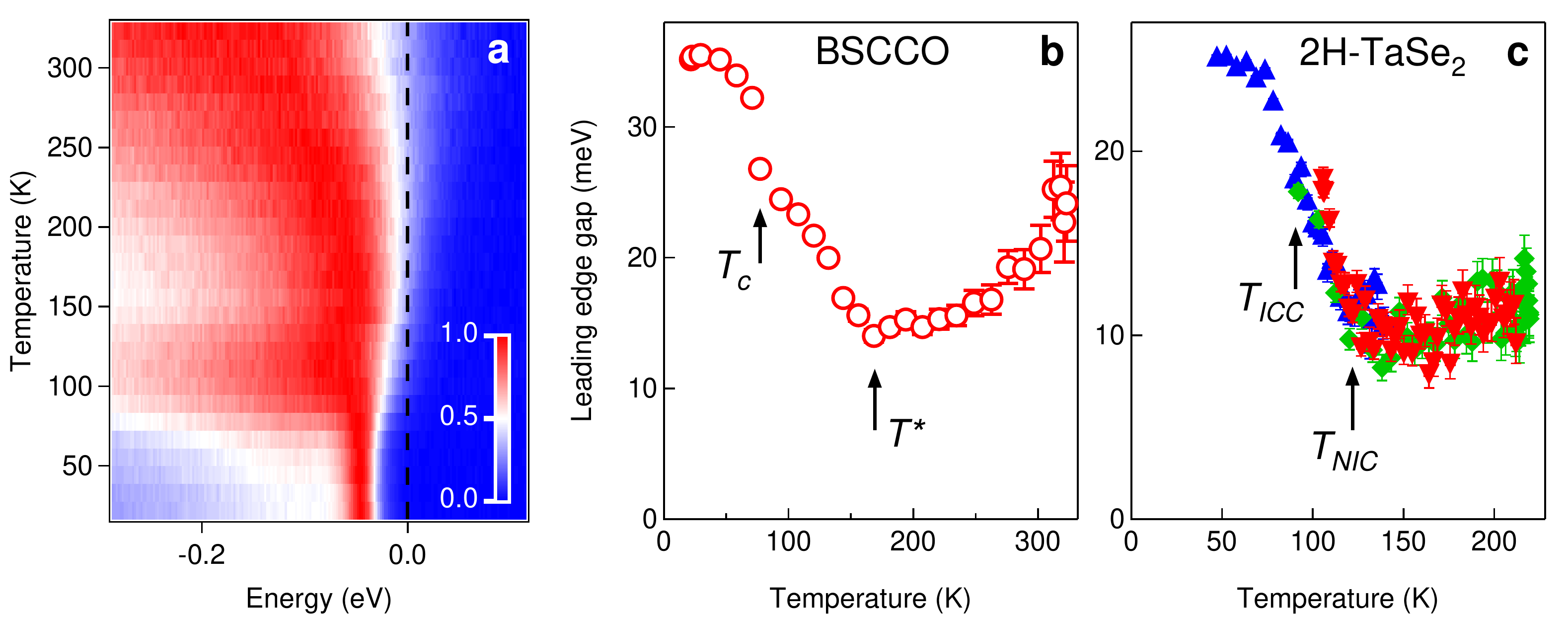}
\caption{Nonmonotonic pseudogap in cuprates. (a) The temperature map which consists of a number of momentum integrated energy distribution curves (EDCs) measured at different temperatures at a `hot spot'. The gap is seen as a shift of the leading edge midpoint (LEM) which corresponds to white color close to the Fermi level. (b) The position of LEM as function of temperature for an underdoped Tb-BSCCO with $T_c$ = 77 K and $T^*$ = 170 K is remarkably similar to the pseudo-gap in a transition-metal dichalcogenide 2H-TaSe$_2$ (c) with the transitions to the commensurate and incommensurate CDW phases at $T_{ICC}$ = 90 K and $T_{NIC}$ = 122 K, respectively. After \protect\ignorecitefornumbering{\cite{KordyukPRB2009, KordyukEPJ2010}}.
\label{NonmonoGap}}
\end{center}
\end{figure*}

Nevertheless, the question is not closed and Fermi arcs remain enigmatic. The initially proposed scenario of hole pockets \cite{1996_PRL_Marshall} is still considered. And while authors of \cite{2009_N_Meng} report on coexistence of both the Fermi arcs and hole pockets, the authors of Ref.\;\onlinecite{2011_PRL_Yang} insist that the Fermi arcs are illusion made by fully enclosed hole pockets with vanishingly small spectral weight at the magnetic zone boundary.

Interestingly that in view of `two gaps' scenario \cite{1999_PRL_Tallon, 2002_PRL_Markiewicz}, now widely accepted \cite{2006_NP_LeTacon, 2006_S_Tanaka, 2007_N_Lee, 2007_PRL_Kondo, Huefner2008, KordyukPRB2009, 2009_N_Kondo, 2010_NP_Hashimoto, 2014_NP_Hashimoto, 2014_xxx_Kaminski}, the Fermi arcs are natural signature of a competing to superconductivity order which is peaked at the antinodal region.

To finish with LEG method one should admit that it is less susceptible, in spite of Fig.\;\ref{ModelLEG} (c), to sudden artificial changes of the derived gap values and the real gap opening can be detected in LEG($T$). Also, LEG is a good quantity for the ARPES map of gaps \cite{2002_PRB_Borisenko}. Moreover, the LEG method works much better if applied to the momentum integrated spectrum (aforementioned partial DOS), since integration along a cut perpendicular to the Fermi surface removes the problem of $k_F$ determination error and, flattening the spectrum, place the leading edge midpoint of non-gapped spectra at the Fermi level \cite{2006_S_Valla, KordyukPRB2009}. In the same way the symmetrization of this partial DOS has much more sense than of single EDC and can be effectively used for visualization of the gap. So, both the LEG and symmetrization methods applied to partial DOS are simple but most robust procedures of gap detection, but to determine the gap value one shout fit it to the appropriate model, such as Eq.(1), for example.

\begin{figure*}
\begin{center}
\includegraphics[width=1\textwidth]{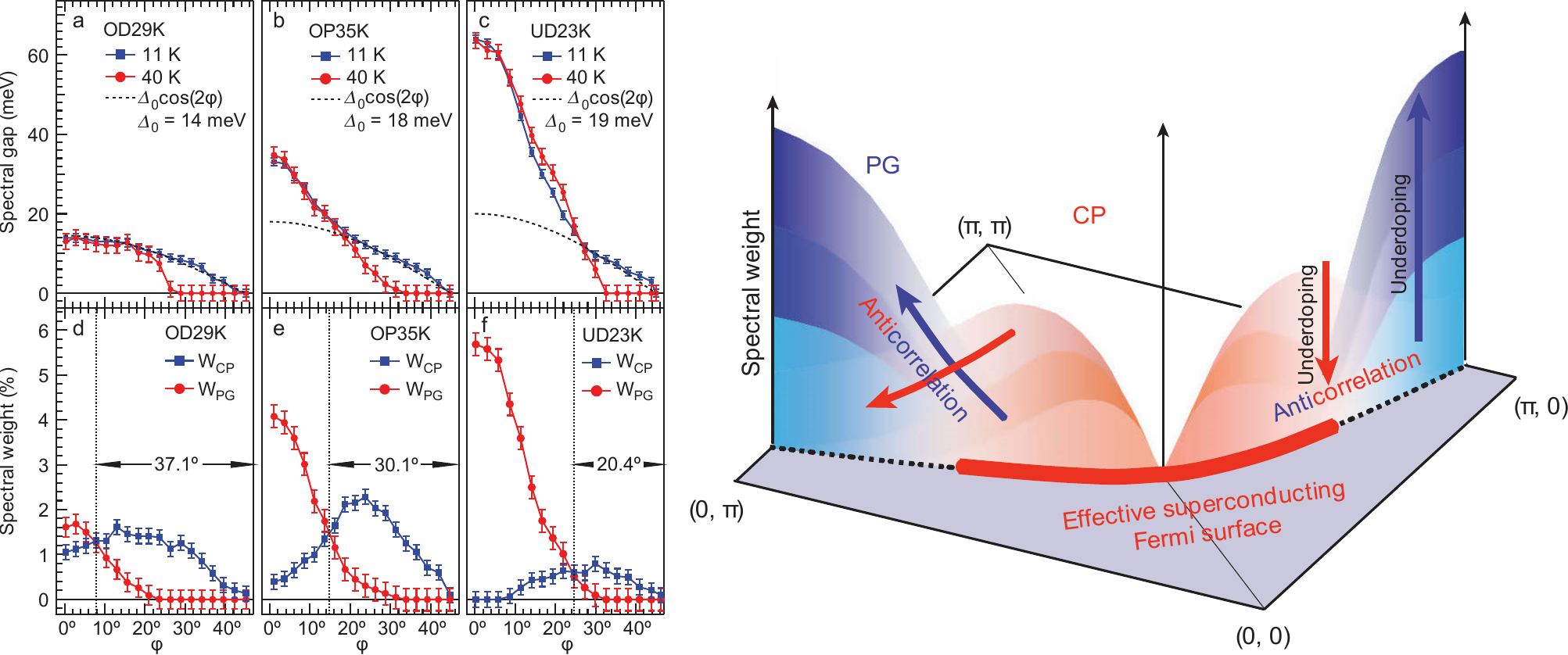}
\caption{The gaps above and below $T_c$ (a-c) and the coherent and pseudogap spectral weights (d-f) over the Fermi surface for different doping levels of Bi-2201. After \protect\ignorecitefornumbering{\cite{2009_N_Kondo}}.
\label{Kondo}}
\end{center}
\end{figure*}

All the said about the gap evaluation from ARPES is valid for a deep band, if it is much deeper than the gap. The \emph{Van Hove singularities} nearby the Fermi level complicate the situation \cite{2002_PRL_Kordyuk, 2003_PRL_Borisenko}. Typical set of EDCs from the antinodal region in superconducting state is shown in Fig.\;\ref{PGfromARPES}(a) \cite{1999_PRL_Campuzano}. In a wide doping range around optimal doping the spectra have so-called `peak-dip-hump' line shape \cite{1991_PRL_Dessau} that has been considered \cite{1999_PRL_Campuzano} as a consequence of interaction with the spin-fluctuations resonance seen by inelastic neutron scattering \cite{2006_AP_Eschrig}. The `superconducting peak' which dominates the overdoped spectra vanishes with underdoping evolving into a kink, which can be defined as the second derivative maximum of the spectra, as shown in panel (d) \cite{2006_S_Tanaka}. The energies of all the features, `peak', `dip', and `hump', scale similarly, increasing with underdoping (see Fig.\;\ref{PGfromARPES}(b,c)), but it is the superconducting peak position that fits the pseudogap values derived from other experiments \cite{Huefner2008, 2006_NP_LeTacon}, as has been shown earlier in Fig.\;\ref{RamanPhD}.

Later it has been shown \cite{2002_PRL_Kordyuk} that the `peak-dip-hump' structure is completely due to the bi-layer splitting (the peak and hump correspond to the VHs's of the antibonding and bonding bands respectively) at the overdoped side, and only with underdoping the $(\pi,0)$-spectra become affected by both the superconducting gap and the spin-fluctuations resonance \cite{2003_PRL_Borisenko}: the latter contributes to the dip while the peak, being sandwiched between the gap and the resonance, becomes narrower and finally looses its spectral weight. One can mention here that besides the spin-fluctuations also the low-energy CDW modes can contribute to the peak-dip-hump structure \cite{2001_PRB_Seibold}.

The asymmetric STM spectra also can be naturally explained by the bi-layer split VHs \cite{2003_PRB_Hoogenboom}. So, the doping dependence of the gaps derived from $(\pi,0)$ ARPES spectra and from tunneling in the superconducting state should be taken with caution. On the other hand, the asymmetry of SIN tunneling spectra can be due to a contribution to the Green function (and tunnel current) that represents the electron-hole pairing and is proportional to the CDW order parameter depending on its phase \cite{2008_JPCM_Ekino, 2014_PC_Gabovich} (as shown in the earlier work \cite{1983_JL_Artemenko}). Also, there are reports that the bi-layer splitting may be vanishing with underdoping \cite{2010_NP_Fournier}, but the most careful spectra for underdoped one-layer Bi-compound \cite{2014_xxx_Kaminski} do not show the `peak-dip-hump' line shape.


Despite all the mentioned complications, one can make the following conclusions. (1) Maximal (for given sample) pseudogap value exhibits similar doping dependence as the temperature at which it starts to develop, i.e. $\Delta$*$(x) \sim T$*$(x)$. (2) This dependence is essentially different from the dependence of the superconducting transition temperature: $T$*$(x) \nsim T_c(x)$. But the relation between $T_c$ and $\Delta_{SC}$ remained controversial since different techniques gave different $\Delta_{SC}(x)$ dependences. One can say that this controversy is now resolved \cite{Huefner2008}.

\subsection{Two gaps in C\lowercase{u}-SC}
\label{twogaps}

The idea that the pseudogap and superconducting gap are two distinct gaps \cite{1999_PRL_Tallon, 2002_PRL_Markiewicz} rather than one is a precursor of another has become started to find wide acceptance when a number of evidences for different doping dependence of the gaps measured in different experiments has reached some critical value (see Fig.\;\ref{RamanPhD} and Refs.\;\onlinecite{Huefner2008, 2006_NP_LeTacon}) and, that may be more important, when those different dependence have been observed in one experiment, first in Raman \cite{2006_NP_LeTacon} and then in ARPES \cite{2006_S_Tanaka, 2007_N_Lee, 2007_PRL_Kondo}. It has been shown that in superconducting state the gap measured around the node does not increase with underdoping as the antinodal gap but scales with $T_c$. Studying the evolution of the spectral weight of some portions of ARPES spectra (the weight under the `coherent peak' and the weight depleted by the pseudogap) it has been concluded that the pseudogap state competes with the superconductivity \cite{2009_N_Kondo}. These results are summarized in Fig.\;\ref{Kondo}.

Another difference between the pseudogap and superconducting gap has come from STM: the superconducting gap is homogeneous while the pseudogap is not \cite{2007_NP_Boyer}.

One may conclude that the pseudogap which opens at $T$* and the superconducting gap have different and competing mechanisms and that $T$* is not the temperature of the preformed pairs: $\Delta_{SC} \sim T_c \sim T_p \nsim T$*$ \sim \Delta$*. This does neither exclude an existence of the preformed pairs nor uncover the $T$* origin. To do the latter, one should find the pseudogap features peculiar for a certain mechanism. And probably this could be done empirically, comparing the pseudogap in cuprates to known cases.

\begin{figure*}
\begin{center}
\includegraphics[width=1\textwidth]{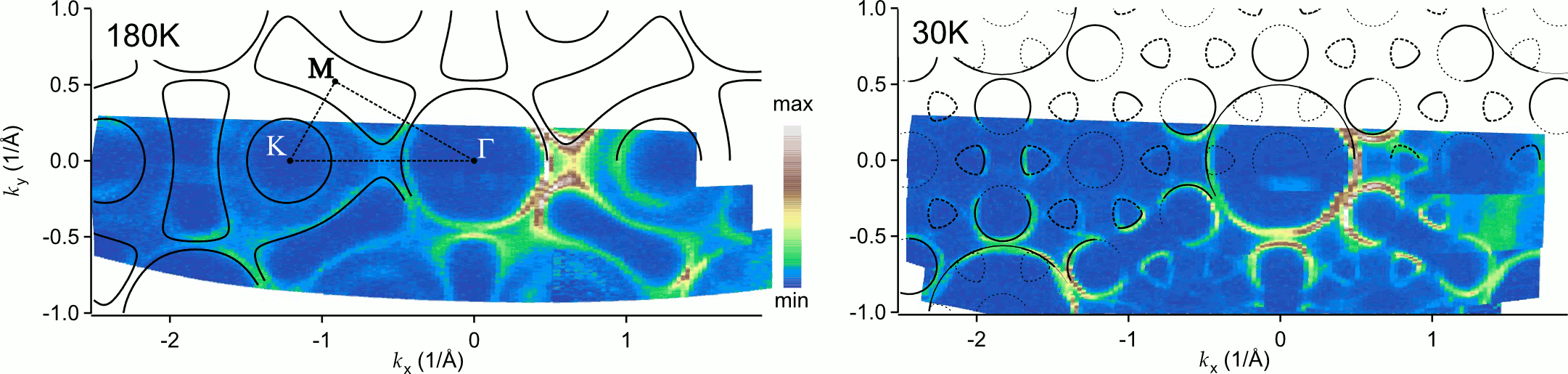}
\includegraphics[width=1\textwidth]{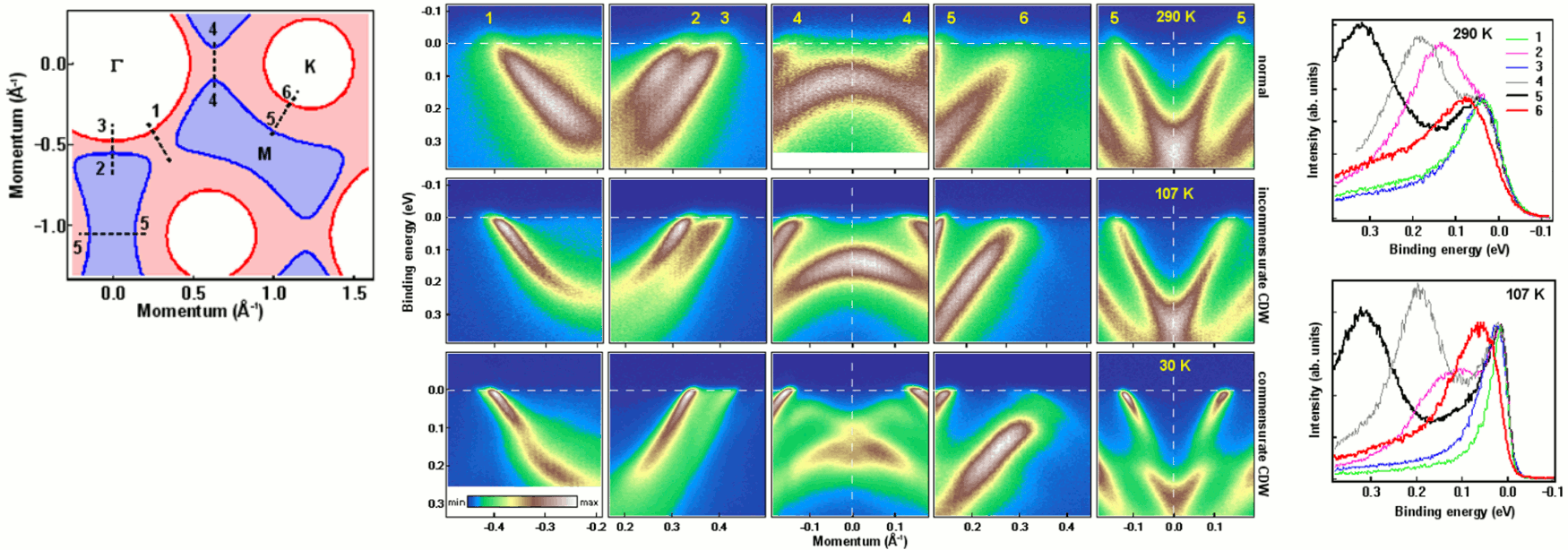}
\caption{Evolution of the Fermi surface (upper row) \protect\ignorecitefornumbering{\cite{2008_PRL_Evtushinsky}} and underlying electronic structure (lower row) of 2H-TaSe$_2$ with temperature. Fermi surface changes topology at 90 K (transition to the commensurate CDW state) while the pseudogap opens on some parts of the Fermi surface at 122 K (incommensurate CDW transition). After \protect\ignorecitefornumbering{\cite{BorisenkoPRL2008}}.
\label{TaSe2}}
\end{center}
\end{figure*}

Indeed, it has been found \cite{KordyukPRB2009} that from ARPES point of view, the pseudogap in BSCCO is remarkably similar to the incommensurate CDW gap in another quasi-2D metal, the transition metal dichalcogenide 2H-TaSe$_2$. Fig.\;\ref{NonmonoGap} shows evolution of the gap with temperature as a temperature map (a) and as the position of the leading edge (b). The temperature dependence of LEG in an underdoped Tb-BSCCO with $T_c$ = 77 K and $T^*$ = 170 K looks identical to the same quantity (c) measured in 2H-TaSe$_2$ with the transitions to the commensurate and incommensurate CDW phases at $T_{ICC}$ = 90 K and $T_{NIC}$ = 122 K, respectively \cite{KordyukPRB2009, KordyukEPJ2010}. Note, that if one plots the peak position from panel (a), it would increase above $T_c$ having a local maximum at about 120 K. Such a behavior has been considered as the most convincing evidence for the existence of two distinct gaps \cite{2014_xxx_Kaminski}.

So, the incommensurate CDW or other density wave could be the main reason for the pseudogap below $T$*, but the spectroscopic consequences of it are not trivial and even difficult to calculate from the first principles \cite{2000_S_Voit}. In this  case, one may try to use TMD as model systems to compare in details the charge ordering gaps to the pseudogap in cuprates.

\subsection{CDW gaps in TMD}
\label{CDW}

\begin{figure*}
\begin{center}
\includegraphics[width=0.84\textwidth]{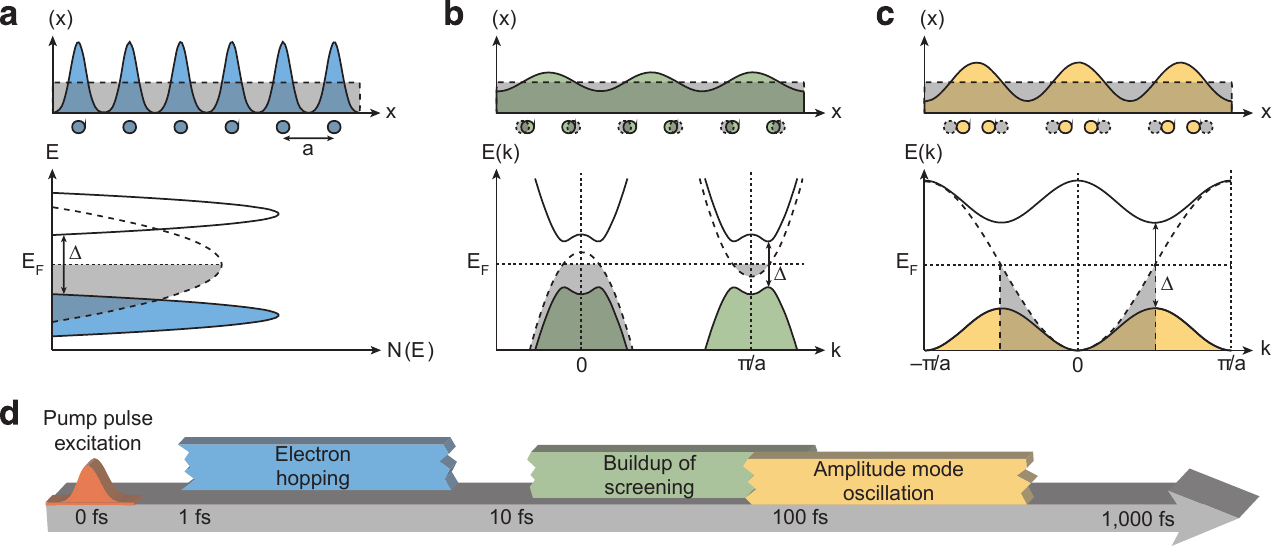}
\caption{Time-domain classification of CDW insulators. (a) Mott insulator. (b) Excitonic insulator. (c) Peierls insulator. (d) Corresponding timescales of the responses to impulsive near-infrared excitation and their assignment to elementary model-specific processes. After \protect\ignorecitefornumbering{\cite{2012_NC_Hellmann}}.
\label{CDW_ins}}
\end{center}
\end{figure*}

Quasi-2D transition metal dichalcogenides in which a number of CDW phases are realized \cite{1975_AiP_Wilson} can be useful model systems to study the spectroscopic manifestations of those phases and their relation to the electronic structure. In general, the quasi-2D electronic systems have a weaker tendency towards the formation of CDW and SDW instabilities than quasi-1D metals because the Fermi surfaces in 2D can be only partially nested and therefore partially gapped, so the system may be metallic even in the CDW state. The 2D character and the existence of an anisotropic gap make these systems similar to the HTSC cuprates \cite{2000b_PRL_Valla}, especially taking into account similarity between $T$* and $T_c$ lines in cuprates and $T_{CDW}$ and $T_c$ lines in the $T$-doping and $T$-pressure phase diagrams of dichalcogenides \cite{2004_PRL_Valla}, see Fig.\;\ref{PhDs}. For topical review on the origin of charge-density waves in layered transition-metal dichalcogenides see Ref.\;\onlinecite{2011_JPCM_Rossnagel}.

2H-TaSe$_2$ \cite{BorisenkoPRL2008} and 2H-NbSe$_2$ \cite{2009_PRL_Borisenko, 2007_NP_Kiss, 2012_PRB_Rahn} seem to be perfect model systems to understand the effect of different CDW on electronic density of states and ARPES spectra. Fig.\;\ref{TaSe2} shows the Fermi surface of 2H-TaSe$_2$ \cite{BorisenkoPRL2008, 2008_PRL_Evtushinsky}, a compound in which there are two phase transitions into the states with incommensurate (122 K) and commensurate $3\times3$ (90 K) CDW. It is the first transition at which a jump in the heat capacity and a kink in the resistance are observed, while the second transition has almost no effect on these properties \cite{1975_AiP_Wilson}. From ARPES point of view the situation is opposite. The Fermi surface (shown in the upper left panel) remains virtually unchanged up to 90 K, and a new order appears just below the commensurate transition. The explanation for this dichotomy comes from the behavior of the spectral weight near the Fermi level on the Fermi surface sheet centered around K-points. Below 122 K the spectral weight starts to decrease sharply, that is the pseudogap opening (see the cross-section 5-6). When passing through 90 K, the pseudogap is transformed into a band gap in the new Brillouin zone, but this transition is not accompanied by such a gain in kinetic energy.

It is a good example when both the commensurate and incommensurate CDW are driven by the Fermi surface nesting, that, as the name implies, is a measure of coincidence of the Fermi surface parts shifted by a `nesting' vector. Numerically, the nesting vectors can be found by autocorrelation of the measured Fermi surface \cite{BorisenkoPRL2008}, or, more physically, from peaks of the imaginary part of electronic susceptibility \cite{2008_NJP_Inosov, 2009_PRB_Inosov}. Interestingly, there is opinion that the Fermi surface nesting is a misconception since it is very sensitive to the Fermi surface geometry while the calculations show that the Fermi surfaces almost never nest at the right CDW vectors \cite{2008_PRB_Johannes}. The mentioned ARPES studies have shown that the nesting, which, of course, is better to discuss in terms of peaks in electron susceptibility, is indeed very sensitive to the Fermi surface geometry \cite{BorisenkoPRL2008}. That is why the nesting vectors coincide with CDW vectors when derived from the experimental band structures rather than from the calculated ones. In fact, the incommensurate CDW in 2H-TaSe$_2$ has appeared to be more complex at some temperature range, consisting of one commensurate and two incommensurate wave vectors \cite{1980_PRL_Fleming, 2011_PRB_Leininger}.

The ARPES data on 2H-TaSe$_2$ and other TMDs prove empirically that the formation of the incommensurate charge density wave, which can be described within the scenario based on short-range-order CDW fluctuations \cite{2012_JETP_Kuchinskii}, leads to depletion of the spectral weight at the Fermi level, while the transition from incommensurate to commensurate order leads rather to a redistribution of the spectral weight in momentum. This is consistent with the sign changing Hall coefficient in this compound \cite{2008_PRL_Evtushinsky}. So, the incommensurate gap in dichalcogenides looks very similar to the pseudogap in cuprates \cite{KordyukPRB2009}.

Among other types of CDW, which are observed in 2H-TMDs and could be similar to CDW in cuprates, I would mention the striped incommensurate CDW \cite{1980_PRL_Fleming, 2013_PNAS_Soumyanarayanan} and nearly commensurate CDW observed in 2H-NbSe$_2$ by STM \cite{2014_PRB_Arguello}. The latter is established in nanoscale regions in the vicinity of defects at temperatures that are several times the bulk transition temperature $T_{CDW}$.

Other analogies may be found between cuprates and 1T-TDMs in VHs nesting and correlation gap, as discussed in Sec.\;\ref{VHsTMD}.

\subsection{CDW in cuprates}
\label{CDWinCuSC}

Until recently, CDW in cuprates remained almost purely theoretical idea, but now one may say that that was due to the dynamical nature of CDW fluctuations \cite{2009_PRB_Grilli, 2009_PRL_Seibold}.

Last years of experimental studies added much to the evidence concerning CDW in cuprates \cite{NormanAP2005, 2013_LTP_Gabovich, 2015_xxx_Chowdhury}. Initially, a copper-oxygen bond-oriented ``checkerboard" pattern has been observed by STM in vortex cores in BSCCO \cite{2002_S_Hoffman}. The proposed explanation was a spin density wave localized surrounding each vortex core, but similar pattern had been observed also in zero field \cite{2003_PRB_Howald}. In BSCCO above $T_c$ there is energy-independent incommensurate periodicity in the pseudogap state close to 1/4 \cite{2004_S_Vershinin} or 1/4.5 \cite{2005_PRL_McElroy}, if measured deep in superconducting state.

Transport measurements for LSCO also find a tendency towards charge ordering at particular rational hole-doping fractions of 1/16, 3/32, 1/8, and 3/16 at which resistivity is peaked \cite{2005_PRL_Komiya}. The charge ordering, in terms of Cooper pairs density waves (PDW), was expected to be particularly pronounced near certain `magic' doping levels, where the charge modulation is commensurate with the underlying lattice \cite{2004_PRB_Chen, 2004_PRL_Tesanovic}.

Raman at higher frequencies on LSCO \cite{2011_PRB_Caprara} has shown that the spin fluctuations are present even in overdoped samples, but their strength tends to decrease substantially upon overdoping, while the charge-ordering fluctuations increase and reach a maximum intensity around $x \approx$ 0.19.

Recent neutron and X-ray scattering experiments on underdoped Bi$_2$Sr$_{2-x}$La$_x$CuO$_{6+\delta}$ \cite{RosenNC2013} point to a surface-enhanced incipient CDW instability, driven by Fermi surface nesting.

Hard x-ray diffraction measurements \cite{2014_PRB_Croft} on LSCO of three compositions ($x=$ 0.11, 0.12, 0.13) revealed CDW order with onset temperatures in the range 51-80 K and ordering wave vectors close to (0.23, 0, 0.5). On entering the superconducting state the CDW is suppressed, demonstrating the strong competition between the charge order and superconductivity. CDW order coexists with incommensurate magnetic order and the wave vector of CDW is twice of the wave vector of SDW. This fluctuating CDW order is strongly coupled to, and competes with, superconductivity, as demonstrated by the observed non-monotonic temperature dependence of the scattering intensity and the correlation length \cite{ChangNP2012, FradkinNP2012}.

\begin{figure*}
\begin{center}
\includegraphics[width=0.84\textwidth]{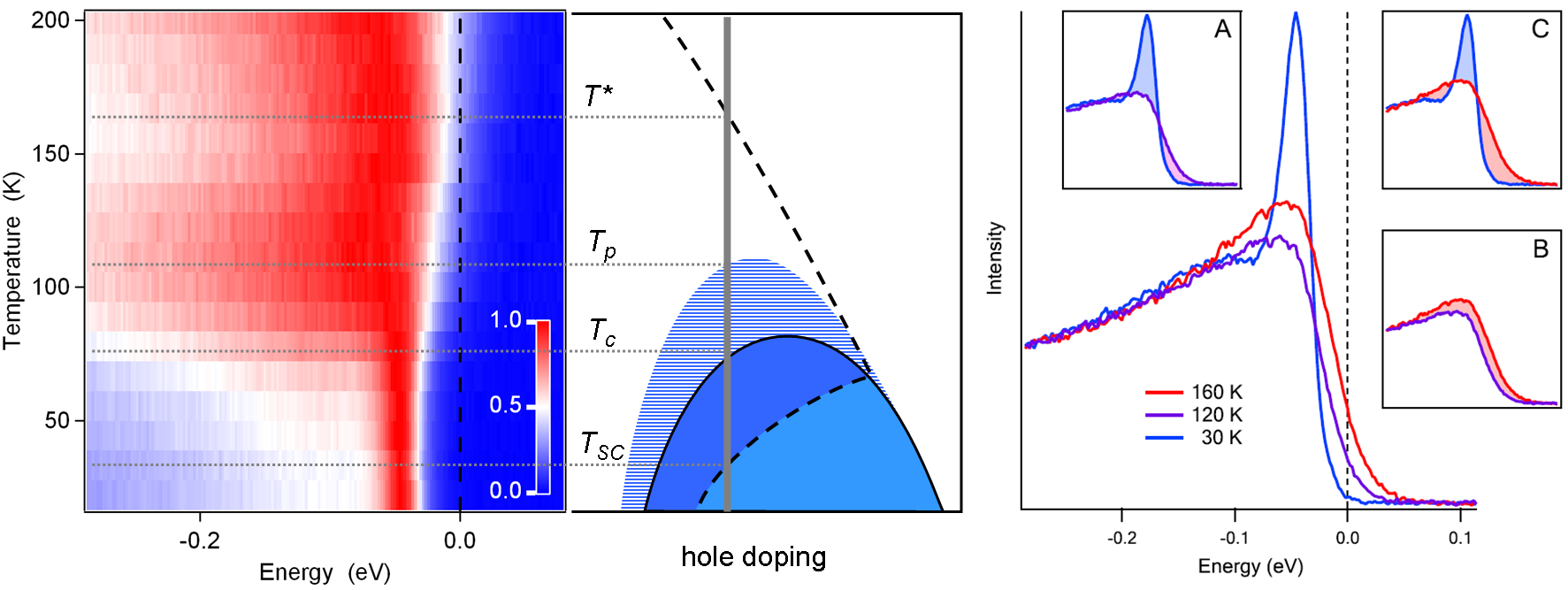}
\caption{Temperature evolution of the hot spot EDC for underdoped BSCCO (77 K). The transition temperatures on the phase diagram (center) correspond to marked changes in EDC evolution as it can be seen from the temperature map (left): at $T$*, the pseudogap starts to increase rapidly, the spectral weight starts to decrease; at $T_p$, the spectral weight starts to increase; at $T_c$, the superconducting gap opens, the spectral weight continues to increase up to $T_{SC}$. The examples of non-normalized EDC's at 160 K, 120 K, and 30 K (right) illustrate the spectral weight evolution. Adopted from \protect\ignorecitefornumbering{\cite{KordyukPRB2009}}.
\label{hotspotEDCvsT}}
\end{center}
\end{figure*}

In many studies the break of four-fold rotational symmetry have detected in the pseudogap state, pointing to stripe or nematic order \cite{2003_RMP_Kivelson, 2009_AP_Vojta}. For example, the uni-directional stripes within the checkerboard has been detected by STM \cite{2004_N_Hanaguri}.
A large in-plane anisotropy of the Nernst effect has been observed in YBCO \cite{2010_N_Daou}. The anisotropy, as reported, sets in precisely at $T$* throughout the doping phase diagram.


So, nowadays there are enough evidences for the CDW ordering in cuprates. These waves are generally consistent with the idea of Fermi surface nesting, thus should gap the straight sections of the Fermi surface, but it is unlikely that they can be responsible alone for the whole pseudogap state bordered by $T$*$(x)$. Then other possible constituents of the pseudogap are SDW due to VHs nesting, AFM order, and Mott gap, each one or all together.

\subsection{VHs nesting and Mott gap in TMD}
\label{VHsTMD}

Let us first consider Van Hove singularity driven CDW in 1T-TMD's. Some of those compounds are known as `excitonic insulators' \cite{2011_JPCM_Rossnagel}. The driving force for new ordering is a win of electron kinetic energy that happens when two VHs's of opposite character (e.g., top and bottom of different bands) residing near the Fermi level are folded to the same momentum, as shown in Fig.\;\ref{CDW_ins} (b). Among a few known examples is 1T-TiSe$_2$ \cite{2007_PRL_Cercellier}. It shows large band renormalizations at high-symmetry points of the Brillouin zone and a very large transfer of spectral weight to backfolded bands.

Another example of VHs nesting has been found recently in $5d$ transition metal compound IrTe$_2$ \cite{2014_NJP_Qian}. It has been shown that the band related to the saddle points at the Fermi level is strongly reconstructed below transition temperature, removing VHs from $E_F$ and the wavevector between the adjacent saddle points is consistent with the in-plane structural modulation vector.

Partial gaps have been reported for other 1T-compounds: 1T-VSe$_2$ \cite{2003_PRB_Terashima} and classical 1T-TaS$_2$, where CDW is called `quasicommensurate' \cite{1999_PRL_Pillo, 2000_PRB_Pillo} or `nearly commensurate' \cite{1977_JPSJ_Nakanishi, 1989_S_Wu, 2004_PRB_Bovet, 2008_NM_Sipos} (domain-like discommensurate \cite{1977_JPSJ_Nakanishi}, i.e., commensurate domains separated by discommensurate areas \cite{1989_S_Wu}). In case of 1T-TaS$_2$, the commensurate CDW phase has been discussed in relation to Mott transition \cite{1999_PRL_Pillo, 2008_NM_Sipos}. It has been suggested that the Mott phase melts into a textured CDW and superconductivity develops within the CDW state, and survives to very high pressures \cite{2008_NM_Sipos}. This compound becomes superconducting when subjected to external pressure \cite{2008_NM_Sipos} or chemical doping or Fe \cite{2012_PRL_Ang}. 1T-TaS$_2$ with Cu intercalation reveals a disorder-induced metallic state; a non-Fermi liquid with a pseudogap that persists at finite temperatures \cite{2014_PRL_Lahoud}. A Mott transition has been found also at the surface of 1T-TaSe$_2$ \cite{2003_PRL_Perfetti, 2003_PRB_Aiura}.

The assignment of the partial gap observed by ARPES to Peierls or Mott type could be controversial \cite{1999_PRL_Pillo, 2004_PRB_Bovet}, but it seems that the time resolved ARPES, measuring the melting times of electronic order parameters, can help to resolve this controversy \cite{2012_NC_Hellmann}. A time-domain classification of charge-density-wave insulators is shown in Fig.\;\ref{CDW_ins} \cite{2012_NC_Hellmann}: the Mott insulator collapses due to an ultrafast rearrangement of the electronic states on the elementary timescale of electron hopping, the excitonic insulator breaks down because the Coulomb attraction causing electrons and holes to form excitons is screened by the added free carriers, and the Peierls insulator melts with atomic rearrangement. In particular, it has been proved that Rb intercalated 1T-TaS$_2$ is a Peierls insulator while the 1T-TiSe$_2$ is an excitonic insulator.

While the mechanism of the Mott transition in TMD is under active consideration now \cite{2011_JPCM_Rossnagel, 2012_NC_Hellmann}, one may think about it in terms of critical depth of the pseudogap derived by Mott in 1969 for liquid metals \cite{MottPM1969}. One can also expect that flattening of the band leads to localization of the band forming electrons.

\subsection{Three gaps in C\lowercase{u}-SC}
\label{threegaps}

From incommensurate CDW one may expect a transfer of the spectral weight from the pseudogap to other momenta while the Mott transition involves the weight transfer to higher binding energies above 1-2 eV \cite{2010_PRB_Das}. So, in order to distinguish between different mechanisms of pseudogap formation, careful temperature dependence of ARPES spectra in the whole Brillouin zone is required, that is a lot of experimental work still to be done.

As an example, Fig.\;\ref{hotspotEDCvsT} shows the same `hot-spot' EDC as in Fig.\;\ref{NonmonoGap} but not normalized. One can see that from 160 K to 120 K the spectral weight disappears. It may be transferred from around $E_F$ either to much higher energies or to other momenta. In the superconducting state the spectral weigh recovers in `coherence peak'. It has been shown while ago \cite{1998_S_Shen} that the weight to the peak is transferred from other momenta and higher binding energy (up to 0.3 eV), so, one may assume that both the incommensurate CDW and localization do affect the `hot-spot' spectrum, but more temperature dependent studies are clearly needed.

The AFM $(\pi,\pi)$ interaction in cuprates is certainly a strong one, taking into account its persistence on electron doped side of the phase diagram (see Fig.\;\ref{Electron_doped_FS}) and the energy transfer involved at Mott transition \cite{2010_PRB_Das}. Based on comparison with TMD, one may speculate that the Mott transition in cuprates occurs due to commensurate SDW gap development (as in the spin-fermion model \cite{2003_AiP_Abanov}, for example) for which the reason is VHs at $(\pi,0)$. Also, due to interaction of two extended saddle points with opposite curvatures, the resulting band flattening is expected. One should note that some evidence for incommensurate SDW has been obtained in neutron experiments on YBCO \cite{HaugNJP2010}. In Refs.\;\cite{2010_NP_Hashimoto, 2014_NP_Hashimoto} it has been shown that temperature evolution of antinodal ARPES spectrum for Bi-2201 is mostly consistent with a commensurate $(\pi,\pi)$ density-wave order, but not with the preformed pairs scenario.

On the other hand, some evidence for the preformed pairs in the underdoped Bi-2212 with $T_c$ = 65 K has been found looking for the particle–hole symmetry in the pseudogap state \cite{2008_N_Yang}. A $d$-wave symmetry of the pseudogap has been observed in non-superconducting La$_{2-x}$Ba$_{x}$CuO$_4$ (LBCO) ($x=1/8$) and concluded that the Cooper pairs form spin-charge–ordered structures instead of becoming superconducting \cite{2006_S_Valla}. Finally, evidence for the preformed pairs state have been found in accurate ARPES experiments by Kaminski \cite{2014_xxx_Kaminski}.

To conclude, now it seems evident that at least three mechanisms form the pseudogap in the hole doped cuprates: the preformed pairing, the incommensurate CDW due to nesting of the straight parallel Fermi surface sections around $(\pi,0)$, and the $(\pi,\pi)$ SDW which is dominant constituent of the pseudogap assosiated with $T$* and is either causing or caused by the Mott localization. These phases occupy different parts of the phase diagram, as shown in Fig.\;\ref{HTSC_PhD}, and gap different parts of the Fermi surface \cite{2014_xxx_Kaminski, 2014_NP_Hashimoto} competing for it.

\subsection{Two sides of the phase diagram}
\label{PhD2}

\begin{figure*}
\begin{center}
\includegraphics[width=1\textwidth]{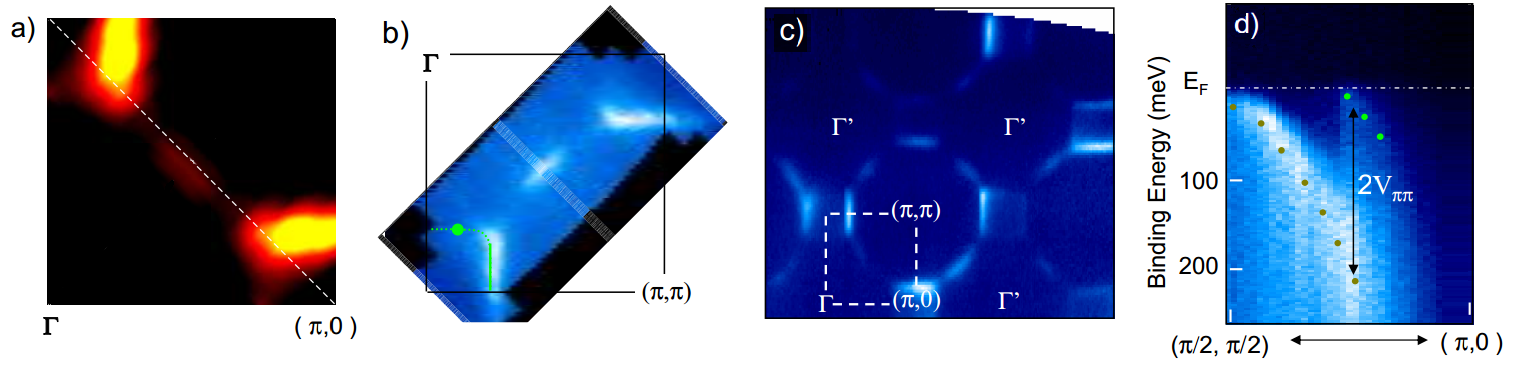}
\caption{ARPES evidence for AFM ordering in superconducting electron doped cuprates: fragmented Fermi surfaces of Nd$_{1.87}$Ce$_{0.13}$CuO$_4$ (a) \protect\ignorecitefornumbering{\cite{2005_PRL_Matsui}} and Sm$_{1.86}$Ce$_{0.14}$CuO$_4$ (b,c) and split `shadow' and main bands along the magnetic zone boundary (d) \protect\ignorecitefornumbering{\cite{2007_PRB_Park}}.
\label{Electron_doped_FS}}
\end{center}
\end{figure*}

It is believed that electron- and hole-doped cuprates represent the Slater and Mott pictures, respectively \cite{2010_PRB_Das, 2010_NP_Weber}. Although the electron-doped cuprates share the same layered structure based on CuO$_2$ planes, their phase diagram differs essentially. In Nd$_{2-x}$Ce$_x$CuO$_{4-y}$ (NCCO), for example, the 3D antiferromagnetic state extends up to $x$ = 0.15, and the superconducting region is confined to a narrow doping range (0.15-0.17) neighboring the AFM state. On the other hand, the superconducting dome of another electron doped compound, La$_{2-x}$Ce$_x$CuO$_{4-y}$, is in a similar position as for the hole doped LSCO \cite{2008_PRB_Krockenberger}. So, one may conclude that universality of the phase diagram at the electron doped side is still an open question.

The presence of the pseudogap phase at the electron doped side is also controversial, but in any case it is not so extended as on the hole side. Some experiments show existence of a pseudogap when superconductivity is suppressed by magnetic field \cite{2003_N_Alff, 2006_PRB_Wang}, that excludes precursor of superconductivity as its origin.

The magnetic excitations are present in both hole- and electron-doped cuprates been even stronger in the latter \cite{2014_NP_Lee}, but it does not correspond to a higher superconducting transition temperature. Thus, it is important to identify which factors, the magnetic excitations, the underlying Fermi-surface topology, or additional effects, are not optimized here.

ARPES confirms that the Brillouin zone is magnetic, i.e. there is clear observations of a gap along the magnetic zone boundary \cite{2005_PRL_Matsui, 2007_PRB_Park, 2009_PRB_Nekrasov}. Fig.\;\ref{Electron_doped_FS} show a fragmented Fermi surface, which suggests that the large Fermi surface is gapped by into electron and hole pockets \cite{2005_PRL_Matsui, 2007_PRB_Park, 2009_PRB_Nekrasov}, and `shadow' and main bands are split along the magnetic BZ boundary \cite{2007_PRB_Park}. This can be described by the generalized dynamical mean-field theory with the $k$-dependent self-energy (LDA + DMFT + $\Sigma_\mathbf{k}$) \cite{2009_PRB_Nekrasov}. Similar $s$-wave like dependence of the pseudogap has been recently suggested based on the analysis of Raman spectra and for hole doped BSCCO \cite{2013_PRL_Sakai}.

One may conclude that the electron-hole asymmetry of the phase diagram of cuprates is a piece of pseudogap puzzle that should be addressed by any consistent model.

\section{Pseudogap in F\lowercase{e}-SC}
\label{Fe-SC}

In the iron based superconductors, the pseudogap is hardly seen by ARPES \cite{2012_LTP_Kordyuk}. It has been reported in several early studies on polycrystalline samples \cite{SatoJPSJ2008, HaiYun2008, 2008_JPSJ_Ishida} and later on Ba$_{1-x}$K$_{x}$Fe$_2$As$_2$ (BKFA) single crystals \cite{XuNC2011}, but that observations are not supported by a majority of ARPES \cite{EvtushinskyPRB2009, RichardPRL2009, Evtushinsky2011, ShimojimaSci2011, BorisenkoSym2012, ZhangNP2012, EvtushinskyPRB2014} and STM \cite{YinPhC2009, MasseeEPL2010} experiments.

It is surprising because from a nearly perfect Fermi surface nesting one would expect the pseudogap due to incommensurate ordering like in transition metal dichalcogenides and cuprates. The absence of the pseudogap in ARPES spectra may be just a consequence of low spectral weight modulation by the magnetic ordering that may question its importance for superconductivity, discussed in previous section. Also, the band gap due to anti-ferromagnetic order, even commensurate, is small and partial, it opens the gap on Fermi surface parts but not even along each direction \cite{2011_AdP_Andersen}.

Meanwhile, a growing evidence for pseudogap comes from other experiments \cite{2012_LTP_Kordyuk}.
NMR on some of 1111 compounds and Ba(Fe$_{1-x}$Co$_{x}$)$_2$As$_2$ (BFCA) \cite{IshidaJPSJ2009} and nuclear spin-lattice relaxation rate on Ca(Fe$_{1-x}$Co$_x$)$_2$As$_2$ \cite{2011_PRB_Baek} reveal a pseudogap-like gradual decrease of $(T1T)^{-1}$ below some temperature above $T_c$ as function of doping, similarly to the spin-gap behavior in cuprates.

The interplane resistivity data for BFCA over a broad doping range also shows a clear correlation with the NMR Knight shift, assigned to the formation of the pseudogap \cite{TanatarPRB2010}. In SmFeAsO$_{1-x}$, the pseudogap was determined from resistivity measurements \cite{SolovjevLTP2009, SolovjovLTP2011}.
The evidence for the superconducting pairs in the normal state (up to temperature $T \approx 1.3 T_c$) has been obtained using point-contact spectroscopy on BFCA film \cite{2010_PRL_Sheet}.

The optical spectroscopies reveal the presence of the low- and high-energy pseudogaps in the Ba122 \cite{2014_PRB_Moon} and FeSe \cite{2012_PRL_Wen}. The former shares striking similarities with the infrared pseudogap in YBCO while the later is similar to features in an electron-doped NCCO. Recently a pseudogap like feature has been observed in LiFeAs above $T_c$ up to 40K by ultrafast optical spectroscopy \cite{2014_PRB_Lin}.

In magnetic torque measurements of the isovalent-doping system BaFe$_2$(As$_{1-x}$P$_x$)$_2$ (BFAP), electronic nematicity has been observed above the structural and superconducting transitions \cite{2012_N_Kasahara}. It has been supported by recent ARPES study of the same compound \cite{2014_PRB_Shimojima} in which a composition-dependent pseudogap formation has been reported. The pseudogap develops a dome on the phase diagram very similar to cuprates and is accompanied by inequivalent energy shifts in the Fe $zx/yz$ orbitals, which are thus responsible for breaking the fourfold rotational symmetry.

The pseudogap related to the fourfold symmetry breaking and electronic nematic fluctuations has been observed by a time-resolved optical study for electron doped BFCA \cite{2012_PRB_Stojchevska} and near optimally doped Sm(Fe,Co)AsO \cite{2013_PRB_Mertelj}. The observed anisotropy persists into the superconducting state, that indicates that the superconductivity is coexisting with nematicity and the pseudogap in these compounds.

Very recently, the pseudogap-like behavior has been found in the novel iron-based superconductor with a triclinic crystal structure (CaFe$_{1-x}$Pt$_x$As)$_{10}$Pt$_3$As$_8$ ($T_c$ = 13 K), containing platinum-arsenide intermediary layers, studied by $\mu$SR, INS, and NMR \cite{2014_xxx_Surmach}. Authors have found two superconducting gaps like in other Fe-SCs, but smaller, about 2 meV and 0.3 meV, and also an unusual peak in the spin-excitation spectrum around 7 meV, which disappears only above $T$* = 45 K. A suppression of the spin-lattice relaxation rate observed by NMR immediately below this temperature indicates that $T$* could mark the onset of a pseudogap, which is likely associated with the emergence of preformed Cooper pairs.

To conclude, there is much less consensus about the pseudogap in the iron based superconductors than in cuprates. The fact that in contrast to cuprates the pseudogap in Fe-SC is not easily seen by ARPES says for its more sophisticated appearance in multi-band superconductors. Thus, at the moment, unlike the CDW bearing dichalcogenides, the ferro-pnictides and ferro-chalcogenides can hardly provide deeper incite into pseudogap origins. On the other hand, due to their multi-band electronic structure, studying these materials may shed some light on the interplay of the pseudogap and superconductivity.

\section{Pseudogap and superconductivity}
\label{PG-SC}

\begin{figure}
\begin{center}
\includegraphics[width=0.48\textwidth]{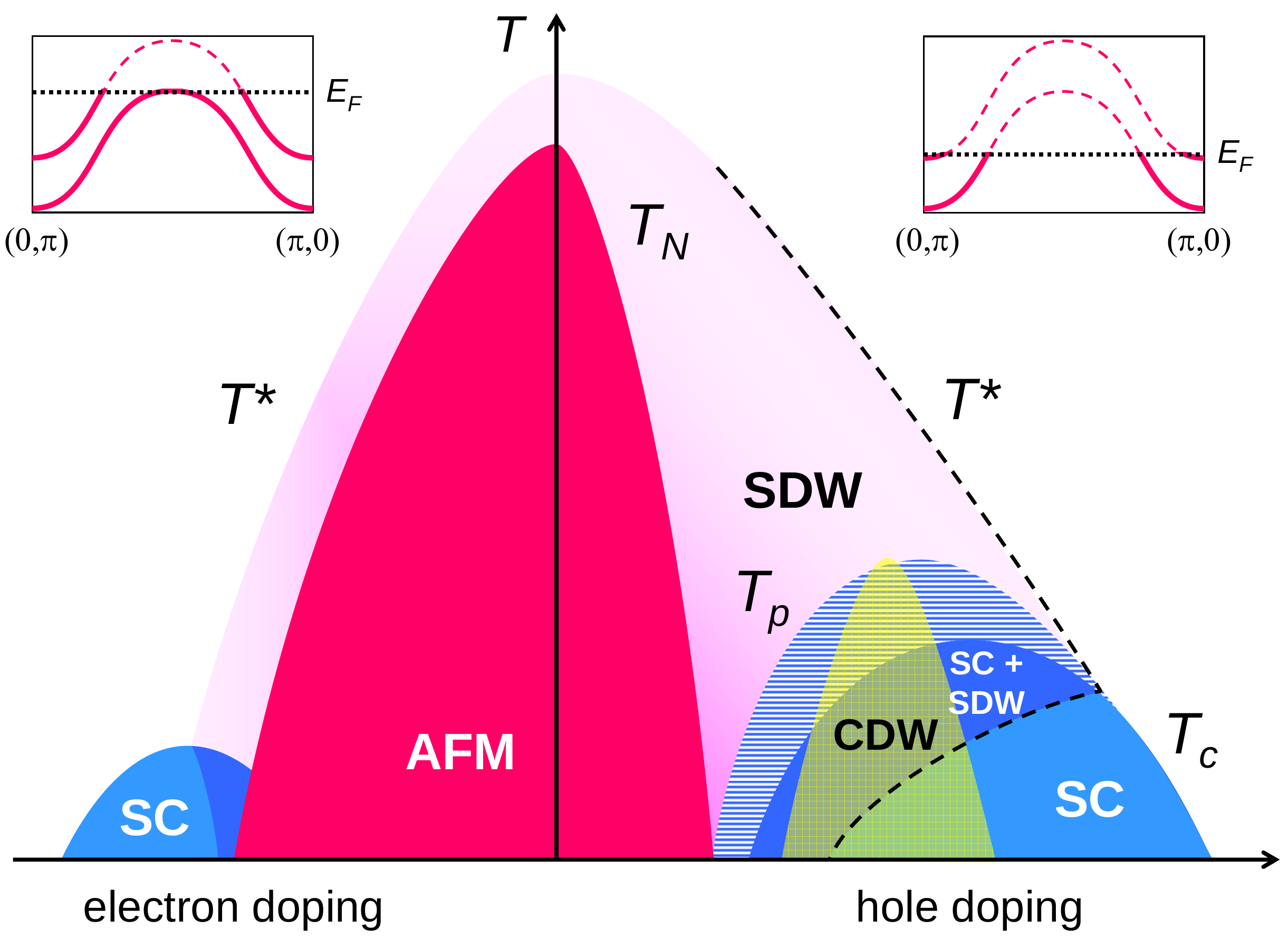}
\caption{Compiled phase diagram of HTSC cuprates. Insets show a sketch of the AFM split conducting band along the magnetic zone boundary illustrating the idea of `topological superconductivity'.
\label{HTSC_PhD}}
\end{center}
\end{figure}

Density wave (SDW or CDW) in cuprates, like CDW in TMD, competes with superconductivity for the phase space and is generally expected to suppress $T_c$. Though the interaction of two orders can be more complex \cite{FradkinNP2012, 2014_PRB_Wang}. At this point, I would like to recall the idea of CDW induced superconductivity \cite{KopaevZETF1970, KopaevPLA1987, 1988_PC_Machida}, in which the superconducting transition temperature can be increased when one of CDW induced peaks in the density of states (due to new VHs) is shifted to the Fermi level. This idea was criticized since it looks unlikely that a self-consistent solution of both orders caused by the same mechanism (competing for the same electronic states) could lead to such situation. On the other hand, if the density wave has different origin, one can imagine the situations when such an enhancement would be possible.

For example, if spin and charge degrees of freedom are decoupled \cite{2014_NP_Ebrahimnejad}, the AFM ordering can enhance the electronic density of states at certain momenta. The VHs nesting scenario in cuprates \cite{1997_JPCS_Markiewicz} is different by origin but should have the same consequences. The situation when the upper split band at $(\pi,0)$ is just touching the Fermi level, as shown in the right inset in Fig.\;\ref{HTSC_PhD}, should be favorable for both $(\pi,\pi)$ density wave and superconductivity. The pessimistic view on such a scenario says that such an increase of DOS in 2D system would not enough to explain HTSC, especially taking into account finite scattering rate \cite{1997_JPCS_Markiewicz, GabovichPR2002}.

The new experience with the iron based superconductors may help to understand the superconducting mechanism in both Fe-SC and Cu-SC. It has been found \cite{2012_LTP_Kordyuk, 2013_JSNM_Kordyuk} that the Fermi surface of every optimally doped Fe-SC compound (the compounds with highest $T_c$) has the Van Hove singularities of the Fe $3d_{xz/yz}$ bands in the vicinity to the Fermi level. The ARPES data for new Fe-SC compounds received thereafter, such as Ca$_{1-x}$Na$_{x}$Fe$_{2}$As$_{2}$ \cite{2013_PRB_Evtushinsky}, Rb-Fe-Se ('245' family) \cite{2013_PRB_Maletz}, and Ca-Pt-Fe-As \cite{2013_PRB_Thirupathaiah} completely support this observation. This suggests that the proximity to an electronic topological transition, known as Lifshitz transition, for one of the multiple Fermi surfaces makes the superconductivity dome at the phase diagram of Fe-SCs  \cite{2012_LTP_Kordyuk}. It seems that new Bi-dichalcogenide layered superconductors follow the same empirical rule: $\mathrm{La}{\mathrm{O}}_{0.54}{\mathrm{F}}_{0.46}\mathrm{Bi}{\mathrm{S}}_{2}$ at optimal doping has the Fermi surface in close proximity to the topological change \cite{2014_PRB_Terashima}. The high-$T_c$ superconductivity driven by `shape-resonance pairing' in a multiband system in the proximity of a Lifshitz topological transition \cite{InnocentiPRB2010, InnocentiSUST2011, 2013_NP_Bianconi} is one of possible models to explain the observed correlation.

With the discussed $(\pi,\pi)$ density wave taken into account, the high-$T_c$ cuprates may share the same `topological' mechanism. If the superconducting dome at the hole side is made by shallow electron pockets around $(\pi,0)$, the dome at the electron side is made by the hole pockets around $(\pi/2,\pi/2)$, as shown in the left inset in Fig.\;\ref{HTSC_PhD}. The role of Lifshitz transition can be twofold here: shaping the spectrum of magnetic fluctuations \cite{2006_JPSJ_Yamaji} and formation of critically slow quasiparticles \cite{2011_NP_Yelland}. Earlier, an enhancement of superconductivity due to proximity to Lifshitz transition has been discussed in connection to the $(\pi,0)$ saddle point \cite{1999_PRL_Onufrieva} (see also \cite{2002_PRB_Angilella} and references therein), but the main objection against the relevance of this scenario for the cuprates was that for optimal doping the saddle point is essentially below the Fermi level.

\section{Conclusions}

The present review represents a contribution dealing with the pseudogap, focusing on ARPES results. Based on the available data, it is tempting to conclude that the pseudogap in cuprates is a complex phenomenon which includes different combinations of density waves (CDW with Fermi surface nesting vector and SDW with AFM vector) and preformed pairs in different parts of the phase diagram. Although the density waves are generally competing to superconductivity, the $(\pi,\pi)$ SDW, the main constituent of the pseudogap phase, may be responsible for a `topological' mechanism of superconducting pairing, that may be similar for high-$T_c$ cuprates, iron base superconductors, and even superconducting transition metal dichalcogenides.

\begin{acknowledgements}
I acknowledge discussions with L. Alff, A. Bianconi, S. V. Borisenko, B. B\"{u}chner, A. V. Chubukov, T. Dahm, I. Eremin, D. V. Evtushinsky, J. Fink, A. M. Gabovich, M. S. Golden, M. Grilli, D. S. Inosov, A. L. Kasatkin, T. K. Kim, Yu. V. Kopaev, M. M. Korshunov, S. A. Kuzmichev, V. M. Loktev, I. A. Nekrasov, S. G. Ovchinnikov, N. M. Plakida, M. V. Sadovskii, A. V. Semenov, S. G. Sharapov, D. J. Scalapino, A. L. Solovjov, M. A. Tanatar, T. Valla, C. M. Varma, A. N. Yaresko, and V. B. Zabolotnyy. The work was supported by the National Academy of Sciences of Ukraine (project 73-02-14) and the State Fund for Fundamental Research (project F50/052).
\end{acknowledgements}

\bibliographystyle{PRL}

\bibliography{FNT_PG_xxx}

\begin{thebibliography}{100}

\bibitem{MottRMP1968}
{N.~F. Mott, \href{http://link.aps.org/doi/10.1103/RevModPhys.40.677}{
\newblock Rev. Mod. Phys. {\bf 40}, 677 (1968)}.}

\bibitem{MottPM1969}
{N.~F. Mott, \href{http://dx.doi.org/10.1080/14786436908216338}{
\newblock Philos. Mag. {\bf 19}, 835 (1969)}.}

\bibitem{1930_AdP_Peierls}
{R.~Peierls, \href{http://dx.doi.org/10.1002/andp.19303960202}{
\newblock Ann. Phys. (Leipzig) {\bf 396}, 121 (1930)}.}

\bibitem{1973_SSC_Rice}
{M.~J. Rice and S.~Str\"{a}ssler,
  \href{http://www.sciencedirect.com/science/article/pii/0038109873901737}{
\newblock Solid State Commun. {\bf 13}, 1389  (1973)}.}

\bibitem{1973_PRL_Lee}
{P.~A. Lee, T.~M. Rice, and P.~W. Anderson,
  \href{http://link.aps.org/doi/10.1103/PhysRevLett.31.462}{
\newblock Phys. Rev. Lett. {\bf 31}, 462 (1973)}.}

\bibitem{1974_SP_Sadovskii}
{M.~Sadovskii,
\newblock Sov. Phys. - Solid State {\bf 16}, 1632 (1974).}

\bibitem{1978_PR_Toombs}
{G.~Toombs,
  \href{http://www.sciencedirect.com/science/article/pii/0370157378901497}{
\newblock Phys. Rep. {\bf 40}, 181  (1978)}.}

\bibitem{Sadovskii1974}
{M.~V. Sadovskii, \href{http://www.jetp.ac.ru/cgi-bin/dn/e_039_05_0845.pdf}{
\newblock Sov. Phys. JETP {\bf 39}, 845 (1974)}.}

\bibitem{1975_AiP_Wilson}
{J.~Wilson, F.~Di~Salvo, and S.~Mahajan,
  \href{http://dx.doi.org/10.1080/00018737500101391}{
\newblock Adv. Phys. {\bf 24}, 117 (1975)}.}

\bibitem{2011_JPCM_Rossnagel}
{K.~Rossnagel, \href{http://stacks.iop.org/0953-8984/23/i=21/a=213001}{
\newblock J. Phys.: Cond. Matter {\bf 23}, 213001 (2011)}.}

\bibitem{BorisenkoPRL2008}
{S.~V. Borisenko \textit{et~al}.,
  \href{http://www.imp.kiev.ua/~kord/papers/box/2008_PRL_Borisenko.pdf}{
\newblock Phys. Rev. Lett. {\bf 100}, 196402 (2008)}.}

\bibitem{2009_PRL_Borisenko}
{S.~V. Borisenko \textit{et~al}.,
  \href{http://www.imp.kiev.ua/~kord/papers/box/2009_PRL_Borisenko.pdf}{
\newblock Phys. Rev. Lett. {\bf 102}, 166402 (2009)}.}

\bibitem{2012_JETP_Kuchinskii}
{E.~Z. Kuchinskii, I.~A. Nekrasov, and M.~V. Sadovskii,
  \href{http://dx.doi.org/10.1134/S1063776112020252}{
\newblock JETP {\bf 114}, 671 (2012)}.}

\bibitem{1977_JPSJ_Nakanishi}
{K.~Nakanishi and H.~Shiba, \href{http://dx.doi.org/10.1143/JPSJ.43.1839}{
\newblock J. Phys. Soc. Japan {\bf 43}, 1839 (1977)}.}

\bibitem{2014_PRB_Arguello}
{C.~J. Arguello \textit{et~al}.,
  \href{http://link.aps.org/doi/10.1103/PhysRevB.89.235115}{
\newblock Phys. Rev. B {\bf 89}, 235115 (2014)}.}

\bibitem{1999_PRL_Pillo}
{T.~Pillo \textit{et~al}.,
  \href{http://link.aps.org/doi/10.1103/PhysRevLett.83.3494}{
\newblock Phys. Rev. Lett. {\bf 83}, 3494 (1999)}.}

\bibitem{1992_PRB_Dardel}
{B.~Dardel \textit{et~al}.,
  \href{http://link.aps.org/doi/10.1103/PhysRevB.45.1462}{
\newblock Phys. Rev. B {\bf 45}, 1462 (1992)}.}

\bibitem{2003_PRL_Perfetti}
{L.~Perfetti \textit{et~al}.,
  \href{http://link.aps.org/doi/10.1103/PhysRevLett.90.166401}{
\newblock Phys. Rev. Lett. {\bf 90}, 166401 (2003)}.}

\bibitem{Gruner}
{G.~Gruner,
\newblock {\em Density Waves In Solids}
\newblock Frontiers in Physics, Westview Press, 2009.}

\bibitem{TimuskRPP1999}
{T.~Timusk and B.~Statt, \href{http://stacks.iop.org/0034-4885/62/i=1/a=002}{
\newblock Rep. Prog. Phys. {\bf 62}, 61 (1999)}.}

\bibitem{LoktevPR2001}
{V.~M. Loktev, R.~M. Quick, and S.~G. Sharapov,
  \href{http://www.sciencedirect.com/science/article/pii/S0370157300001149}{
\newblock Phys. Rep. {\bf 349}, 1  (2001)}.}

\bibitem{2001_P_Sadovskii}
{M.~V. Sadovskii, \href{http://stacks.iop.org/1063-7869/44/i=5/a=R03}{
\newblock Physics-Uspekhi {\bf 44}, 515 (2001)}.}

\bibitem{GabovichSST2001}
{A.~M. Gabovich \textit{et~al}.,
  \href{http://stacks.iop.org/0953-2048/14/i=4/a=201}{
\newblock Supercond. Sci. Technol. {\bf 14}, R1 (2001)}.}

\bibitem{GabovichPR2002}
{A.~M. Gabovich, A.~I. Voitenko, and M.~Ausloos,
  \href{http://www.sciencedirect.com/science/article/pii/S0370157302000297}{
\newblock Phys. Rep. {\bf 367}, 583  (2002)}.}

\bibitem{2004_xxx_Sadovskii}
{M.~V. {Sadovskii}, \href{http://arxiv.org/abs/cond-mat/0408489}{
\newblock arXiv:cond-mat/0408489  (2004)}.}

\bibitem{NormanAP2005}
{M.~R. Norman, D.~Pines, and C.~Kallin,
  \href{http://www.tandfonline.com/doi/abs/10.1080/00018730500459906}{
\newblock Adv. Phys. {\bf 54}, 715 (2005)}.}

\bibitem{Huefner2008}
{S.~Huefner \textit{et~al}.,
  \href{http://stacks.iop.org/0034-4885/71/i=6/a=062501}{
\newblock Rep. Prog. Phys. {\bf 71}, 062501 (2008)}.}

\bibitem{plakida2010high}
{N.~Plakida,
\newblock {\em High-Temperature Cuprate Superconductors: Experiment, Theory,
  and Applications}
\newblock Springer Series in Solid-State Sciences, Springer, 2010.}

\bibitem{1996_PT_Levi}
{B.~G. Levi, \href{http://dx.doi.org/10.1063/1.2807647}{
\newblock Physics Today {\bf 49}, 17 (1996)}.}

\bibitem{2014_LTP_Ekino}
{T.~Ekino \textit{et~al}.,
  \href{http://scitation.aip.org/content/aip/journal/ltp/40/10/10.1063/1.48974%
15}{
\newblock Low Temp. Phys. {\bf 40}, 925 (2014)}.}

\bibitem{1998_PRL_Dessau}
{D.~S. Dessau \textit{et~al}.,
  \href{http://link.aps.org/doi/10.1103/PhysRevLett.81.192}{
\newblock Phys. Rev. Lett. {\bf 81}, 192 (1998)}.}

\bibitem{2005_N_Mannella}
{N.~Mannella \textit{et~al}., \href{http://dx.doi.org/10.1038/nature04273}{
\newblock Nature {\bf 438}, 474 (2005)}.}

\bibitem{1996_PRL_Susaki}
{T.~Susaki \textit{et~al}.,
  \href{http://link.aps.org/doi/10.1103/PhysRevLett.77.4269}{
\newblock Phys. Rev. Lett. {\bf 77}, 4269 (1996)}.}

\bibitem{2014_xxx_Okawa}
{M.~{Okawa} \textit{et~al}., \href{http://arxiv.org/abs/1407.0578}{
\newblock arXiv:1407.0578  (2014)}.}

\bibitem{2010_NC_Sacepe}
{B.~Sac\'{e}p\'{e} \textit{et~al}.,
  \href{http://dx.doi.org/10.1038/ncomms1140}{
\newblock Nat. Commun. {\bf 1}, 140 (2010)}.}

\bibitem{2009_PRL_Wang}
{K.~Wang \textit{et~al}.,
  \href{http://link.aps.org/doi/10.1103/PhysRevLett.102.076801}{
\newblock Phys. Rev. Lett. {\bf 102}, 076801 (2009)}.}

\bibitem{2013_SST_Liu}
{J.~Liu \textit{et~al}.,
  \href{http://stacks.iop.org/0953-2048/26/i=8/a=085009}{
\newblock Supercond. Sci. Technol. {\bf 26}, 085009 (2013)}.}

\bibitem{2012_PRX_Mann}
{A.~Mann \textit{et~al}.,
  \href{http://link.aps.org/doi/10.1103/PhysRevX.2.041008}{
\newblock Phys. Rev. X {\bf 2}, 041008 (2012)}.}

\bibitem{2011_PRL_Magierski}
{P.~Magierski, G.~Wlaz\l{}owski, and A.~Bulgac,
  \href{http://link.aps.org/doi/10.1103/PhysRevLett.107.145304}{
\newblock Phys. Rev. Lett. {\bf 107}, 145304 (2011)}.}

\bibitem{2008_PRB_Wagner}
{K.~E. Wagner \textit{et~al}.,
  \href{http://link.aps.org/doi/10.1103/PhysRevB.78.104520}{
\newblock Phys. Rev. B {\bf 78}, 104520 (2008)}.}

\bibitem{2014_PRB_Croft}
{T.~P. Croft \textit{et~al}.,
  \href{http://link.aps.org/doi/10.1103/PhysRevB.89.224513}{
\newblock Phys. Rev. B {\bf 89}, 224513 (2014)}.}

\bibitem{2013_NJP_Wang}
{A.~F. Wang \textit{et~al}.,
  \href{http://stacks.iop.org/1367-2630/15/i=4/a=043048}{
\newblock New J. Phys. {\bf 15}, 043048 (2013)}.}

\bibitem{lynch1999}
{D.~W. Lynch and C.~G. Olson,
\newblock {\em Photoemission Studies of High-Temperature Superconductors}
\newblock Cambridge Studies in Low Temperature Physics, Cambridge University
  Press, 1999.}

\bibitem{DamascelliRMP2003}
{A.~Damascelli, Z.~Hussain, and Z.-X. Shen,
  \href{http://link.aps.org/doi/10.1103/RevModPhys.75.473}{
\newblock Rev. Mod. Phys. {\bf 75}, 473 (2003)}.}

\bibitem{2014_LTP_Kordyuk}
{A.~A. Kordyuk,
  \href{http://www.imp.kiev.ua/~kord/papers/box/2014_LTP_Kordyuk.pdf}{
\newblock Low Temp. Phys. {\bf 40}, 286 (2014)}.}

\bibitem{2012_LTP_Kordyuk}
{A.~A. Kordyuk,
  \href{http://www.imp.kiev.ua/~kord/papers/box/2012_LTP_Kordyuk.pdf}{
\newblock Low Temp. Phys. {\bf 38}, 888 (2012)}.}

\bibitem{2006_RMP_Lee}
{P.~Lee, N.~Nagaosa, and X.-G. Wen,
  \href{http://link.aps.org/doi/10.1103/RevModPhys.78.17}{
\newblock Rev. Mod. Phys. {\bf 78}, 17 (2006)}.}

\bibitem{2006_LTP_Tremblay}
{A.-M.~S. Tremblay, B.~Kyung, and D.~S\'en\'echal,
  \href{http://scitation.aip.org/content/aip/journal/ltp/32/4/10.1063/1.219944%
6}{
\newblock Low Temp. Phys. {\bf 32}, 424 (2006)}.}

\bibitem{2012_RPP_Rice}
{T.~M. Rice, K.-Y. Yang, and F.~C. Zhang,
  \href{http://stacks.iop.org/0034-4885/75/i=1/a=016502}{
\newblock Rep. Prog. Phys. {\bf 75}, 016502 (2012)}.}

\bibitem{2014_PRB_Wang}
{Y.~Wang and A.~Chubukov,
  \href{http://link.aps.org/doi/10.1103/PhysRevB.90.035149}{
\newblock Phys. Rev. B {\bf 90}, 035149 (2014)}.}

\bibitem{manske2004theory}
{D.~Manske,
\newblock {\em Theory of Unconventional Superconductors: Cooper-Pairing
  Mediated by Spin Excitations}
\newblock Number no. 202 in Physics and Astronomy Online Library, Springer,
  2004.}

\bibitem{2005_Larkin}
{A.~Larkin and A.~Varlamov,
\newblock {\em Theory of Fluctuations in Superconductors}
\newblock Oxford: Oxford University Press, 2005.}

\bibitem{1968_PLA_Aslamasov}
{L.~G. Aslamasov and A.~I. Larkin,
  \href{http://www.sciencedirect.com/science/article/pii/0375960168906233}{
\newblock Phys. Lett. A {\bf 26}, 238  (1968)}.}

\bibitem{1994_RPP_Alexandrov}
{A.~S. Alexandrov and N.~F. Mott,
  \href{http://stacks.iop.org/0034-4885/57/i=12/a=001}{
\newblock Rep. Prog. Phys. {\bf 57}, 1197 (1994)}.}

\bibitem{FriedelPC1988}
{J.~Friedel,
  \href{http://www.sciencedirect.com/science/article/pii/0921453488904315}{
\newblock Physica C {\bf 153-155, Part 3}, 1610  (1988)}.}

\bibitem{LoktevLTP2000}
{V.~M. Loktev \textit{et~al}.,
  \href{http://scitation.aip.org/content/aip/journal/ltp/26/6/10.1063/1.593917%
}{
\newblock Low Temp. Phys. {\bf 26}, 414 (2000)}.}

\bibitem{1994_PC_Chakraverty}
{B.~K. Chakraverty, A.~Taraphder, and M.~Avignon,
  \href{http://www.sciencedirect.com/science/article/pii/0921453494923833}{
\newblock Physica C {\bf 235-240, Part 4}, 2323  (1994)}.}

\bibitem{1995_N_Emery}
{V.~J. Emery and S.~A. Kivelson, \href{http://dx.doi.org/10.1038/374434a0}{
\newblock Nature {\bf 374}, 434 (1995)}.}

\bibitem{1989_PRL_Uemura}
{Y.~J. Uemura \textit{et~al}.,
  \href{http://link.aps.org/doi/10.1103/PhysRevLett.62.2317}{
\newblock Phys. Rev. Lett. {\bf 62}, 2317 (1989)}.}

\bibitem{2006_PRL_Rufenacht}
{A.~R\"ufenacht \textit{et~al}.,
  \href{http://link.aps.org/doi/10.1103/PhysRevLett.96.227002}{
\newblock Phys. Rev. Lett. {\bf 96}, 227002 (2006)}.}

\bibitem{1998_JPCS_Ariosa}
{D.~Ariosa, H.~Beck, and M.~Capezzali,
  \href{http://www.sciencedirect.com/science/article/pii/S0022369798001000}{
\newblock J. Phys. Chem. Solids {\bf 59}, 1783  (1998)}.}

\bibitem{1999_PC_Junod}
{A.~Junod, A.~Erb, and C.~Renner,
  \href{http://www.sciencedirect.com/science/article/pii/S0921453499000775}{
\newblock Physica C {\bf 317-318}, 333  (1999)}.}

\bibitem{2002_S_Hoffman}
{J.~E. Hoffman \textit{et~al}.,
  \href{http://www.sciencemag.org/content/295/5554/466.abstract}{
\newblock Science {\bf 295}, 466 (2002)}.}

\bibitem{2004_S_Vershinin}
{M.~Vershinin \textit{et~al}.,
  \href{http://www.sciencemag.org/content/303/5666/1995.abstract}{
\newblock Science {\bf 303}, 1995 (2004)}.}

\bibitem{2005_PRL_McElroy}
{K.~McElroy \textit{et~al}.,
  \href{http://link.aps.org/doi/10.1103/PhysRevLett.94.197005}{
\newblock Phys. Rev. Lett. {\bf 94}, 197005 (2005)}.}

\bibitem{2014_PRB_Meier}
{H.~Meier \textit{et~al}.,
  \href{http://link.aps.org/doi/10.1103/PhysRevB.89.195115}{
\newblock Phys. Rev. B {\bf 89}, 195115 (2014)}.}

\bibitem{1992_PRB_Nagaosa}
{N.~Nagaosa and P.~A. Lee,
  \href{http://link.aps.org/doi/10.1103/PhysRevB.45.966}{
\newblock Phys. Rev. B {\bf 45}, 966 (1992)}.}

\bibitem{1987_S_Anderson}
{P.~W. Anderson,
  \href{http://www.sciencemag.org/content/235/4793/1196.abstract}{
\newblock Science {\bf 235}, 1196 (1987)}.}

\bibitem{2004_RMP_Demler}
{E.~Demler, W.~Hanke, and S.-C. Zhang,
  \href{http://link.aps.org/doi/10.1103/RevModPhys.76.909}{
\newblock Rev. Mod. Phys. {\bf 76}, 909 (2004)}.}

\bibitem{1988_SST_Zhang}
{F.~C. Zhang \textit{et~al}.,
  \href{http://stacks.iop.org/0953-2048/1/i=1/a=009}{
\newblock Supercond. Sci. Technol. {\bf 1}, 36 (1988)}.}

\bibitem{1976_PRB_Hertz}
{J.~A. Hertz, \href{http://link.aps.org/doi/10.1103/PhysRevB.14.1165}{
\newblock Phys. Rev. B {\bf 14}, 1165 (1976)}.}

\bibitem{1993_PRL_Sokol}
{A.~Sokol and D.~Pines,
  \href{http://link.aps.org/doi/10.1103/PhysRevLett.71.2813}{
\newblock Phys. Rev. Lett. {\bf 71}, 2813 (1993)}.}

\bibitem{1995_PRL_Varma}
{C.~M. Varma, \href{http://link.aps.org/doi/10.1103/PhysRevLett.75.898}{
\newblock Phys. Rev. Lett. {\bf 75}, 898 (1995)}.}

\bibitem{1996_ZPBCM_Castellani}
{C.~Castellani, C.~Di~Castro, and M.~Grilli,
  \href{http://dx.doi.org/10.1007/s002570050347}{
\newblock Z. Phys. B Cond. Mat. {\bf 103}, 137 (1996)}.}

\bibitem{1995_PRL_Castellani}
{C.~Castellani, C.~Di~Castro, and M.~Grilli,
  \href{http://link.aps.org/doi/10.1103/PhysRevLett.75.4650}{
\newblock Phys. Rev. Lett. {\bf 75}, 4650 (1995)}.}

\bibitem{1996_PRB_Perali}
{A.~Perali \textit{et~al}.,
  \href{http://link.aps.org/doi/10.1103/PhysRevB.54.16216}{
\newblock Phys. Rev. B {\bf 54}, 16216 (1996)}.}

\bibitem{2001_PRL_Andergassen}
{S.~Andergassen \textit{et~al}.,
  \href{http://link.aps.org/doi/10.1103/PhysRevLett.87.056401}{
\newblock Phys. Rev. Lett. {\bf 87}, 056401 (2001)}.}

\bibitem{2008_PRL_Ortix}
{C.~Ortix, J.~Lorenzana, and C.~Di~Castro,
  \href{http://link.aps.org/doi/10.1103/PhysRevLett.100.246402}{
\newblock Phys. Rev. Lett. {\bf 100}, 246402 (2008)}.}

\bibitem{Sachdev2000}
{S.~Sachdev, \href{http://www.sciencemag.org/content/288/5465/475.abstract}{
\newblock Science {\bf 288}, 475 (2000)}.}

\bibitem{1995_PRB_Barzykin}
{V.~Barzykin and D.~Pines,
  \href{http://link.aps.org/doi/10.1103/PhysRevB.52.13585}{
\newblock Phys. Rev. B {\bf 52}, 13585 (1995)}.}

\bibitem{1996_ZPBCM_Pines}
{D.~Pines, \href{http://dx.doi.org/10.1007/s002570050346}{
\newblock Z. Phys. B Cond. Matt. {\bf 103}, 129 (1996)}.}

\bibitem{1999_PRB_Schmalian}
{J.~Schmalian, D.~Pines, and B.~Stojkovi\ifmmode~\acute{c}\else \'{c}\fi{},
  \href{http://link.aps.org/doi/10.1103/PhysRevB.60.667}{
\newblock Phys. Rev. B {\bf 60}, 667 (1999)}.}

\bibitem{1999_JETP_Kuchinskii}
{E.~Z. Kuchinskii and M.~V. Sadovskii,
  \href{http://dx.doi.org/10.1134/1.558879}{
\newblock JETP {\bf 88}, 968 (1999)}.}

\bibitem{2003_AiP_Abanov}
{A.~Abanov, A.~V. Chubukov, and J.~Schmalian,
  \href{http://dx.doi.org/10.1080/0001873021000057123}{
\newblock Adv. Phys. {\bf 52}, 119 (2003)}.}

\bibitem{2013_NP_Efetov}
{K.~B. Efetov, H.~Meier, and C.~Pepin,
  \href{http://dx.doi.org/10.1038/nphys2641}{
\newblock Nat. Phys. {\bf 9}, 442 (2013)}.}

\bibitem{2010_PRB_Metlitski}
{M.~Metlitski and S.~Sachdev,
  \href{http://link.aps.org/doi/10.1103/PhysRevB.82.075128}{
\newblock Phys. Rev. B {\bf 82}, 075128 (2010)}.}

\bibitem{2014_PNAS_Fujita}
{K.~Fujita \textit{et~al}.,
  \href{http://www.pnas.org/content/111/30/E3026.abstract}{
\newblock Proc. Natl. Acad. Sci. U.S.A. {\bf 111}, E3026 (2014)}.}

\bibitem{2004_PRL_Senechal}
{D.~S\'en\'echal and A.-M. Tremblay,
  \href{http://link.aps.org/doi/10.1103/PhysRevLett.92.126401}{
\newblock Phys. Rev. Lett. {\bf 92}, 126401 (2004)}.}

\bibitem{2005_PRB_Sadovskii}
{M.~Sadovskii \textit{et~al}.,
  \href{http://link.aps.org/doi/10.1103/PhysRevB.72.155105}{
\newblock Phys. Rev. B {\bf 72}, 155105 (2005)}.}

\bibitem{2008_JL_Kuchinskii}
{E.~Z. Kuchinskii and M.~V. Sadovskii,
  \href{http://dx.doi.org/10.1134/S0021364008150101}{
\newblock JETP Letters {\bf 88}, 192 (2008)}.}

\bibitem{1999_PRL_Varma}
{C.~Varma, \href{http://link.aps.org/doi/10.1103/PhysRevLett.83.3538}{
\newblock Phys. Rev. Lett. {\bf 83}, 3538 (1999)}.}

\bibitem{1990_PRB_Wang}
{Z.~Wang, G.~Kotliar, and X.-F. Wang,
  \href{http://link.aps.org/doi/10.1103/PhysRevB.42.8690}{
\newblock Phys. Rev. B {\bf 42}, 8690 (1990)}.}

\bibitem{2001_PRB_Chakravarty}
{S.~Chakravarty \textit{et~al}.,
  \href{http://link.aps.org/doi/10.1103/PhysRevB.63.094503}{
\newblock Phys. Rev. B {\bf 63}, 094503 (2001)}.}

\bibitem{1989_PRB_Zaanen}
{J.~Zaanen and O.~Gunnarsson,
  \href{http://link.aps.org/doi/10.1103/PhysRevB.40.7391}{
\newblock Phys. Rev. B {\bf 40}, 7391 (1989)}.}

\bibitem{1989_PC_Machida}
{K.~Machida,
  \href{http://www.sciencedirect.com/science/article/pii/092145348990316X}{
\newblock Physica C {\bf 158}, 192 (1989)}.}

\bibitem{1990_JPSJ_Kato}
{M.~Kato \textit{et~al}., \href{http://dx.doi.org/10.1143/JPSJ.59.1047}{
\newblock J. Phys. Soc. Jpn. {\bf 59}, 1047 (1990)}.}

\bibitem{1995_N_Tranquada}
{J.~M. Tranquada \textit{et~al}., \href{http://dx.doi.org/10.1038/375561a0}{
\newblock Nature {\bf 375}, 561 (1995)}.}

\bibitem{1997_JS_Eremin}
{I.~Eremin and M.~Eremin, \href{http://dx.doi.org/10.1007/BF02765738}{
\newblock J. Supercond. {\bf 10}, 459 (1997)}.}

\bibitem{1997_PRB_Eremin}
{I.~Eremin \textit{et~al}.,
  \href{http://link.aps.org/doi/10.1103/PhysRevB.56.11305}{
\newblock Phys. Rev. B {\bf 56}, 11305 (1997)}.}

\bibitem{1997_JPCM_Gabovich}
{A.~M. Gabovich and A.~I. Voitenko,
  \href{http://stacks.iop.org/0953-8984/9/i=19/a=011}{
\newblock J. Phys.: Condens. Matter {\bf 9}, 3901 (1997)}.}

\bibitem{2003_RMP_Kivelson}
{S.~A. Kivelson \textit{et~al}.,
  \href{http://link.aps.org/doi/10.1103/RevModPhys.75.1201}{
\newblock Rev. Mod. Phys. {\bf 75}, 1201 (2003)}.}

\bibitem{1998_N_Kivelson}
{S.~A. Kivelson, E.~Fradkin, and V.~J. Emery,
  \href{http://dx.doi.org/10.1038/31177}{
\newblock Nature {\bf 393}, 550 (1998)}.}

\bibitem{2010_N_Daou}
{R.~Daou \textit{et~al}., \href{http://dx.doi.org/10.1038/nature08716}{
\newblock Nature {\bf 463}, 519 (2010)}.}

\bibitem{2009_AP_Vojta}
{M.~Vojta, \href{http://dx.doi.org/10.1080/00018730903122242}{
\newblock Adv. Phys. {\bf 58}, 699 (2009)}.}

\bibitem{1975_PRL_Rice}
{T.~Rice and G.~Scott,
  \href{http://link.aps.org/doi/10.1103/PhysRevLett.35.120}{
\newblock Phys. Rev. Lett. {\bf 35}, 120 (1975)}.}

\bibitem{1997_PRB_Markiewicz}
{R.~S. Markiewicz, \href{http://link.aps.org/doi/10.1103/PhysRevB.56.9091}{
\newblock Phys. Rev. B {\bf 56}, 9091 (1997)}.}

\bibitem{1997_JPCS_Markiewicz}
{R.~S. Markiewicz,
  \href{http://www.sciencedirect.com/science/article/pii/S0022369797000255}{
\newblock J. Phys. Chem. Solids {\bf 58}, 1179  (1997)}.}

\bibitem{2001_S_Chuang}
{Y.-D. Chuang \textit{et~al}.,
  \href{http://www.sciencemag.org/content/292/5521/1509.abstract}{
\newblock Science {\bf 292}, 1509 (2001)}.}

\bibitem{2010_PRL_Evtushinsky}
{D.~V. Evtushinsky \textit{et~al}.,
  \href{http://www.imp.kiev.ua/~kord/papers/box/2010_PRL_Evtushinsky.pdf}{
\newblock Phys. Rev. Lett. {\bf 105}, 147201 (2010)}.}

\bibitem{1989_ACR_Whangbo}
{M.~H. Whangbo and E.~Canadell, \href{http://dx.doi.org/10.1021/ar00167a001}{
\newblock Acc. Chem. Res. {\bf 22}, 375 (1989)}.}

\bibitem{2012_PRB_Usui}
{H.~Usui, K.~Suzuki, and K.~Kuroki,
  \href{http://link.aps.org/doi/10.1103/PhysRevB.86.220501}{
\newblock Phys. Rev. B {\bf 86}, 220501 (2012)}.}

\bibitem{2013_JL_Shein}
{I.~R. Shein and A.~L. Ivanovskii,
  \href{http://dx.doi.org/10.1134/S0021364012240101}{
\newblock JETP Letters {\bf 96}, 769 (2013)}.}

\bibitem{1976_PRB_Bilbro}
{G.~Bilbro and W.~L. McMillan,
  \href{http://link.aps.org/doi/10.1103/PhysRevB.14.1887}{
\newblock Phys. Rev. B {\bf 14}, 1887 (1976)}.}

\bibitem{2015_JMMM_Regueiro}
{M.~N\'{u}\~{n}ez Regueiro,
  \href{http://www.sciencedirect.com/science/article/pii/S0304885314003655}{
\newblock J. Magn. Magn. Mater. {\bf 376}, 25  (2015)}.}

\bibitem{2000_PC_Kresin}
{V.~Kresin, Y.~Ovchinnikov, and S.~Wolf,
  \href{http://www.sciencedirect.com/science/article/pii/S0921453400004068}{
\newblock Physica C {\bf 341-348, Part 1}, 103  (2000)}.}

\bibitem{2007_PRB_Yamase}
{H.~Yamase and W.~Metzner,
  \href{http://link.aps.org/doi/10.1103/PhysRevB.75.155117}{
\newblock Phys. Rev. B {\bf 75}, 155117 (2007)}.}

\bibitem{2008_JETP_Mitsen}
{K.~V. Mitsen and O.~M. Ivanenko,
  \href{http://dx.doi.org/10.1134/S106377610812008X}{
\newblock JETP {\bf 107}, 984 (2008)}.}

\bibitem{1996_S_Loeser}
{A.~G. Loeser \textit{et~al}.,
  \href{http://www.sciencemag.org/content/273/5273/325.abstract}{
\newblock Science {\bf 273}, 325 (1996)}.}

\bibitem{1996_N_Ding}
{H.~Ding \textit{et~al}., \href{http://dx.doi.org/10.1038/382051a0}{
\newblock Nature {\bf 382}, 51 (1996)}.}

\bibitem{1997_PCS_Tao}
{H.~J. Tao, F.~Lu, and E.~L. Wolf,
  \href{http://www.sciencedirect.com/science/article/pii/S0921453497008629}{
\newblock Physica C {\bf 282-287, Part 3}, 1507  (1997)}.}

\bibitem{1998_PRL_Renner}
{C.~Renner \textit{et~al}.,
  \href{http://link.aps.org/doi/10.1103/PhysRevLett.80.149}{
\newblock Phys. Rev. Lett. {\bf 80}, 149 (1998)}.}

\bibitem{1989_PRL_Warren}
{W.~Warren \textit{et~al}.,
  \href{http://link.aps.org/doi/10.1103/PhysRevLett.62.1193}{
\newblock Phys. Rev. Lett. {\bf 62}, 1193 (1989)}.}

\bibitem{1990_PRB_Walstedt}
{R.~Walstedt \textit{et~al}.,
  \href{http://link.aps.org/doi/10.1103/PhysRevB.41.9574}{
\newblock Phys. Rev. B {\bf 41}, 9574 (1990)}.}

\bibitem{1998_PRB_Ishida}
{K.~Ishida \textit{et~al}.,
  \href{http://link.aps.org/doi/10.1103/PhysRevB.58.R5960}{
\newblock Phys. Rev. B {\bf 58}, R5960 (1998)}.}

\bibitem{2004_JPSJ_Matsuzaki}
{T.~Matsuzaki \textit{et~al}., \href{http://dx.doi.org/10.1143/JPSJ.73.2232}{
\newblock Journal of the Physical Society of Japan {\bf 73}, 2232 (2004)}.}

\bibitem{1994_PC_Loram}
{J.~W. Loram \textit{et~al}.,
  \href{http://www.sciencedirect.com/science/article/pii/0921453494920893}{
\newblock Physica C {\bf 235-240, Part 3}, 1735  (1994)}.}

\bibitem{1999_pssb_Tallon}
{J.~L. Tallon \textit{et~al}.,
  \href{http://dx.doi.org/10.1002/(SICI)1521-3951(199909)215:1<531::AID-PSSB53%
1>3.0.CO;2-W}{
\newblock Phys. Status Solidi (b) {\bf 215}, 531 (1999)}.}

\bibitem{1999_PRL_Tallon}
{J.~Tallon and G.~Williams,
  \href{http://link.aps.org/doi/10.1103/PhysRevLett.82.3725}{
\newblock Phys. Rev. Lett. {\bf 82}, 3725 (1999)}.}

\bibitem{2002_PRL_Markiewicz}
{R.~Markiewicz, \href{http://link.aps.org/doi/10.1103/PhysRevLett.89.229703}{
\newblock Phys. Rev. Lett. {\bf 89}, 229703 (2002)}.}

\bibitem{1989_S_Wu}
{X.~L. Wu and C.~M. Lieber,
  \href{http://www.sciencemag.org/content/243/4899/1703.abstract}{
\newblock Science {\bf 243}, 1703 (1989)}.}

\bibitem{2000_PRB_Pillo}
{T.~Pillo \textit{et~al}.,
  \href{http://link.aps.org/doi/10.1103/PhysRevB.62.4277}{
\newblock Phys. Rev. B {\bf 62}, 4277 (2000)}.}

\bibitem{2004_PRL_Ando}
{Y.~Ando \textit{et~al}.,
  \href{http://link.aps.org/doi/10.1103/PhysRevLett.93.267001}{
\newblock Phys. Rev. Lett. {\bf 93}, 267001 (2004)}.}

\bibitem{1987_PRL_Gurvitch}
{M.~Gurvitch and A.~Fiory,
  \href{http://link.aps.org/doi/10.1103/PhysRevLett.59.1337}{
\newblock Phys. Rev. Lett. {\bf 59}, 1337 (1987)}.}

\bibitem{1996_PRL_Wen}
{X.-G. Wen and P.~Lee,
  \href{http://link.aps.org/doi/10.1103/PhysRevLett.76.503}{
\newblock Phys. Rev. Lett. {\bf 76}, 503 (1996)}.}

\bibitem{2003_RMP_Gunnarsson}
{O.~Gunnarsson, M.~Calandra, and J.~Han,
  \href{http://link.aps.org/doi/10.1103/RevModPhys.75.1085}{
\newblock Rev. Mod. Phys. {\bf 75}, 1085 (2003)}.}

\bibitem{2002_EPL_Lavrov}
{{Lavrov, A. N.}, {Ando, Y.}, and {Ono, S.},
  \href{http://dx.doi.org/10.1209/epl/i2002-00571-0}{
\newblock Europhys. Lett. {\bf 57}, 267 (2002)}.}

\bibitem{2005_PRB_Naqib}
{S.~Naqib \textit{et~al}.,
  \href{http://link.aps.org/doi/10.1103/PhysRevB.71.054502}{
\newblock Phys. Rev. B {\bf 71}, 054502 (2005)}.}

\bibitem{2006_PRB_Wang}
{Y.~Wang, L.~Li, and N.~P. Ong,
  \href{http://link.aps.org/doi/10.1103/PhysRevB.73.024510}{
\newblock Phys. Rev. B {\bf 73}, 024510 (2006)}.}

\bibitem{2001_PRB_Wang}
{Y.~Wang \textit{et~al}.,
  \href{http://link.aps.org/doi/10.1103/PhysRevB.64.224519}{
\newblock Phys. Rev. B {\bf 64}, 224519 (2001)}.}

\bibitem{2013_LTP_Dyachenko}
{A.~I. Dyachenko \textit{et~al}.,
  \href{http://scitation.aip.org/content/aip/journal/ltp/39/4/10.1063/1.480198%
9}{
\newblock Low Temp. Phys. {\bf 39}, 323 (2013)}.}

\bibitem{1996_JPCM_Puchkov}
{A.~V. Puchkov, D.~N. Basov, and T.~Timusk,
  \href{http://stacks.iop.org/0953-8984/8/i=48/a=023}{
\newblock J. Phys.: Cond. Matter {\bf 8}, 10049 (1996)}.}

\bibitem{2005_RMP_Basov}
{D.~Basov and T.~Timusk,
  \href{http://link.aps.org/doi/10.1103/RevModPhys.77.721}{
\newblock Rev. Mod. Phys. {\bf 77}, 721 (2005)}.}

\bibitem{2004_S_Boris}
{A.~V. Boris \textit{et~al}.,
  \href{http://www.sciencemag.org/content/304/5671/708.abstract}{
\newblock Science {\bf 304}, 708 (2004)}.}

\bibitem{1990_PRB_Orenstein}
{J.~Orenstein \textit{et~al}.,
  \href{http://link.aps.org/doi/10.1103/PhysRevB.42.6342}{
\newblock Phys. Rev. B {\bf 42}, 6342 (1990)}.}

\bibitem{1991_PRL_Rotter}
{L.~Rotter \textit{et~al}.,
  \href{http://link.aps.org/doi/10.1103/PhysRevLett.67.2741}{
\newblock Phys. Rev. Lett. {\bf 67}, 2741 (1991)}.}

\bibitem{1991_PRB_Uchida}
{S.~Uchida \textit{et~al}.,
  \href{http://link.aps.org/doi/10.1103/PhysRevB.43.7942}{
\newblock Phys. Rev. B {\bf 43}, 7942 (1991)}.}

\bibitem{2004_PRB_Onose}
{Y.~Onose \textit{et~al}.,
  \href{http://link.aps.org/doi/10.1103/PhysRevB.69.024504}{
\newblock Phys. Rev. B {\bf 69}, 024504 (2004)}.}

\bibitem{2010_PRB_Das}
{T.~Das, R.~S. Markiewicz, and A.~Bansil,
  \href{http://link.aps.org/doi/10.1103/PhysRevB.81.174504}{
\newblock Phys. Rev. B {\bf 81}, 174504 (2010)}.}

\bibitem{2004_N_Hwang}
{J.~Hwang, T.~Timusk, and G.~D. Gu,
  \href{http://dx.doi.org/10.1038/nature02347}{
\newblock Nature {\bf 427}, 714 (2004)}.}

\bibitem{2005_PRL_Pimenov}
{A.~Pimenov \textit{et~al}.,
  \href{http://link.aps.org/doi/10.1103/PhysRevLett.94.227003}{
\newblock Phys. Rev. Lett. {\bf 94}, 227003 (2005)}.}

\bibitem{2008_PRL_Yu}
{L.~Yu \textit{et~al}.,
  \href{http://link.aps.org/doi/10.1103/PhysRevLett.100.177004}{
\newblock Phys. Rev. Lett. {\bf 100}, 177004 (2008)}.}

\bibitem{2006_NP_LeTacon}
{M.~Le~Tacon \textit{et~al}., \href{http://dx.doi.org/10.1038/nphys362}{
\newblock Nat. Phys. {\bf 2}, 537 (2006)}.}

\bibitem{2002_JPCS_Venturini}
{F.~Venturini \textit{et~al}.,
  \href{http://www.sciencedirect.com/science/article/pii/S0022369702002391}{
\newblock J. Phys. Chem. Solids {\bf 63}, 2345  (2002)}.}

\bibitem{2003_PRB_Sugai}
{S.~Sugai \textit{et~al}.,
  \href{http://link.aps.org/doi/10.1103/PhysRevB.68.184504}{
\newblock Phys. Rev. B {\bf 68}, 184504 (2003)}.}

\bibitem{2002_PRB_Borisenko}
{S.~V. Borisenko \textit{et~al}.,
  \href{http://www.imp.kiev.ua/~kord/papers/box/2002_PRB_Borisenko.pdf}{
\newblock Phys. Rev. B {\bf 66}, 140509 (2002)}.}

\bibitem{1998_N_Norman}
{M.~R. Norman \textit{et~al}., \href{http://dx.doi.org/10.1038/32366}{
\newblock Nature {\bf 392}, 157 (1998)}.}

\bibitem{1999_PRL_Campuzano}
{J.~Campuzano \textit{et~al}.,
  \href{http://link.aps.org/doi/10.1103/PhysRevLett.83.3709}{
\newblock Phys. Rev. Lett. {\bf 83}, 3709 (1999)}.}

\bibitem{1998_PRL_Miyakawa}
{N.~Miyakawa \textit{et~al}.,
  \href{http://link.aps.org/doi/10.1103/PhysRevLett.80.157}{
\newblock Phys. Rev. Lett. {\bf 80}, 157 (1998)}.}

\bibitem{1998_PRL_DeWilde}
{Y.~DeWilde \textit{et~al}.,
  \href{http://link.aps.org/doi/10.1103/PhysRevLett.80.153}{
\newblock Phys. Rev. Lett. {\bf 80}, 153 (1998)}.}

\bibitem{2007_RMP_Devereaux}
{T.~Devereaux and R.~Hackl,
  \href{http://link.aps.org/doi/10.1103/RevModPhys.79.175}{
\newblock Rev. Mod. Phys. {\bf 79}, 175 (2007)}.}

\bibitem{2001_RMP_Kotani}
{A.~Kotani and S.~Shin,
  \href{http://link.aps.org/doi/10.1103/RevModPhys.73.203}{
\newblock Rev. Mod. Phys. {\bf 73}, 203 (2001)}.}

\bibitem{2011_RMP_Ament}
{L.~Ament \textit{et~al}.,
  \href{http://link.aps.org/doi/10.1103/RevModPhys.83.705}{
\newblock Rev. Mod. Phys. {\bf 83}, 705 (2011)}.}

\bibitem{2013_JESRP_Schmitt}
{T.~Schmitt \textit{et~al}.,
  \href{http://www.sciencedirect.com/science/article/pii/S0368204813000030}{
\newblock J. Electron Spectrosc. Relat. Phenom. {\bf 188}, 38  (2013)}.}

\bibitem{2012_PRB_Basak}
{S.~Basak \textit{et~al}.,
  \href{http://link.aps.org/doi/10.1103/PhysRevB.85.075104}{
\newblock Phys. Rev. B {\bf 85}, 075104 (2012)}.}

\bibitem{2005_NP_Abbamonte}
{P.~Abbamonte \textit{et~al}., \href{http://dx.doi.org/10.1038/nphys178}{
\newblock Nat. Phys. {\bf 1}, 155 (2005)}.}

\bibitem{2011_PRB_Fink}
{J.~Fink \textit{et~al}.,
  \href{http://link.aps.org/doi/10.1103/PhysRevB.83.092503}{
\newblock Phys. Rev. B {\bf 83}, 092503 (2011)}.}

\bibitem{2012_S_Ghiringhelli}
{G.~Ghiringhelli \textit{et~al}.,
  \href{http://www.sciencemag.org/content/337/6096/821.abstract}{
\newblock Science {\bf 337}, 821 (2012)}.}

\bibitem{2014_NP_LeTacon}
{M.~Le~Tacon \textit{et~al}., \href{http://dx.doi.org/10.1038/nphys2805}{
\newblock Nat. Phys. {\bf 10}, 52 (2014)}.}

\bibitem{ChangNP2012}
{J.~Chang \textit{et~al}., \href{http://dx.doi.org/10.1038/nphys2456}{
\newblock Nat. Phys. {\bf 8}, 871 (2012)}.}

\bibitem{2006_AP_Eschrig}
{M.~Eschrig, \href{http://dx.doi.org/10.1080/00018730600645636}{
\newblock Adv. Phys. {\bf 55}, 47 (2006)}.}

\bibitem{2004_N_Tranquada}
{J.~M. Tranquada \textit{et~al}., \href{http://dx.doi.org/10.1038/nature02574}{
\newblock Nature {\bf 429}, 534 (2004)}.}

\bibitem{2004_PRL_Pailhes}
{S.~Pailh\`es \textit{et~al}.,
  \href{http://link.aps.org/doi/10.1103/PhysRevLett.93.167001}{
\newblock Phys. Rev. Lett. {\bf 93}, 167001 (2004)}.}

\bibitem{2001_PRB_Dai}
{P.~Dai \textit{et~al}.,
  \href{http://link.aps.org/doi/10.1103/PhysRevB.63.054525}{
\newblock Phys. Rev. B {\bf 63}, 054525 (2001)}.}

\bibitem{1999_S_Dai}
{P.~Dai \textit{et~al}.,
  \href{http://www.sciencemag.org/content/284/5418/1344.abstract}{
\newblock Science {\bf 284}, 1344 (1999)}.}

\bibitem{2000_PRB_Fong}
{H.~Fong \textit{et~al}.,
  \href{http://link.aps.org/doi/10.1103/PhysRevB.61.14773}{
\newblock Phys. Rev. B {\bf 61}, 14773 (2000)}.}

\bibitem{1997_PRL_Tranquada}
{J.~Tranquada \textit{et~al}.,
  \href{http://link.aps.org/doi/10.1103/PhysRevLett.78.338}{
\newblock Phys. Rev. Lett. {\bf 78}, 338 (1997)}.}

\bibitem{2001_PRL_Sidis}
{Y.~Sidis \textit{et~al}.,
  \href{http://link.aps.org/doi/10.1103/PhysRevLett.86.4100}{
\newblock Phys. Rev. Lett. {\bf 86}, 4100 (2001)}.}

\bibitem{2006_PRL_Fauque}
{B.~Fauqu\'e \textit{et~al}.,
  \href{http://link.aps.org/doi/10.1103/PhysRevLett.96.197001}{
\newblock Phys. Rev. Lett. {\bf 96}, 197001 (2006)}.}

\bibitem{2008_N_Li}
{Y.~Li \textit{et~al}., \href{http://dx.doi.org/10.1038/nature07251}{
\newblock Nature {\bf 455}, 372 (2008)}.}

\bibitem{2010_N_Li}
{Y.~Li \textit{et~al}., \href{http://dx.doi.org/10.1038/nature09477}{
\newblock Nature {\bf 468}, 283 (2010)}.}

\bibitem{2014_PRB_Mangin-Thro}
{L.~Mangin-Thro \textit{et~al}.,
  \href{http://link.aps.org/doi/10.1103/PhysRevB.89.094523}{
\newblock Phys. Rev. B {\bf 89}, 094523 (2014)}.}

\bibitem{2000_PRL_Krasnov}
{V.~M. Krasnov \textit{et~al}.,
  \href{http://link.aps.org/doi/10.1103/PhysRevLett.84.5860}{
\newblock Phys. Rev. Lett. {\bf 84}, 5860 (2000)}.}

\bibitem{2009_PRB_Krasnov}
{V.~Krasnov, \href{http://link.aps.org/doi/10.1103/PhysRevB.79.214510}{
\newblock Phys. Rev. B {\bf 79}, 214510 (2009)}.}

\bibitem{2000_RPP_Kashiwaya}
{S.~Kashiwaya and Y.~Tanaka,
  \href{http://stacks.iop.org/0034-4885/63/i=10/a=202}{
\newblock Rep. Prog. Phys. {\bf 63}, 1641 (2000)}.}

\bibitem{2005_RMP_Deutscher}
{G.~Deutscher, \href{http://link.aps.org/doi/10.1103/RevModPhys.77.109}{
\newblock Rev. Mod. Phys. {\bf 77}, 109 (2005)}.}

\bibitem{2001_PRL_Zasadzinski}
{J.~Zasadzinski \textit{et~al}.,
  \href{http://link.aps.org/doi/10.1103/PhysRevLett.87.067005}{
\newblock Phys. Rev. Lett. {\bf 87}, 067005 (2001)}.}

\bibitem{1999_PRL_Miyakawa}
{N.~Miyakawa \textit{et~al}.,
  \href{http://link.aps.org/doi/10.1103/PhysRevLett.83.1018}{
\newblock Phys. Rev. Lett. {\bf 83}, 1018 (1999)}.}

\bibitem{2007_RMP_Fischer}
{{\O}.~Fischer \textit{et~al}.,
  \href{http://link.aps.org/doi/10.1103/RevModPhys.79.353}{
\newblock Rev. Mod. Phys. {\bf 79}, 353 (2007)}.}

\bibitem{2002_S_Hoffman_b}
{J.~E. Hoffman \textit{et~al}.,
  \href{http://www.sciencemag.org/content/297/5584/1148.abstract}{
\newblock Science {\bf 297}, 1148 (2002)}.}

\bibitem{2003_N_McElroy}
{K.~McElroy \textit{et~al}., \href{http://dx.doi.org/10.1038/nature01496}{
\newblock Nature {\bf 422}, 592 (2003)}.}

\bibitem{2003_PRB_Wang}
{Q.-H. Wang and D.-H. Lee,
  \href{http://link.aps.org/doi/10.1103/PhysRevB.67.020511}{
\newblock Phys. Rev. B {\bf 67}, 020511 (2003)}.}

\bibitem{2004_PRB_Markiewicz}
{R.~Markiewicz, \href{http://link.aps.org/doi/10.1103/PhysRevB.69.214517}{
\newblock Phys. Rev. B {\bf 69}, 214517 (2004)}.}

\bibitem{2007_JESRP_Kordyuk}
{A.~A. Kordyuk \textit{et~al}.,
  \href{http://www.imp.kiev.ua/~kord/papers/box/2007_JESRP_Kordyuk.pdf}{
\newblock J. Electron Spectrosc. Relat. Phenom. {\bf 159}, 91 (2007)}.}

\bibitem{2002_PRB_Krasnov}
{V.~Krasnov, \href{http://link.aps.org/doi/10.1103/PhysRevB.65.140504}{
\newblock Phys. Rev. B {\bf 65}, 140504 (2002)}.}

\bibitem{2003_PRL_Yurgens}
{A.~Yurgens \textit{et~al}.,
  \href{http://link.aps.org/doi/10.1103/PhysRevLett.90.147005}{
\newblock Phys. Rev. Lett. {\bf 90}, 147005 (2003)}.}

\bibitem{Tersoff}
{J.~Tersoff and D.~R. Hamann,
\newblock Theory of the scanning tunneling microscope
\newblock in {\em Perspectives in Condensed Matter Physics}, Springer
  Netherlands, 1993.}

\bibitem{2002_N_Lang}
{K.~M. Lang \textit{et~al}., \href{http://dx.doi.org/10.1038/415412a}{
\newblock Nature {\bf 415}, 412 (2002)}.}

\bibitem{2007_NP_Boyer}
{M.~C. Boyer \textit{et~al}., \href{http://dx.doi.org/10.1038/nphys725}{
\newblock Nat. Phys. {\bf 3}, 802 (2007)}.}

\bibitem{2003_PRB_Howald}
{C.~Howald \textit{et~al}.,
  \href{http://link.aps.org/doi/10.1103/PhysRevB.67.014533}{
\newblock Phys. Rev. B {\bf 67}, 014533 (2003)}.}

\bibitem{2004_N_Hanaguri}
{T.~Hanaguri \textit{et~al}., \href{http://dx.doi.org/10.1038/nature02861}{
\newblock Nature {\bf 430}, 1001 (2004)}.}

\bibitem{2007_S_Kohsaka}
{Y.~Kohsaka \textit{et~al}.,
  \href{http://www.sciencemag.org/content/315/5817/1380.abstract}{
\newblock Science {\bf 315}, 1380 (2007)}.}

\bibitem{1996_PRL_Marshall}
{D.~Marshall \textit{et~al}.,
  \href{http://link.aps.org/doi/10.1103/PhysRevLett.76.4841}{
\newblock Phys. Rev. Lett. {\bf 76}, 4841 (1996)}.}

\bibitem{2003_PRB_Kordyuk}
{A.~A. Kordyuk \textit{et~al}.,
  \href{http://www.imp.kiev.ua/~kord/papers/box/2003_PRB_Kordyuk.pdf}{
\newblock Phys. Rev. B {\bf 67}, 064504 (2003)}.}

\bibitem{2004_N_Borisenko}
{S.~V. Borisenko \textit{et~al}.,
  \href{http://www.imp.kiev.ua/~kord/papers/box/2004_N_Borisenko.pdf}{
\newblock Nature {\bf 431},  (2004)}.}

\bibitem{2001_PRB_Borisenko}
{S.~V. Borisenko \textit{et~al}.,
  \href{http://www.imp.kiev.ua/~kord/papers/box/2001_PRB_Borisenko.pdf}{
\newblock Phys. Rev. B {\bf 64}, 094513 (2001)}.}

\bibitem{2004_PC_Borisenko}
{S.~V. Borisenko \textit{et~al}.,
  \href{http://www.imp.kiev.ua/~kord/papers/box/2004_PC_Borisenko.pdf}{
\newblock Physica C {\bf 417}, 1 (2004)}.}

\bibitem{2002_PRL_Kordyuk}
{A.~A. Kordyuk \textit{et~al}.,
  \href{http://www.imp.kiev.ua/~kord/papers/box/2002_PRL_Kordyuk.pdf}{
\newblock Phys. Rev. Lett. {\bf 89}, 077003 (2002)}.}

\bibitem{EvtushinskyPRB2009}
{D.~V. Evtushinsky \textit{et~al}.,
  \href{http://www.imp.kiev.ua/~kord/papers/box/2009_PRB_Evtushinsky.pdf}{
\newblock Phys. Rev. B {\bf 79}, 054517 (2009)}.}

\bibitem{1978_PRL_Dynes}
{R.~Dynes, V.~Narayanamurti, and J.~Garno,
  \href{http://link.aps.org/doi/10.1103/PhysRevLett.41.1509}{
\newblock Phys. Rev. Lett. {\bf 41}, 1509 (1978)}.}

\bibitem{2001_PRL_Tsuda}
{S.~Tsuda \textit{et~al}.,
  \href{http://link.aps.org/doi/10.1103/PhysRevLett.87.177006}{
\newblock Phys. Rev. Lett. {\bf 87}, 177006 (2001)}.}

\bibitem{2007_N_Lee}
{W.~S. Lee \textit{et~al}., \href{http://dx.doi.org/10.1038/nature06219}{
\newblock Nature {\bf 450}, 81 (2007)}.}

\bibitem{2006_NP_Kanigel}
{A.~Kanigel \textit{et~al}., \href{http://dx.doi.org/10.1038/nphys334}{
\newblock Nat. Phys. {\bf 2}, 447 (2006)}.}

\bibitem{2007_PRB_Chubukov}
{A.~Chubukov \textit{et~al}.,
  \href{http://link.aps.org/doi/10.1103/PhysRevB.76.180501}{
\newblock Phys. Rev. B {\bf 76}, 180501 (2007)}.}

\bibitem{2007_PRB_Norman}
{M.~Norman \textit{et~al}.,
  \href{http://link.aps.org/doi/10.1103/PhysRevB.76.174501}{
\newblock Phys. Rev. B {\bf 76}, 174501 (2007)}.}

\bibitem{2009_N_Meng}
{J.~Meng \textit{et~al}., \href{http://dx.doi.org/10.1038/nature08521}{
\newblock Nature {\bf 462}, 335 (2009)}.}

\bibitem{2011_PRL_Yang}
{H.-B. Yang \textit{et~al}.,
  \href{http://link.aps.org/doi/10.1103/PhysRevLett.107.047003}{
\newblock Phys. Rev. Lett. {\bf 107}, 047003 (2011)}.}

\bibitem{2006_S_Tanaka}
{K.~Tanaka \textit{et~al}.,
  \href{http://www.sciencemag.org/content/314/5807/1910.abstract}{
\newblock Science {\bf 314}, 1910 (2006)}.}

\bibitem{2007_PRL_Kondo}
{T.~Kondo \textit{et~al}.,
  \href{http://link.aps.org/doi/10.1103/PhysRevLett.98.267004}{
\newblock Phys. Rev. Lett. {\bf 98}, 267004 (2007)}.}

\bibitem{KordyukPRB2009}
{A.~A. Kordyuk \textit{et~al}.,
  \href{http://www.imp.kiev.ua/~kord/papers/box/2009_PRB_Kordyuk.pdf}{
\newblock Phys. Rev. B {\bf 79}, 020504 (2009)}.}

\bibitem{2009_N_Kondo}
{T.~Kondo \textit{et~al}., \href{http://dx.doi.org/10.1038/nature07644}{
\newblock Nature {\bf 457}, 296 (2009)}.}

\bibitem{2010_NP_Hashimoto}
{M.~Hashimoto \textit{et~al}., \href{http://dx.doi.org/10.1038/nphys1632}{
\newblock Nat. Phys. {\bf 6}, 414 (2010)}.}

\bibitem{2014_NP_Hashimoto}
{M.~Hashimoto \textit{et~al}., \href{http://dx.doi.org/10.1038/nphys3009}{
\newblock Nat. Phys. {\bf 10}, 483 (2014)}.}

\bibitem{2014_xxx_Kaminski}
{A.~{Kaminski} \textit{et~al}., \href{http://arxiv.org/abs/arXiv:1403.0492}{
\newblock arXiv:1403.0492v1  (2014)}.}

\bibitem{2006_S_Valla}
{T.~Valla \textit{et~al}.,
  \href{http://www.sciencemag.org/content/314/5807/1914.abstract}{
\newblock Science {\bf 314}, 1914 (2006)}.}

\bibitem{2003_PRL_Borisenko}
{S.~V. Borisenko \textit{et~al}.,
  \href{http://www.imp.kiev.ua/~kord/papers/box/2003_PRL_Borisenko.pdf}{
\newblock Phys. Rev. Lett. {\bf 90}, 207001 (2003)}.}

\bibitem{1991_PRL_Dessau}
{D.~Dessau \textit{et~al}.,
  \href{http://link.aps.org/doi/10.1103/PhysRevLett.66.2160}{
\newblock Phys. Rev. Lett. {\bf 66}, 2160 (1991)}.}

\bibitem{2001_PRB_Seibold}
{G.~Seibold and M.~Grilli,
  \href{http://link.aps.org/doi/10.1103/PhysRevB.63.224505}{
\newblock Phys. Rev. B {\bf 63}, 224505 (2001)}.}

\bibitem{2003_PRB_Hoogenboom}
{B.~W. Hoogenboom \textit{et~al}.,
  \href{http://link.aps.org/abstract/PRB/v67/e224502}{
\newblock Phys. Rev. B {\bf 67}, 224502 (2003)}.}

\bibitem{2008_JPCM_Ekino}
{T.~Ekino \textit{et~al}.,
  \href{http://stacks.iop.org/0953-8984/20/i=42/a=425218}{
\newblock J. Phys.: Condens. Matter {\bf 20}, 425218 (2008)}.}

\bibitem{2014_PC_Gabovich}
{A.~M. Gabovich and A.~I. Voitenko,
  \href{http://www.sciencedirect.com/science/article/pii/S0921453414001634}{
\newblock Physica C {\bf 503}, 7  (2014)}.}

\bibitem{1983_JL_Artemenko}
{S.~N. Artemenko and A.~F. Volkov,
  \href{http://www.jetpletters.ac.ru/ps/1494/article_22809.pdf}{
\newblock JETP Letters {\bf 37}, 368 (1983)}.}

\bibitem{2010_NP_Fournier}
{D.~Fournier \textit{et~al}., \href{http://dx.doi.org/10.1038/nphys1763}{
\newblock Nat. Phys. {\bf 6}, 905 (2010)}.}

\bibitem{KordyukEPJ2010}
{A.~A. Kordyuk \textit{et~al}.,
  \href{http://www.imp.kiev.ua/~kord/papers/box/2010_EPJST_Kordyuk.pdf}{
\newblock Eur. Phys. J. Special Topics {\bf 188}, 153 (2010)}.}

\bibitem{2000_S_Voit}
{J.~Voit \textit{et~al}.,
  \href{http://www.sciencemag.org/content/290/5491/501.abstract}{
\newblock Science {\bf 290}, 501 (2000)}.}

\bibitem{2000b_PRL_Valla}
{T.~Valla \textit{et~al}.,
  \href{http://link.aps.org/doi/10.1103/PhysRevLett.85.4759}{
\newblock Phys. Rev. Lett. {\bf 85}, 4759 (2000)}.}

\bibitem{2004_PRL_Valla}
{T.~Valla \textit{et~al}.,
  \href{http://link.aps.org/doi/10.1103/PhysRevLett.92.086401}{
\newblock Phys. Rev. Lett. {\bf 92}, 086401 (2004)}.}

\bibitem{2007_NP_Kiss}
{T.~Kiss \textit{et~al}., \href{http://dx.doi.org/10.1038/nphys699}{
\newblock Nat. Phys. {\bf 3}, 720 (2007)}.}

\bibitem{2012_PRB_Rahn}
{D.~J. Rahn \textit{et~al}.,
  \href{http://link.aps.org/doi/10.1103/PhysRevB.85.224532}{
\newblock Phys. Rev. B {\bf 85}, 224532 (2012)}.}

\bibitem{2008_PRL_Evtushinsky}
{D.~V. Evtushinsky \textit{et~al}.,
  \href{http://www.imp.kiev.ua/~kord/papers/box/2008_PRL_Evtushinsky.pdf}{
\newblock Phys. Rev. Lett. {\bf 100}, 236402 (2008)}.}

\bibitem{2008_NJP_Inosov}
{D.~S. Inosov \textit{et~al}.,
  \href{http://www.imp.kiev.ua/~kord/papers/box/2008_NJP_Inosov.pdf}{
\newblock New J. Phys. {\bf 10}, 125027 (2008)}.}

\bibitem{2009_PRB_Inosov}
{D.~S. Inosov \textit{et~al}.,
  \href{http://www.imp.kiev.ua/~kord/papers/box/2009_PRB_Inosov.pdf}{
\newblock Phys. Rev. B {\bf 79}, 125112 (2009)}.}

\bibitem{2008_PRB_Johannes}
{M.~D. Johannes and I.~I. Mazin,
  \href{http://link.aps.org/doi/10.1103/PhysRevB.77.165135}{
\newblock Phys. Rev. B {\bf 77}, 165135 (2008)}.}

\bibitem{1980_PRL_Fleming}
{R.~M. Fleming \textit{et~al}.,
  \href{http://link.aps.org/doi/10.1103/PhysRevLett.45.576}{
\newblock Phys. Rev. Lett. {\bf 45}, 576 (1980)}.}

\bibitem{2011_PRB_Leininger}
{P.~Leininger \textit{et~al}.,
  \href{http://link.aps.org/doi/10.1103/PhysRevB.83.233101}{
\newblock Phys. Rev. B {\bf 83}, 233101 (2011)}.}

\bibitem{2013_PNAS_Soumyanarayanan}
{A.~Soumyanarayanan \textit{et~al}.,
  \href{http://www.pnas.org/content/110/5/1623.abstract}{
\newblock Proc. Natl. Acad. Sci. U.S.A. {\bf 110}, 1623 (2013)}.}

\bibitem{2009_PRB_Grilli}
{M.~Grilli \textit{et~al}.,
  \href{http://link.aps.org/doi/10.1103/PhysRevB.79.125111}{
\newblock Phys. Rev. B {\bf 79}, 125111 (2009)}.}

\bibitem{2009_PRL_Seibold}
{G.~Seibold, M.~Grilli, and J.~Lorenzana,
  \href{http://link.aps.org/doi/10.1103/PhysRevLett.103.217005}{
\newblock Phys. Rev. Lett. {\bf 103}, 217005 (2009)}.}

\bibitem{2013_LTP_Gabovich}
{A.~M. Gabovich and A.~I. Voitenko,
  \href{http://scitation.aip.org/content/aip/journal/ltp/39/3/10.1063/1.479520%
2}{
\newblock Low Temp. Phys. {\bf 39}, 232 (2013)}.}

\bibitem{2015_xxx_Chowdhury}
{D.~{Chowdhury} and S.~{Sachdev}, \href{http://arxiv.org/abs/1501.00002}{
\newblock arXiv:1501.00002  (2015)}.}

\bibitem{2005_PRL_Komiya}
{S.~Komiya \textit{et~al}.,
  \href{http://link.aps.org/doi/10.1103/PhysRevLett.94.207004}{
\newblock Phys. Rev. Lett. {\bf 94}, 207004 (2005)}.}

\bibitem{2004_PRB_Chen}
{H.-D. Chen \textit{et~al}.,
  \href{http://link.aps.org/doi/10.1103/PhysRevB.70.024516}{
\newblock Phys. Rev. B {\bf 70}, 024516 (2004)}.}

\bibitem{2004_PRL_Tesanovic}
{Z.~Te\v{s}anovi\'{c},
  \href{http://link.aps.org/doi/10.1103/PhysRevLett.93.217004}{
\newblock Phys. Rev. Lett. {\bf 93}, 217004 (2004)}.}

\bibitem{2011_PRB_Caprara}
{S.~Caprara \textit{et~al}.,
  \href{http://link.aps.org/doi/10.1103/PhysRevB.84.054508}{
\newblock Phys. Rev. B {\bf 84}, 054508 (2011)}.}

\bibitem{RosenNC2013}
{J.~A. Rosen \textit{et~al}., \href{http://dx.doi.org/10.1038/ncomms2977}{
\newblock Nat. Commun. {\bf 4}, 1977 (2013)}.}

\bibitem{FradkinNP2012}
{E.~Fradkin and S.~A. Kivelson, \href{http://dx.doi.org/10.1038/nphys2498}{
\newblock Nat. Phys. {\bf 8}, 864 (2012)}.}

\bibitem{2007_PRL_Cercellier}
{H.~Cercellier \textit{et~al}.,
  \href{http://link.aps.org/doi/10.1103/PhysRevLett.99.146403}{
\newblock Phys. Rev. Lett. {\bf 99}, 146403 (2007)}.}

\bibitem{2014_NJP_Qian}
{T.~Qian \textit{et~al}.,
  \href{http://stacks.iop.org/1367-2630/16/i=12/a=123038}{
\newblock New J. Phys. {\bf 16}, 123038 (2014)}.}

\bibitem{2003_PRB_Terashima}
{K.~Terashima \textit{et~al}.,
  \href{http://link.aps.org/doi/10.1103/PhysRevB.68.155108}{
\newblock Phys. Rev. B {\bf 68}, 155108 (2003)}.}

\bibitem{2004_PRB_Bovet}
{M.~Bovet \textit{et~al}.,
  \href{http://link.aps.org/doi/10.1103/PhysRevB.69.125117}{
\newblock Phys. Rev. B {\bf 69}, 125117 (2004)}.}

\bibitem{2008_NM_Sipos}
{B.~Sipos \textit{et~al}., \href{http://dx.doi.org/10.1038/nmat2318}{
\newblock Nat. Mater. {\bf 7}, 960 (2008)}.}

\bibitem{2012_PRL_Ang}
{R.~Ang \textit{et~al}.,
  \href{http://link.aps.org/doi/10.1103/PhysRevLett.109.176403}{
\newblock Phys. Rev. Lett. {\bf 109}, 176403 (2012)}.}

\bibitem{2014_PRL_Lahoud}
{E.~Lahoud \textit{et~al}.,
  \href{http://link.aps.org/doi/10.1103/PhysRevLett.112.206402}{
\newblock Phys. Rev. Lett. {\bf 112}, 206402 (2014)}.}

\bibitem{2003_PRB_Aiura}
{Y.~Aiura \textit{et~al}.,
  \href{http://link.aps.org/doi/10.1103/PhysRevB.68.073408}{
\newblock Phys. Rev. B {\bf 68}, 073408 (2003)}.}

\bibitem{2012_NC_Hellmann}
{S.~Hellmann \textit{et~al}., \href{http://dx.doi.org/10.1038/ncomms2078}{
\newblock Nat. Commun. {\bf 3}, 1069 (2012)}.}

\bibitem{1998_S_Shen}
{Z.-X. Shen \textit{et~al}.,
  \href{http://www.sciencemag.org/content/280/5361/259.abstract}{
\newblock Science {\bf 280}, 259 (1998)}.}

\bibitem{HaugNJP2010}
{D.~Haug \textit{et~al}.,
  \href{http://stacks.iop.org/1367-2630/12/i=10/a=105006}{
\newblock New J. Phys. {\bf 12}, 105006 (2010)}.}

\bibitem{2008_N_Yang}
{H.-B. Yang \textit{et~al}., \href{http://dx.doi.org/10.1038/nature07400}{
\newblock Nature {\bf 456}, 77 (2008)}.}

\bibitem{2010_NP_Weber}
{C.~Weber, K.~Haule, and G.~Kotliar,
  \href{http://dx.doi.org/10.1038/nphys1706}{
\newblock Nat. Phys. {\bf 6}, 574 (2010)}.}

\bibitem{2008_PRB_Krockenberger}
{Y.~Krockenberger \textit{et~al}.,
  \href{http://link.aps.org/doi/10.1103/PhysRevB.77.060505}{
\newblock Phys. Rev. B {\bf 77}, 060505 (2008)}.}

\bibitem{2003_N_Alff}
{L.~Alff \textit{et~al}., \href{http://dx.doi.org/10.1038/nature01488}{
\newblock Nature {\bf 422}, 698 (2003)}.}

\bibitem{2014_NP_Lee}
{W.~S. Lee \textit{et~al}., \href{http://dx.doi.org/10.1038/nphys3117}{
\newblock Nat. Phys. {\bf 10}, 883 (2014)}.}

\bibitem{2005_PRL_Matsui}
{H.~Matsui \textit{et~al}.,
  \href{http://link.aps.org/doi/10.1103/PhysRevLett.94.047005}{
\newblock Phys. Rev. Lett. {\bf 94}, 047005 (2005)}.}

\bibitem{2007_PRB_Park}
{S.~R. Park \textit{et~al}.,
  \href{http://link.aps.org/doi/10.1103/PhysRevB.75.060501}{
\newblock Phys. Rev. B {\bf 75}, 060501 (2007)}.}

\bibitem{2009_PRB_Nekrasov}
{I.~A. Nekrasov \textit{et~al}.,
  \href{http://link.aps.org/doi/10.1103/PhysRevB.80.140510}{
\newblock Phys. Rev. B {\bf 80}, 140510 (2009)}.}

\bibitem{2013_PRL_Sakai}
{S.~Sakai \textit{et~al}.,
  \href{http://link.aps.org/doi/10.1103/PhysRevLett.111.107001}{
\newblock Phys. Rev. Lett. {\bf 111}, 107001 (2013)}.}

\bibitem{SatoJPSJ2008}
{T.~Sato \textit{et~al}., \href{http://jpsj.ipap.jp/link?JPSJ/77/063708/}{
\newblock J. Phys. Soc. Jpn. {\bf 77}, 063708 (2008)}.}

\bibitem{HaiYun2008}
{H.-Y. Liu \textit{et~al}.,
  \href{http://159.226.36.45/Jwk_cpl/EN/abstract/article_42665.shtml}{
\newblock Chin. Phys. Lett. {\bf 25}, 3761 (2008)}.}

\bibitem{2008_JPSJ_Ishida}
{Y.~Ishida \textit{et~al}., \href{http://dx.doi.org/10.1143/JPSJS.77SC.61}{
\newblock J. Phys. Soc. Japan {\bf 77}, 61 (2008)}.}

\bibitem{XuNC2011}
{Y.-M. Xu \textit{et~al}., \href{http://dx.doi.org/10.1038/ncomms1394}{
\newblock Nat. Commun. {\bf 2}, 392 (2011)}.}

\bibitem{RichardPRL2009}
{P.~Richard \textit{et~al}.,
  \href{http://link.aps.org/doi/10.1103/PhysRevLett.102.047003}{
\newblock Phys. Rev. Lett. {\bf 102}, 047003 (2009)}.}

\bibitem{Evtushinsky2011}
{D.~V. {Evtushinsky} \textit{et~al}., \href{http://arxiv.org/abs/1106.4584}{
\newblock arXiv:1106.4584v1  (2011)}.}

\bibitem{ShimojimaSci2011}
{T.~Shimojima \textit{et~al}.,
  \href{http://www.sciencemag.org/content/332/6029/564.abstract}{
\newblock Science {\bf 332}, 564 (2011)}.}

\bibitem{BorisenkoSym2012}
{S.~V. Borisenko \textit{et~al}.,
  \href{http://www.imp.kiev.ua/~kord/papers/box/2012_S_Borisenko.pdf}{
\newblock Symmetry {\bf 4}, 251 (2012)}.}

\bibitem{ZhangNP2012}
{Y.~Zhang \textit{et~al}., \href{http://dx.doi.org/10.1038/nphys2248}{
\newblock Nat. Phys. {\bf 8}, 371 (2012)}.}

\bibitem{EvtushinskyPRB2014}
{D.~V. Evtushinsky \textit{et~al}.,
  \href{http://www.imp.kiev.ua/~kord/papers/box/2014_PRB_Evtushinsky.pdf}{
\newblock Phys. Rev. B {\bf 89}, 064514 (2014)}.}

\bibitem{YinPhC2009}
{Y.~Yin \textit{et~al}.,
  \href{http://www.sciencedirect.com/science/article/pii/S0921453409000914}{
\newblock Physica C {\bf 469}, 535  (2009)}.}

\bibitem{MasseeEPL2010}
{F.~Massee \textit{et~al}.,
  \href{http://stacks.iop.org/0295-5075/92/i=5/a=57012}{
\newblock EPL {\bf 92}, 57012 (2010)}.}

\bibitem{2011_AdP_Andersen}
{O.~K. Andersen and L.~Boeri, \href{http://dx.doi.org/10.1002/andp.201000149}{
\newblock Annalen der Physik {\bf 523}, 8 (2011)}.}

\bibitem{IshidaJPSJ2009}
{K.~Ishida, Y.~Nakai, and H.~Hosono,
  \href{http://jpsj.ipap.jp/link?JPSJ/78/062001/}{
\newblock J. Phys. Soc. Jpn. {\bf 78}, 062001 (2009)}.}

\bibitem{2011_PRB_Baek}
{S.-H. Baek \textit{et~al}.,
  \href{http://link.aps.org/doi/10.1103/PhysRevB.84.094510}{
\newblock Phys. Rev. B {\bf 84}, 094510 (2011)}.}

\bibitem{TanatarPRB2010}
{M.~A. Tanatar \textit{et~al}.,
  \href{http://link.aps.org/doi/10.1103/PhysRevB.82.134528}{
\newblock Phys. Rev. B {\bf 82}, 134528 (2010)}.}

\bibitem{SolovjevLTP2009}
{A.~L. Solovjev \textit{et~al}., \href{http://link.aip.org/link/?LTP/35/826/1}{
\newblock Low Temp. Phys. {\bf 35}, 826 (2009)}.}

\bibitem{SolovjovLTP2011}
{A.~L. Solovjov \textit{et~al}.,
  \href{http://scitation.aip.org/content/aip/journal/ltp/37/10/10.1063/1.36700%
27}{
\newblock Low Temp. Phys. {\bf 37}, 840 (2011)}.}

\bibitem{2010_PRL_Sheet}
{G.~Sheet \textit{et~al}.,
  \href{http://link.aps.org/doi/10.1103/PhysRevLett.105.167003}{
\newblock Phys. Rev. Lett. {\bf 105}, 167003 (2010)}.}

\bibitem{2014_PRB_Moon}
{S.~J. Moon \textit{et~al}.,
  \href{http://link.aps.org/doi/10.1103/PhysRevB.90.014503}{
\newblock Phys. Rev. B {\bf 90}, 014503 (2014)}.}

\bibitem{2012_PRL_Wen}
{Y.-C. Wen \textit{et~al}.,
  \href{http://link.aps.org/doi/10.1103/PhysRevLett.108.267002}{
\newblock Phys. Rev. Lett. {\bf 108}, 267002 (2012)}.}

\bibitem{2014_PRB_Lin}
{K.-H. Lin \textit{et~al}.,
  \href{http://link.aps.org/doi/10.1103/PhysRevB.90.174502}{
\newblock Phys. Rev. B {\bf 90}, 174502 (2014)}.}

\bibitem{2012_N_Kasahara}
{S.~Kasahara \textit{et~al}., \href{http://dx.doi.org/10.1038/nature11178}{
\newblock Nature {\bf 486}, 382 (2012)}.}

\bibitem{2014_PRB_Shimojima}
{T.~Shimojima \textit{et~al}.,
  \href{http://link.aps.org/doi/10.1103/PhysRevB.89.045101}{
\newblock Phys. Rev. B {\bf 89}, 045101 (2014)}.}

\bibitem{2012_PRB_Stojchevska}
{L.~Stojchevska \textit{et~al}.,
  \href{http://link.aps.org/doi/10.1103/PhysRevB.86.024519}{
\newblock Phys. Rev. B {\bf 86}, 024519 (2012)}.}

\bibitem{2013_PRB_Mertelj}
{T.~Mertelj \textit{et~al}.,
  \href{http://link.aps.org/doi/10.1103/PhysRevB.87.174525}{
\newblock Phys. Rev. B {\bf 87}, 174525 (2013)}.}

\bibitem{2014_xxx_Surmach}
{M.~A. {Surmach} \textit{et~al}., \href{http://arxiv.org/abs/1411.7858}{
\newblock arXiv:1411.7858v1  (2014)}.}

\bibitem{KopaevZETF1970}
{Y.~V. Kopaev,
\newblock Zh. Eksp. Teor. Fiz. {\bf 58}, 1012 (1970)
\newblock [Sov. Phys. JETP. 31544 (1970)].}

\bibitem{KopaevPLA1987}
{Y.~V. Kopaev and A.~I. Rusinov,
  \href{http://www.sciencedirect.com/science/article/pii/0375960187905330}{
\newblock Phys. Lett. A {\bf 121}, 300  (1987)}.}

\bibitem{1988_PC_Machida}
{K.~Machida,
  \href{http://www.sciencedirect.com/science/article/pii/0921453488908222}{
\newblock Physica C {\bf 156}, 276 (1988)}.}

\bibitem{2014_NP_Ebrahimnejad}
{H.~Ebrahimnejad, G.~A. Sawatzky, and M.~Berciu,
  \href{http://dx.doi.org/10.1038/nphys3130}{
\newblock Nat. Phys. {\bf 10}, 951 (2014)}.}

\bibitem{2013_JSNM_Kordyuk}
{A.~A. Kordyuk \textit{et~al}.,
  \href{http://www.imp.kiev.ua/~kord/papers/box/2013_JSNM_Kordyuk.pdf}{
\newblock J. Supercond. Nov. Magn. {\bf 26}, 2837 (2013)}.}

\bibitem{2013_PRB_Evtushinsky}
{D.~V. Evtushinsky \textit{et~al}.,
  \href{http://www.imp.kiev.ua/~kord/papers/box/2013_PRB_Evtushinsky.pdf}{
\newblock Phys. Rev. B {\bf 87}, 094501 (2013)}.}

\bibitem{2013_PRB_Maletz}
{J.~Maletz \textit{et~al}.,
  \href{http://link.aps.org/doi/10.1103/PhysRevB.88.134501}{
\newblock Phys. Rev. B {\bf 88}, 134501 (2013)}.}

\bibitem{2013_PRB_Thirupathaiah}
{S.~Thirupathaiah \textit{et~al}.,
  \href{http://link.aps.org/doi/10.1103/PhysRevB.88.140505}{
\newblock Phys. Rev. B {\bf 88}, 140505 (2013)}.}

\bibitem{2014_PRB_Terashima}
{K.~Terashima \textit{et~al}.,
  \href{http://link.aps.org/doi/10.1103/PhysRevB.90.220512}{
\newblock Phys. Rev. B {\bf 90}, 220512 (2014)}.}

\bibitem{InnocentiPRB2010}
{D.~Innocenti \textit{et~al}.,
  \href{http://link.aps.org/doi/10.1103/PhysRevB.82.184528}{
\newblock Phys. Rev. B {\bf 82}, 184528 (2010)}.}

\bibitem{InnocentiSUST2011}
{D.~Innocenti \textit{et~al}.,
  \href{http://stacks.iop.org/0953-2048/24/i=1/a=015012}{
\newblock Supercond. Sci. Technol. {\bf 24}, 015012 (2011)}.}

\bibitem{2013_NP_Bianconi}
{A.~Bianconi, \href{http://dx.doi.org/10.1038/nphys2738}{
\newblock Nat. Phys. {\bf 9}, 536 (2013)}.}

\bibitem{2006_JPSJ_Yamaji}
{Y.~Yamaji, T.~Misawa, and M.~Imada,
  \href{http://dx.doi.org/10.1143/JPSJ.75.094719}{
\newblock J. Phys. Soc. Japan {\bf 75}, 094719 (2006)}.}

\bibitem{2011_NP_Yelland}
{E.~A. Yelland \textit{et~al}., \href{http://dx.doi.org/10.1038/nphys2073}{
\newblock Nat. Phys. {\bf 7}, 890 (2011)}.}

\bibitem{1999_PRL_Onufrieva}
{F.~Onufrieva, P.~Pfeuty, and M.~Kiselev,
  \href{http://link.aps.org/doi/10.1103/PhysRevLett.82.2370}{
\newblock Phys. Rev. Lett. {\bf 82}, 2370 (1999)}.}

\bibitem{2002_PRB_Angilella}
{G.~Angilella, E.~Piegari, and A.~Varlamov,
  \href{http://link.aps.org/doi/10.1103/PhysRevB.66.014501}{
\newblock Phys. Rev. B {\bf 66}, 014501 (2002)}.}

\end{thebibliography}

\end{document}